\documentclass{jpp}
\usepackage{graphicx}

\usepackage[utf8]{inputenc}
\usepackage[T1]{fontenc}
\usepackage{amsmath}

\usepackage[dvipsnames]{xcolor}

\usepackage[colorlinks=true,
            linkcolor=blue,
            citecolor=blue]{hyperref}

\newcommand{\romand}{\textrm{d}}
\newcommand{\energyspecE}{\mathcal{E}}
\newcommand{\energyspecEtilde}{\tilde{\energyspecE}}
\newcommand{\energyspecEtildetwoD}{\energyspecEtilde_{\rm 2D}}
\newcommand{\Bvector}[1]{\boldsymbol{#1}}
\newcommand{\Bvectorgreek}[1]{\boldsymbol{#1}}
\newcommand{\norm}[1]{|#1|}
\newcommand{\unitvector}[1]{\hat{\Bvector{#1}}}
\newcommand{\unitvectorgreek}[1]{\hat{\Bvectorgreek{#1}}}
\newcommand{\sech}{\textrm{sech}}
\newcommand{\totfrac}[2]{\frac{\romand#1}{\romand #2}}
\newcommand{\parfrac}[2]{\frac{\partial #1}{\partial #2}}
\newcommand{\totfracil}[2]{\romand #1 / \romand #2}
\newcommand{\parfracil}[2]{\partial #1 / \partial #2}
\newcommand{\intsingle}[3]{\int_{#2}^{#3} \romand #1 \ }
\newcommand{\intfrac}[4]{\int_{#3}^{#4} \frac{\romand #1}{#2} \ }
\newcommand{\intvol}[1]{\int \romand V \ #1}
\newcommand{\dotprod}{\boldsymbol{\cdot}}
\newcommand{\crossprod}{\boldsymbol{\times}}
\newcommand{\boldnabla}{\nabla}
\newcommand{\gradprp}{\boldnabla_\perp}
\newcommand{\Icplx}{\textrm{i}}
\newcommand{\kvec}{\Bvector{k}}     
\newcommand{\kprl}{k_\|}
\newcommand{\kprltilde}{\tilde{k}_\|}
\newcommand{\kprltildeCB}{\kprltilde^{\rm CB}}

\newcommand{\kprp}{k_\perp}
\newcommand{\kprptilde}{\tilde{k}_\perp}
\newcommand{\ion}[1]{#1_{\rm i}}

\newcommand{\sone}{s_1}
\newcommand{\stwo}{s_2}
\newcommand{\sthree}{s_3}
\newcommand{\sfour}{s_4}
\newcommand{\sCB}{s_{\rm CB}}
\newcommand{\rhoi}{\rho_{\rm i}}
\newcommand{\rhoiscales}{\kprp\rhoi}
\newcommand{\Lprp}{L_\perp}
\newcommand{\Lprl}{L_\|}
\newcommand{\lprp}{\lambda}
\newcommand{\lprl}{\ell_\|}
\newcommand{\betai}{\ion{\beta}}
\newcommand{\Omegai}{\Omega_{\rm i}}
\newcommand{\Omegap}{\Omega_{\rm p}}
\newcommand{\omegaA}{\omega_{\rm A}}
\newcommand{\omeganl}{\omega_{\rm nl}}
\newcommand{\omeganltilde}{\tilde{\omega}_{\rm nl}}
\newcommand{\tautilde}{\tilde{\tau}}
\newcommand{\tauA}{\tau_{\rm A}}
\newcommand{\tauAouter}{\tau_{\textrm{A}0}}
\newcommand{\taunl}{\tau_{\rm nl}}

\newcommand{\vA}{v_{\rm A}}
\newcommand{\vthi}{v_{\rm th, i}}
\newcommand{\vph}{v_{\rm ph}}
\newcommand{\Evec}{\Bvector{E}}
\newcommand{\B}{\Bvector{B}}
\newcommand{\rB}{\Bvector{b}}

\newcommand{\Bmean}{\B_0}
\newcommand{\Bmeanmag}{B_0}
\newcommand{\ExB}{\Evec\crossprod\B}
\newcommand{\Eveck}{\Evec_{\kvec}}
\newcommand{\Eveckcomp}[1]{E_{\kvec #1}}
\newcommand{\Bveck}{\B_{\kvec}}
\newcommand{\Bveckcomp}[1]{B_{\kvec #1}}
\newcommand{\veckcomp}[2]{#1_{\kvec #2}}
\newcommand{\ppos}{\Bvector{x}}
\newcommand{\pvel}{\Bvector{v}}
\newcommand{\vprl}{v_\|}
\newcommand{\vprp}{v_\perp}
\newcommand{\vel}{\Bvector{u}}
\newcommand{\zpm}{\Bvector{z}^\pm}
\newcommand{\zmp}{\Bvector{z}^\mp}
\newcommand{\zp}{\Bvector{z}^+}
\newcommand{\zm}{\Bvector{z}^-}
\newcommand{\deltaz}{\delta z}
\newcommand{\deltazpm}{\deltaz^\pm}
\newcommand{\deltazmp}{\deltaz^\mp}
\newcommand{\deltazp}{\deltaz^+}
\newcommand{\deltazm}{\deltaz^-}
\newcommand{\scapot}{\varphi}
\newcommand{\stocfrac}[1]{\xi_{#1}}
\newcommand{\stocfracth}{\stocfrac{\rho,\textrm{th}}}
\newcommand{\deltau}[1]{\delta u_{#1}}
\newcommand{\cascadeE}{\varepsilon}
\newcommand{\forcep}{\varepsilon^+}
\newcommand{\forcem}{\varepsilon^-}
\newcommand{\forcepm}{\varepsilon^\pm}
\newcommand{\forceratio}{\alpha}
\newcommand{\injectratio}{\alpha_\varepsilon}
\newcommand{\injectimb}{\sigma_\varepsilon}
\newcommand{\crosshel}{\sigma_{\rm c}}
\newcommand{\BesselJ}[1]{\textrm{J}_{#1}}
\newcommand{\vprlhat}{\hat{v}_{\|}}
\newcommand{\vprphat}{\hat{v}_\perp}
\newcommand{\vprltilde}{\tilde{v}_\|}
\newcommand{\vprptilde}{\tilde{v}_\perp}
\newcommand{\Dprpprp}{D_{\perp\perp}}
\newcommand{\Dprpprl}{D_{\perp\|}}
\newcommand{\Dprlprp}{D_{\|\perp}}
\newcommand{\Dprlprl}{D_{\|\|}}
\newcommand{\Qprp}{Q_\perp}

\shorttitle{Perpendicular ion heating by critically balanced turbulence}
\shortauthor{Z.~Johnston and J.~Squire}

\title{Quasi-linear theory of perpendicular ion heating by critically balanced turbulence}

\author{Zade Johnston\aff{1}
\corresp{\email{zade.t.johnston@gmail.com}},
  Jonathan Squire\aff{1}}

\affiliation{\aff{1}Physics Department, University of Otago, Dunedin 9010, New Zealand}

\begin{document}

\maketitle

\begin{abstract}
In collisionless astrophysical plasmas, turbulence mediates the partitioning of free energy among cascade channels and its dissipation into ion and electron heat. The resulting ion heating is often anisotropic, with ions observed to be preferentially heated perpendicular to the local magnetic field; understanding the mechanisms responsible for this heating is a key step in understanding the evolution of such plasmas.
In this paper, we use the framework of quasi-linear theory to compute analytically the heating rates of ions interacting with turbulent, large-scale Alfv\'enic fluctuations.
We show how the imbalance of the turbulence (the difference in energies between Alfv\'enic fluctuations travelling parallel and antiparallel to the magnetic field) modifies the spatiotemporal spectrum of these fluctuations, allowing the heating mechanism to smoothly transition between stochastic heating in balanced turbulence and cyclotron-resonant heating in imbalanced turbulence.
The resultant heating rate is found to have a general form regardless of the level of imbalance, exhibiting a suppression related to the conservation of the ions' magnetic moment at small turbulent amplitudes and recovering previous empirical results in a formal calculation.
The results of this work help to consolidate our qualitative understanding of ion heating within astrophysical plasmas, as well as yielding specific quantitative predictions to analyse simulations and observations.
\end{abstract}

\section{Introduction}

Plasma turbulence is a defining feature of many astrophysical environments, including the solar corona and wind \citep{Cranmer2005-qg,Bruno2013-iw,Cranmer2015-oc,Chen2016-mq,Cranmer2019-om}, the magnetospheres surrounding the Earth and the outer planets \citep{Rakhmanova2021-fk,Saur2021-sz}, the interstellar medium \citep{Lazio2004-fh}, accretion disks surrounding black holes \citep{Quataert1999-rq}, and within the intracluster medium between galaxy clusters \citep{Zhuravleva2014-ck}.
Unlike a collisional fluid, where turbulent energy can be dissipated at small scales due to collisions (e.g., viscosity and resistivity), astrophysical plasmas are generally weakly collisional, which complicates turbulent energy dissipation; instead, there are multiple channels by which energy can cascade to small scales \citep{Schekochihin2009-qo}.
The partitioning of turbulent energy among cascade channels and its ultimate dissipation into ion and electron heating is fundamental in determining the behaviour of such plasmas. A precise understanding of the mechanism that causes this partitioning is therefore essential to model accurately the plasma's thermal structure, radiative efficiency, and large-scale dynamical evolution.

The proximity of the solar corona and wind to Earth makes them convenient laboratories for studying the dynamics of turbulent plasmas over a wide range of spatial and temporal scales \citep{Bruno2013-iw}.
At large scales, turbulence in the near-Sun environment is predominantly Alfvénic and highly anisotropic, with structures elongated along the magnetic field such that the ratio of field-parallel and field-perpendicular wavenumbers satisfies $\kprl / \kprp \ll 1$ \citep{Belcher1969-mq,DePontieu2007-bn,Chen2016-mq}.
These two properties motivate the use of the framework of reduced magnetohydrodynamics (RMHD; \citealp{Strauss1976-vu,Schekochihin2009-qo,Schekochihin2022-nn}), which captures the essential dynamics of anisotropic, incompressible Alfvénic turbulence.
The equations of RMHD can be written in a physically transparent form using the Elsasser variables $\zpm \equiv \vel_\perp \pm \rB_\perp$ \citep{Elsasser1950-rx}:
\begin{equation}
        \left(\parfrac{}{t} \mp \vA\parfrac{}{z}\right)\zpm = - \zmp\dotprod\gradprp\zpm - \frac{\gradprp p_{\rm tot}}{\rho}. \label{eq:ElsasserRMHD}
\end{equation}
Here, the quantities $\vel_\perp$ and $\rB_\perp \equiv (\vA/\Bmeanmag)\delta\B_\perp $ are perturbations of the velocity and magnetic field perpendicular to the background magnetic field $\Bmean=\Bmeanmag\unitvector{z}$, with $\vA$ the Alfv\'en speed.
The gradient $\gradprp$ is taken in the plane perpendicular to $\Bmean$, and $p_{\rm tot}$ is the total (thermal plus magnetic) pressure of the plasma.
As the plasma is incompressible due to the RMHD ordering, we also have $\gradprp\dotprod\zpm = 0$; the pressure is then completely determined at any instant by enforcing incompressibility.
Additionally, turbulence within the solar corona and wind is measured to be imbalanced, with more Alfv\'enic fluctuations propagating away from the Sun than towards it.
A common measure of this imbalance is the normalised cross-helicity
\begin{equation}
    \crosshel \equiv \frac{\langle (\zp)^2 - (\zm)^2\rangle}{\langle (\zp)^2 + (\zm)^2\rangle},\label{eq:CrossHelicity}
\end{equation}
where $\langle \cdot \rangle$ denotes a volume average.
When the turbulence is balanced, $\crosshel=0$; fully imbalanced turbulence is denoted by $\crosshel=\pm 1$ with the sign corresponding to the dominant $\zpm$ fluctuations.

Protons have been observed to be heated strongly perpendicular to the magnetic field within regions of low-$\beta_{\rm i}$ solar wind \citep{Marsch1982-vu,Marsch2004-fw,Hellinger2006-ff}, with heavy ions observed to have far hotter temperatures than protons \citep{Kohl1998-cq,Esser1999-qd}. Here, $\beta_{\rm i}\equiv \vthi^2/\vA^2$ is the plasma beta, where $\vthi \equiv \sqrt{2k_{\rm B}T_{\rm i}/m_{\rm i}}$ is the thermal velocity of ions with mass $m_{\rm i}$ and temperature $T_{\rm i}$.
An understanding of the mechanisms that heat protons, as well as the partitioning of turbulent energy between protons and electrons, is critical to understanding the global properties of the heliosphere such as the heating of the solar corona and acceleration of the solar wind \citep{Parker1965-ab}. 
Measurements of heavy ions offer additional sensitive diagnostics that help identify the heating mechanisms at work \citep{Zhang2025-yd}.

There have been many theories proposed to describe turbulent ion heating within the solar corona, of which we focus on two in this work.
Stochastic heating \citep{McChesney1987-dp,Chen2001-wd,Johnson2001-ui,Chaston2004-jn,Fiksel2009-nm,Chandran2010-ow} arises when ions are kicked by disordered turbulent fluctuations, disrupting their smooth gyromotion, which breaks the conservation of their magnetic moment.
For a single ion, this leads to a random walk in kinetic energy; for an ion velocity distribution that is monotonically decreasing with respect to $\vprp$ (such as a Maxwellian), this can lead to preferential perpendicular heating at low-$\betai$.
\citet{Chandran2010-ow} show the stochastic heating rate per unit mass can be modelled empirically as
\begin{align}
\Qprp &= c_1 \frac{\delta u^3_{\rho_{\rm i}}}{\rhoi}e^{-c_2 / \stocfracth}\nonumber\\
&= \Omegai\vthi^2 c_1 \stocfracth^3 e^{-c_2 / \stocfracth},\label{eq:SHRate}
\end{align}
where $\Omegai$ and $\rhoi = \vthi/\Omegai$ are the ion gyrofrequency and thermal gyroradius, $\delta u_\rho$ is the r.m.s. velocity of $\rhoi$-scale $\ExB$ fluctuations, and $c_1$ and $c_2$ are empirical constants that are determined by the strength of the heating.
The parameter
\begin{equation}
        \stocfracth \equiv \frac{\deltau{\rho}}{\vthi}\label{eq:SHeps}
\end{equation}
is a measure of the amplitude of $\rhoi$-scale velocity fluctuations and controls the strength of the heating.
The exponential suppression captures the conservation of the magnetic moment of the ions in small-amplitude turbulence.
In addition to stochastic heating, a second mechanism capable of transferring energy to ions is cyclotron-resonant heating with waves \citep{Hollweg2002-dw,Chandran2010-ny,Isenberg2011-wt,Isenberg2019-oy}.
As described by quasi-linear theory \citep{Kennel1966-rx,Stix1992-bo,Schlickeiser1993-if} this occurs when the waves' Doppler-shifted frequencies are resonant with $\ion{\Omega}$, enabling a transfer of energy from waves to particles via strong ion-wave interactions and leading to ion heating.
The boundary between stochastic and cyclotron-resonant heating is not always clear: in critically balanced turbulence, the frequency of turbulent fluctuations that cause stochastic heating can reach frequencies near $\Omegai$ (similarly to cyclotron-resonant heating), despite its usual association with low-frequency turbulence \citep{Cerri2021-xo,Johnston2025-ss}.

In this paper, we use the framework of quasi-linear theory to calculate analytically the perpendicular heating rate of ions interacting with a model wavevector-frequency spectrum of critically balanced RMHD turbulence that captures the dependence of the turbulence on imbalance.
In the balanced case, strong nonlinear interactions between $\zpm$ fluctuations broaden the spectrum in frequency, while in the fully imbalanced limit the spectrum becomes sharply peaked along the dispersion relation of Alfvénic fluctuations.
This change in the behaviour of the spectrum shifts the heating mechanism from a stochastic-heating-like process in the balanced case---corresponding to interactions with random fluctuations across a wide range of frequencies---to a cyclotron-resonant-like process in the imbalanced case, where the sharply-peaked frequency spectrum enables resonant ion-wave interactions.
Regardless of the level of imbalance and corresponding heating mechanism, we derive that this heating rate has the generic form 
\begin{equation}
    \Qprp\propto \stocfracth^3 F(\stocfracth;\crosshel),\label{eq:GeneralHeatingSuppression}
\end{equation}
where $F(\stocfracth;\crosshel)$ is an imbalance-dependent suppression factor that goes to zero as $\stocfracth\to 0$.
In the case of balanced turbulence, we show analytically that this suppression factor closely resembles the empirical exponential factor in \eqref{eq:SHRate}, capturing the key ideas of the empirical stochastic heating model while extending its applicability to general levels of imbalance.

Our approach in this work is similar to \citet{Isenberg2019-oy} in using quasi-linear theory to study the heating of ions interacting with critically balanced fluctuations, but differs in several key respects: we assume large-scale Alfv\'enic fluctuations with their simple dispersion relation and include temporal correlations arising from nonlinear interactions, whereas \citet{Isenberg2019-oy} adopt the more complex kinetic-Alfv\'en-wave dispersion relation using the \citet{Kennel1966-rx} quasi-linear formalism of wave-particle interactions that ignores these temporal correlations.
Despite these differences, both models predict ion heating with properties consistent with previous observations and theory, underscoring the utility of quasi-linear theory in studying ion heating mechanisms.
The results of this paper are complementary to recent work by \citet{Johnston2025-ss}, who use high-resolution numerical simulations of balanced and imbalanced turbulence to study the heating of ions and show their measured heating rates are of the form \eqref{eq:GeneralHeatingSuppression}.
It also complements \citet{Mallet2026-ei}, who analytically study the interactions of ions with coherent electromagnetic fluctuations and also show that their predicted heating rate exhibits an exponential suppression factor as in \eqref{eq:SHRate}.

The paper is organised as follows.
In Section~\ref{sec:QLTheory}, we describe quasi-linear theory and how it can be applied to general small-amplitude electromagnetic fluctuations, presenting the quasi-linear diffusion coefficients that describe the interaction of a collection of ions with large-scale Alfv\'enic fluctuations.
In Section~\ref{sec:RMHDTurbulence}, we discuss the phenomenology of RMHD turbulence, including how the wavevector–frequency spectrum of fluctuations depends on the turbulence imbalance and thereby allows the heating mechanism to transition between a stochastic-heating-like and a cyclotron-resonant-like process. To support the subsequent calculations, we also introduce a novel model of the RMHD wavevector-frequency spectrum at general imbalance, which captures these key properties.
Then, in Section~\ref{sec:HeatingRates} we use the quasi-linear framework with the model RMHD spectrum to calculate the heating rate of ions interacting with RMHD turbulence, and study its dependence on imbalance.
We summarise our results and discuss their implications and caveats in Section~\ref{sec:Conclusion}.

We also include several appendices with supporting information and subsidiary results. 
In Appendix~\ref{app:RMHDModel}, we test the applicability of the model wavevector–frequency spectrum by comparing it with spectra calculated from numerical simulations of RMHD turbulence.
Appendix~\ref{app:QLTheoryDerivation} presents a derivation of the general quasi-linear diffusion coefficients used in Section~\ref{sec:HeatingRates}.
Appendices~\ref{app:SubRhoHeating} and \ref{app:TimeCorrCompare} explore the sensitivity of the calculated heating rates to small-scale ($\kprp\rhoi \gg 1$) fluctuations and to changes in the temporal correlation properties of the turbulence, respectively.
Finally, in Appendix~\ref{app:ICWHeating} we discuss the discrepancies in the heating of ions interacting with fluctuations with an Alfv\'en-wave dispersion relation (as in this work) compared to a more realistic ion-cyclotron-wave dispersion relation.

\section{Quasi-linear diffusion in a general space-time spectrum}\label{sec:QLTheory}

The evolution of the distribution function $f(\ppos,\pvel,t)$ of a collisionless plasma is described by the Vlasov equation for each species
\begin{equation}
    \parfrac{f}{t}+\pvel\dotprod\boldnabla f + \frac{q}{m}\left(\Evec+\frac{\pvel}{c}\crossprod\B\right)\dotprod\parfrac{f}{\pvel}=0.
\end{equation}
As this equation is strongly coupled to Maxwell's equations and therefore nonlinear, an analytic solution is generally not possible.
However, if the plasma is assumed to consist of a uniform background populated by small-amplitude fluctuations, the Vlasov-Maxwell system can be solved approximately through the use of quasi-linear theory \citep{Vedenov1963-qj,Drummond1964-ln,Kennel1966-rx,Lerche1968-qv}.
For a plasma with a background magnetic field $\Bmean=\Bmeanmag\unitvector{z}$ and equilibrium ion distribution function $f_0$, quasi-linear theory first approximates the orbits of ions by the unperturbed helical motion along $\Bmean$ that would occur in the absence of the fluctuations.
The distribution function and fields are split into background and fluctuating components,
\begin{align}
        f(\ppos,\pvel,t)&=f_0(\pvel,t)+f_1(\ppos,\pvel,t),\\
        \Evec(\ppos,t)&=\Evec_1(\ppos,t),\\
        \B(\ppos,t)&=\Bmean+\B_1(\ppos,t);
\end{align}
the evolution of $f_1$, varying on timescales faster than $f_0$, is calculated and inserted into a volume-average of the Vlasov equation to determine the back-reaction of this perturbation on $f_0$.
The result of this calculation shows that, due to their interaction with the fluctuations, ions diffuse in velocity space:
\begin{equation}
    \parfrac{f_0}{t} = \frac{1}{\vprp}\parfrac{}{\vprp}\left[\vprp\left(\Dprpprp\parfrac{f_0}{\vprp}+\Dprlprp\parfrac{f_0}{\vprl}\right)\right] + \parfrac{}{\vprl}\left[\Dprpprl\parfrac{f_0}{\vprp}+\Dprlprl\parfrac{f_0}{\vprl}\right],\label{eq:QLDiffusionForm}
\end{equation}
where cylindrical coordinates $(\vprp,\vprl)$ aligned along $\Bmean$ in velocity space have been used.
The diffusion coefficients $D_{a b}$ in \eqref{eq:QLDiffusionForm}, which are given in detail in Appendix~\ref{app:QLTheoryDerivation} and in a simplified form below, depend on quadratic products of the fluctuations and in principle specify the heating rate of an arbitrary $f_0$.

By assuming that the fluctuations are small-amplitude waves following a prescribed dispersion relation $\omega(\kvec)$, the quasi-linear framework has been used to study strong ion-wave interactions \citep{Vedenov1963-qj,Drummond1964-ln,Kennel1966-rx,Stix1992-bo}.
In the limit of weak wave damping, the evolution of $f_0$ is controlled by the resonance condition
\begin{equation}
    \omega(\kvec)-\kprl\vprl = n\Omegai,\label{eq:ResonanceCondition}
\end{equation} 
which occurs when the frequency of the wave, Doppler-shifted due to the particle's motion along the magnetic field, is an integer ($n$) multiple of the particle's gyrofrequency.
When this condition is satisfied, $f_0$ diffuses along contours of constant energy within the frame of the wave.
This approach has been used for investigating the perpendicular heating of ions within the solar corona \citep{Isenberg2007-nf,Isenberg2009-or,Chandran2010-ny,Isenberg2011-wt,Isenberg2019-oy}.

Compared to waves, which have a single frequency for a given wavevector $\kvec$, the approach above can be extended to general small-amplitude fluctuations within a turbulent system that have a spread of frequencies at a given $\kvec$ \citep{Hall1967-cg,Lerche1968-qv,Schlickeiser1993-if,Chandran2000-pq}.
This approach has been used to investigate the pitch-angle scattering of relativistic cosmic rays in slab turbulence \citep{Schlickeiser1989-fd,Schlickeiser1993-if,Dung1990-nu,Dung1990-ks,Dung1992-ga,Dung1994-bp,Weidl2015-sd} and strong, critically balanced MHD turbulence \citep{Chandran2000-pq,Yan2004-qd}.
The general diffusion coefficients obtained from this approach (whose derivation is presented in Appendix~\ref{app:QLTheoryDerivation}) are intractably complex for general use.
To allow for physical insight into how they affect the heating of ions, and to allow the use of models of RMHD turbulence, we assume Alfv\'enic fluctuations and take the large-scale limit of these coefficients.
The resultant coefficients, presented below, are used to study how ion heating varies with the imbalance of the turbulence in Section~\ref{sec:HeatingRates}.

\subsection{Large-scale Alfv\'enic limit of diffusion coefficients}\label{sec:RMHDLimitCoeffs}

To simplify the diffusion coefficients derived in Appendix~\ref{app:QLTheoryDerivation}, \eqref{eq:OuterDiffusionCoeff} and \eqref{eq:InnerDiffusionCoeffs}, we assume that the turbulence is purely Alfv\'enic, and that there exist equal populations of left- and right-hand polarised fluctuations so that the turbulence is mirror symmetric \citep{Chandran2000-pq}.
For fluctuations at scales $\rhoiscales\ll 1$, we can write the electric-field components in the diffusion coefficients \eqref{eq:InnerDiffusionCoeffs} in terms of the turbulent velocity field $\vel_1$ via Ohm's law, $\Evec_1 = -(\Bmeanmag/c)\vel_1 \crossprod \unitvector{z}$; this choice ensures that  $E_z$, the electric-field component parallel to $\Bmean$, vanishes.
To ensure the components of $\vel_1$ and $\B_1$ perpendicular to $\Bmean$ are Alfv\'enic, we assume fluctuations with perpendicular (unit) wavevector $\unitvector{k}_\perp$ are polarised in the $\unitvector{k}_\perp\crossprod\unitvector{z}$ direction.
Further details on this process are given in Appendix~\ref{app:QLTheoryDerivation}.

With this, the diffusion coefficients \eqref{eq:OuterDiffusionCoeff} can be written as\footnote{These are equivalent to the shear-Alfv\'en-mode coefficients of \citet{Chandran2000-pq} after transforming to pitch-angle coordinates $p = m_{\rm i}(\vprp^2+\vprl^2)^{1/2}$ and $\zeta=\vprl(\vprp^2+\vprl^2)^{-1/2}$.}
\begin{equation}
    \begin{pmatrix} \Dprpprp\\ \Dprlprp\\ \Dprlprl\end{pmatrix} = \lim_{V\to\infty}\Omegai^2\vA^2\sum_{n=-\infty}^{\infty}\intfrac{\kvec}{V}{}{}\frac{n^2\BesselJ{n}^2(\kappa)}{\kappa^2}\intsingle{\tau}{0}{\infty}e^{-\Icplx(\kprl\vprl + n\Omegai)\tau}\begin{pmatrix}\mathcal{K}+2\vprltilde\mathcal{C}+\vprltilde^2\mathcal{M}\\-\vprptilde(\mathcal{C}+\vprltilde\mathcal{M})\\\vprptilde^2\mathcal{M}\end{pmatrix}\label{eq:QLDiffusionCoefficientsAlfvenicPol}
\end{equation}
with $\tilde{v}_{\perp,\|}\equiv v_{\perp,\|}/\vA$ and $\kappa\equiv\kprp\vprp/\Omegai$ the argument of the Bessel functions $\BesselJ{n}$.
The normalised space-time power spectra are defined as
\begin{subequations}\label{eq:TemporalCorrelators}
\begin{align}
    \mathcal{K}(\kvec,\tau)&\equiv\frac{\langle \vel(\kvec, t)\dotprod\vel^*(\kvec, t+\tau) \rangle}{\vA^2},\label{eq:KCorrelator}\\
    \mathcal{C}(\kvec,\tau)&\equiv\frac{\langle \vel(\kvec, t)\dotprod\B^*(\kvec, t+\tau) \rangle}{\vA \Bmeanmag},\label{eq:CCorrelator}\\
    \mathcal{M}(\kvec,\tau)&\equiv\frac{\langle \B(\kvec, t)\dotprod\B^*(\kvec, t+\tau) \rangle}{B^2_0}.\label{eq:MCorrelator}
\end{align}
\end{subequations}
Because $\langle \B(\kvec, t)\dotprod\vel^*(\kvec, t+\tau) \rangle= \langle \vel(\kvec, t)\dotprod\B^*(\kvec, t+\tau) \rangle$ (which follows from the Wiener--Khinchin theorem for real fields; \citealp{Thorne2017-oy}), we have $\Dprpprl=\Dprlprp$.

The diffusion coefficients \eqref{eq:QLDiffusionCoefficientsAlfvenicPol} depend on the wavevector–frequency spectra of the turbulent fluctuations, which describes the distribution of the energy of fluctuations in terms of their frequency $\omega$ and wavevector $\kvec$.
These spectra show up implicitly as the temporal Fourier transform of the normalised power spectra \eqref{eq:TemporalCorrelators} (approximated by the integral over $\tau$ in \ref{eq:QLDiffusionCoefficientsAlfvenicPol}; \citealp{Hall1967-cg,Schlickeiser1993-if}), and contribute to the strength of the diffusion coefficients and to the diffusion of the ion distribution function $f_0$ through velocity space.
The properties of the turbulence with which the ions interact, as described by the wavevector-frequency spectra, thus play a key role in determining how the ions are heated and energised; we give more details on how the turbulence's imbalance affects these spectra in Section~\ref{sec:HeatingMechanismImbalance}. 

\section{RMHD turbulence}\label{sec:RMHDTurbulence}

The large-scale ($\kprp\rhoi\ll 1$) limit of Alfv\'enic fluctuations used in the description of the quasi-linear diffusion coefficients in Section~\ref{sec:RMHDLimitCoeffs} is naturally consistent with an RMHD description of turbulence.
The goal of this section is to develop a model wavevector-frequency spectrum that can be used for the calculation of the diffusion coefficients and heating rates in Section~\ref{sec:HeatingRates}.
We first outline the theory of RMHD turbulence in Section~\ref{sec:ImbalancedTurbulence}, where we discuss the concept of critical balance, phenomenological descriptions of balanced and imbalanced turbulence, and how the wavevector-frequency spectrum of the turbulence is dependent on imbalance.
Using this theoretical description as well as numerical simulations of RMHD turbulence (presented in Appendix~\ref{app:RMHDModel}), we then develop and present a model of this spectrum for RMHD turbulence in Section~\ref{sec:RMHDModel} with a general dependence on $\crosshel$, and show that it qualitatively reproduces the features of RMHD turbulence.

\subsection{Turbulence phenomenology}\label{sec:ImbalancedTurbulence}

In this subsection, we review the phenomenology of RMHD turbulence insofar as it is needed for later calculations, focusing on how the imbalance affects various turbulence properties \citep{Schekochihin2022-nn}.
We assume that energy is injected into the $\zpm$ fluctuations at large scales at a rate $\forcepm$; for the turbulence to be in steady-state, this rate is statistically equivalent to the energy flux between inertial-range scales and the rate of dissipation of energy at small scales.
The magnetic field $\Bmean$ introduces an anisotropy into the system, giving rise to different properties of the turbulence parallel and perpendicular to $\Bmean$.
To account for this, the quantities $\deltazpm_{\lprl}$ and $\deltazpm_{\lambda}$ are defined as r.m.s. values of the $\zpm$ fluctuations measured over increments parallel ($\lprl$) and perpendicular ($\lambda$) to $\Bmean$.

A key feature of the RMHD equations \eqref{eq:ElsasserRMHD} is that the nonlinear term $\zmp\dotprod\gradprp\zpm$ requires both $\zp$ and $\zm$ to be nonzero in order to be activated.
This means that counter-propagating AWs are needed in order for a turbulent cascade to develop.
By inspection of the nonlinear term, the timescale of these nonlinear interactions is
\begin{equation}
    \taunl^\pm \sim \frac{\lambda}{\deltazmp_\lambda}.\label{eq:taunl}
\end{equation}
In addition to the nonlinear timescales \eqref{eq:taunl} over which fluctuations interact, the magnetic field, via Alfv\'en waves, introduces another timescale,
\begin{equation}
    \tauA \sim \frac{\lprl}{\vA},
\end{equation}
corresponding to Alfv\'en-wave propagation along magnetic-field lines over a distance $\lprl$.
Although $\taunl$ and $\tauA$ are not related \emph{a priori}, due to causality it can be argued that $\taunl\sim\tauA$ \citep{Boldyrev2005-yz}; this conjecture is termed \emph{critical balance} \citep[CB;][]{Goldreich1995-fv}.\footnote{The \citet{Goldreich1995-fv} theory of critical balance assumes that fluctuations are isotropic in the plane perpendicular to the magnetic field.
However, numerical simulations show that fields become aligned at smaller scales in a process called dynamic alignment \citep{Boldyrev2006-xd,Beresnyak2011-or,Beresnyak2012-eb,Schekochihin2022-nn}.
This effect leads to a reduction in the nonlinear time and causes fluctuations to become anisotropic in the perpendicular plane.
Because these effects are not well understood in imbalanced turbulence, we only assume the CB phenomenology of \citet{Goldreich1995-fv} in this work; however, we note that using scalings derived assuming dynamic alignment modifies the exponent of $\stocfracth$ within the wavevector energy spectrum \eqref{eq:RMHDSpecNorm}, giving rise to a different scaling of the heating rate with $\stocfracth$ when $\stocfracth\ll 1$ in the calculations of Section~\ref{sec:HeatingRates}.}

To show this, consider two regions of turbulence fluctuations of scale $\lambda$ perpendicular to $\Bmean$ evolving over a timescale $\taunl$, separated by a distance $\lprl$ along $\Bmean$.
For these regions to remain correlated, they must be separated by no more than the distance Alfv\'enic fluctuations can travel along $\Bmean$ in a time $\taunl$, which is the time over which fluctuations decorrelate due to nonlinear interactions:
\begin{equation}
    \lprl \lesssim \vA\taunl \sim \vA\tauA \implies \tauA\sim\frac{\lprl}{\vA}\sim\taunl\sim\frac{\lambda}{\delta z_\lambda}.
\end{equation}
Regions separated by a distance greater than $\lprl$ are thus unable to remain correlated.

In the case of balanced turbulence, with $\forcep\sim\forcem \equiv \cascadeE$ and $\deltazp\sim \deltazm \equiv \deltaz$, one can use the usual assumption that the flux across scales is constant to yield the \citet{Goldreich1995-fv} energy spectrum, $\energyspecE(\kprp) \propto \kprp^{-5/3}$ and $\energyspecE(\kprl)\propto \kprl^{-2}$ (or equivalently $\delta z_\lambda \sim (\cascadeE \lambda)^{1/3}$ and $\delta z_{\lprl} \sim (\cascadeE \lprl/\vA)^{1/2}$).

\subsubsection{Imbalanced turbulence}

Although balanced turbulence simplifies the description of turbulence, many systems are imbalanced with $\forcep\gg\forcem$.
While there is currently no theory that describes imbalanced turbulence fully, phenomenological arguments can be made to explain qualitatively aspects seen in observations and numerical simulations.
Here we cover a simple version, as needed for later, and refer the reader to \citet{Schekochihin2022-nn} for a complete description of the subtleties involved.

Using the constant-flux argument with \eqref{eq:taunl} for $\taunl^\pm$, at each scale $\lambda$ one assumes
\begin{equation}
    \frac{(\deltazpm_\lambda)^2}{\taunl^\pm}\sim \frac{(\deltazpm_\lambda)^2\deltazmp_
    \lambda}{\lambda}\sim \forcepm ,\label{eq:KolmogorovConstantFluxArgumentMHD}
\end{equation}
where $\forcepm$ are the energy injection rates at the outer scales of the turbulence for the $\zpm$ fluctuations.
The nonlinear timescales \eqref{eq:taunl} also imply that
\begin{equation}
    \frac{\taunl^+}{\taunl^-}\sim \frac{\deltazp_\lambda}{\deltazm_\lambda}\gg 1;
\end{equation}
in other words, the cascade of the stronger field becomes less efficient with increasing imbalance as it is dependent on the weaker field to advect it.
Although $\deltazm_\lambda\ll\deltazp_\lambda$, the cascades can still be assumed strong, with $\taunl^\pm \lesssim \tauA$ individually.

The constant-flux argument aligns with results from numerical simulations that the ratio of the energies of the individual fields scales with imbalance as $\forceratio \sim \injectratio^2$ \citep{Schekochihin2022-nn}, where 
\begin{equation}
    \injectratio \equiv \frac{\forcep}{\forcem}\label{eq:InjectionRatio}
\end{equation}
is the injection ratio and
\begin{equation}
   \forceratio \equiv \frac{E^+}{E^-}=\frac{1+\crosshel}{1-\crosshel}\label{eq:ZpmEnergyRatio}
\end{equation}
is the ratio of Elsasser energies (assumed to be independent of scale such that $\alpha \sim (\deltazp_\lambda)^2/(\deltazm_\lambda)^2$).

Despite this agreement with numerics, the fact that the weaker $\zm$ field is able to advect and distort the stronger $\zp$ field over the long timescale $\taunl^+$ requires a separation of its correlation and spatial-distortion timescales \citep{Lithwick2007-ao}.
The stronger field $\zp$ distorts the weaker $\zm$ in space over a timescale $\taunl^-$.
However, $\zm$ is ``swept up'' by the stronger $\zp$ fluctuations: in the frame in which $\zp$ fluctuations are stationary, the $\zm$ fluctuations develop fine structures due to the spatial distortion, but remain approximately constant over $\taunl^+$ in order to remain correlated with the $\zp$ fluctuations.

The separation of these timescales means that the causality argument for critical balance in the balanced case (dubbed ``causal CB'') no longer holds.
Directly applying the assumptions of causal CB in the imbalanced case would imply that the parallel correlation lengths of the $\zpm$ fluctuations are disparate with $\lprl^+/\lprl^- \gg 1$.
However, \citet{Lithwick2007-ao} argue that, because $\zm$ perturbations are spatially distorted over a time $\taunl^-$ (such that $\lprl^-\sim\vA\taunl^-$), $\zp$ fluctuations separated by a distance $\lprl^-$ are advected by the now spatially decorrelated $\zm$ and thus themselves decorrelate over a similar distance: $\lprl^+\sim\lprl^-\sim\vA\taunl^- \sim \kprp \deltazp$.
This effect, termed ``propagation CB'' \citep{Beresnyak2008-aw}, allows a CB argument to be used to balance linear and nonlinear times using the fact that the $\zm$ fluctuations are swept along by $\zp$ fluctuations, rather than information travelling between points.

Although the theory outlined above does well in explaining some results observed in simulations of imbalanced turbulence, it is far from a complete picture.
The scaling $\forceratio\sim\injectratio^2$ appears to be robust, observed in multiple simulations of RMHD turbulence \citep{Schekochihin2022-nn}, but does not agree with alternative theories of imbalanced turbulence \citep{Perez2009-ad}.
The recent theory of \citet{Schekochihin2022-nn}, which assumes non-local nonlinear interactions, appears to account for some of these discrepancies, but further work is needed to test this thoroughly.
Regardless, the phenomenology of propagation CB presented above is adequate for the model we use to investigate ion heating, with the simulations of Appendix~\ref{app:RMHDModel} showing modest support for this theory.

\subsubsection{Imbalance dependence of the wavevector-frequency spectrum}\label{sec:HeatingMechanismImbalance}

The form of the wavevector-frequency spectrum depends on the type of fluctuations present in the turbulence and the strength of the nonlinear interactions between them.
Unbroadened fluctuations---waves following a dispersion relation $\omega(\kvec)$ with a single frequency $\omega$ at a particular $\kvec$---are sharply peaked along this dispersion relation in the sense of a delta function $\delta[\omega-\omega(\kvec)]$ in the wavevector-frequency spectrum.
In contrast, due to the nonlinear interactions between different wavenumbers in the energy cascade, fluctuations in a strongly turbulent system take on a spread of frequencies of width $\Delta\omega$ at a given $\kvec$.
This broadening is directly related to the nonlinear turnover timescale of the fluctuations $\taunl$, which itself is on the order of the lifetime of the eddies before they break apart: the nonlinear interactions generating the cascade are precisely what cause the fluctuations to decorrelate over this time.
For general levels of imbalance, the broadening in fluctuations at a given $\kvec$ must then be $\Delta\omega\sim\omeganl=\taunl^{-1}$.

The dependence of the behaviour of the wavevector-frequency spectrum on the level of imbalance reveals a deep connection to the mechanism by which ions are heated.
In imbalanced turbulence, the linear frequencies $\omegaA\sim \vA / \lprl^+ \sim \kprp\deltazp$ of $\zp$ fluctuations far exceed their nonlinear rate $\omeganl\sim \kprp\deltazm$ \citep{Lithwick2007-ao,Schekochihin2022-nn}, leading the wavevector-frequency spectrum to become sharply peaked around the dispersion relation of the fluctuations (because $\Delta\omega\sim\omeganl\ll\omegaA$).
This is the \emph{cyclotron-resonant-heating limit}, where ions are able to resonate with exactly one frequency at a given $\kvec$.
In contrast, using critical balance we have $\omegaA\sim\vA/\lprl \sim \omeganl\sim \kprp\deltazpm\sim \Delta\omega$ in balanced turbulence \citep{Goldreich1995-fv}.
This is the \emph{stochastic-heating limit}, where fluctuations are sufficiently broadened and decorrelated allowing for ions to be heated by $\rhoi$-scale fluctuations.
A smooth transition between these limits is expected as the imbalance is adjusted and one of the Elsasser fields dominates over the other; this is confirmed in Appendix~\ref{app:RMHDModel}, where the wavevector-frequency spectra of simulations of RMHD turbulence with varying imbalance are investigated.

\subsection{Model wavevector-frequency spectrum of RMHD turbulence}\label{sec:RMHDModel}

The wavevector-frequency spectrum of a turbulent system describes the distribution of the energy of fluctuations in terms of their frequency $\omega$ and wavevector $\kvec$.
The simulations of RMHD turbulence presented in Appendix~\ref{app:RMHDModel} show this spectrum is composed of $\zpm$ fluctuations whose bandwidth in $\omega$ increases due to nonlinear broadening effects, decreasing with increasing $\crosshel$.
The linear physics of these fluctuations is also significant, with peaks centred on the Alfv\'en dispersion relation of the $\zp$ fluctuations ($-\omegaA$).

The wavevector-frequency spectra can be decomposed into two separate components: the \emph{wavevector energy spectrum} $\energyspecE_{\rm 2D}(\kprp,\kprl)$ that controls the amplitude of the frequency spectrum at every $\kvec$, and the \emph{temporal correlation function} $f(\kvec,\omega)$, a general function of $\kvec$ and $\omega$ that encodes the correlations and broadening of the turbulent Alfv\'enic fluctuations.

\subsubsection{Two-dimensional wavevector energy spectrum}\label{sec:SchekochihinSpectrum}

The two-dimensional wavevector energy spectrum in RMHD is defined as
\begin{equation}
    \energyspecE_{\rm 2D}(\kprp,\kprl) = 2\upi\kprp \langle \norm{\zpm(\kprp, \kprl)}^2 \rangle,\label{eq:E2DDef}
\end{equation}
where $\zpm(\kprp, \kprl)$ is the complex Fourier amplitude of the $\zpm$ fields (which are assumed to be statistically cylindrically symmetric about $\Bmean$) at a given $\kvec$; the $2\upi\kprp$ term comes from the Jacobian of cylindrical coordinates in $\kvec$-space.
By analogy to the one-dimensional spectrum, where $\energyspecE(k)\romand k$ measures the energy contained in modes with $k<k'<k+\romand k$, this spectrum measures the energy in modes around a given parallel and perpendicular wavenumber.
Via the Wiener--Khinchin theorem, \eqref{eq:E2DDef} can likewise be interpreted as a measure of how correlated are turbulent fluctuations of a given scale along ($\lprl$) and perpendicular ($\lprp$) to the background magnetic field \citep{Thorne2017-oy}.\footnote{As written, with $\kprl$, information about the local magnetic field is incorporated into the definition of \eqref{eq:E2DDef}, meaning that spectrum measures correlations along the local field \citep{Cho2000-bi,Maron2001-vz,Squire2022-dm}.}

\citet{Schekochihin2022-nn} argues that $\energyspecE_{\rm 2D}(\kprp,\kprl)$ should be a product of power laws of $\kprp$ and $\kprl$:
\begin{equation}\label{eq:RMHD2DSpec}
    \energyspecE_{\rm 2D}(\kprp,\kprl)\sim 
    \begin{cases}
        |\kprl\Lprl|^{-\sone}(\kprp\lambda_{\rm CB})^{\stwo}, & |\kprl\Lprl| > (\kprp\lambda_{\rm CB})^{\sCB},\\
        |\kprl\Lprl|^{\sfour}(\kprp\lambda_{\rm CB})^{-\sthree}, & |\kprl\Lprl| \leq (\kprp\lambda_{\rm CB})^{\sCB}.
    \end{cases}
\end{equation}
Here, $\Lprl$ is the outer scale parallel to $\Bmean$ and $\lambda_{\rm CB}$ is the perpendicular scale at which weak turbulence becomes strong \citep{Schekochihin2022-nn}.
Due to critical balance (CB) we have $\kprl \sim \kprp^{\sCB}$.
For the standard \citet{Goldreich1995-fv} CB argument covered above in Section~\ref{sec:ImbalancedTurbulence}, $\sCB = 2/3$, whereas for dynamic alignment $\sCB = 1/2$ \citep{Boldyrev2006-xd}.

The exponents in \eqref{eq:RMHD2DSpec} can be determined based on the phenomenology of CB \citep{Schekochihin2022-nn}.
At scales where $k_\parallel \leq k_\perp^{\sCB}$, the spectrum measures correlations between points along magnetic-field lines that are separated by a distance $\lprl\gtrsim\vA\tauA$.
Due to CB, these points are causally disconnected and thus uncorrelated; the correlation thus measures white noise and the spectrum must be flat along $\kprl$ \citep{Thorne2017-oy}, so $\sfour = 0$.
The perpendicular exponent $\sthree$ is fixed by requiring that the two-dimensional spectrum reproduces the one-dimensional perpendicular scaling implied by the constant-flux condition and CB (which imply $\delta z_\lambda^2 \sim k_\perp^{-\sCB}$), yielding $\sthree = 2\sCB + 1$.
Continuity across the CB line then requires $\sone = (2\sCB + 1 + \stwo)/\sCB$.
Finally, $\stwo$ depends on whether a kinematic ($\stwo = 3$) or thermodynamic (equal energy in all modes; $\stwo = 1$) regime is assumed \citep{Schekochihin2022-nn}; in this work we adopt the thermodynamic viewpoint, which better matches our simulation results in Appendix~\ref{app:RMHDModel}.

\begin{figure}
    \centering
    \includegraphics[width=0.85\textwidth]{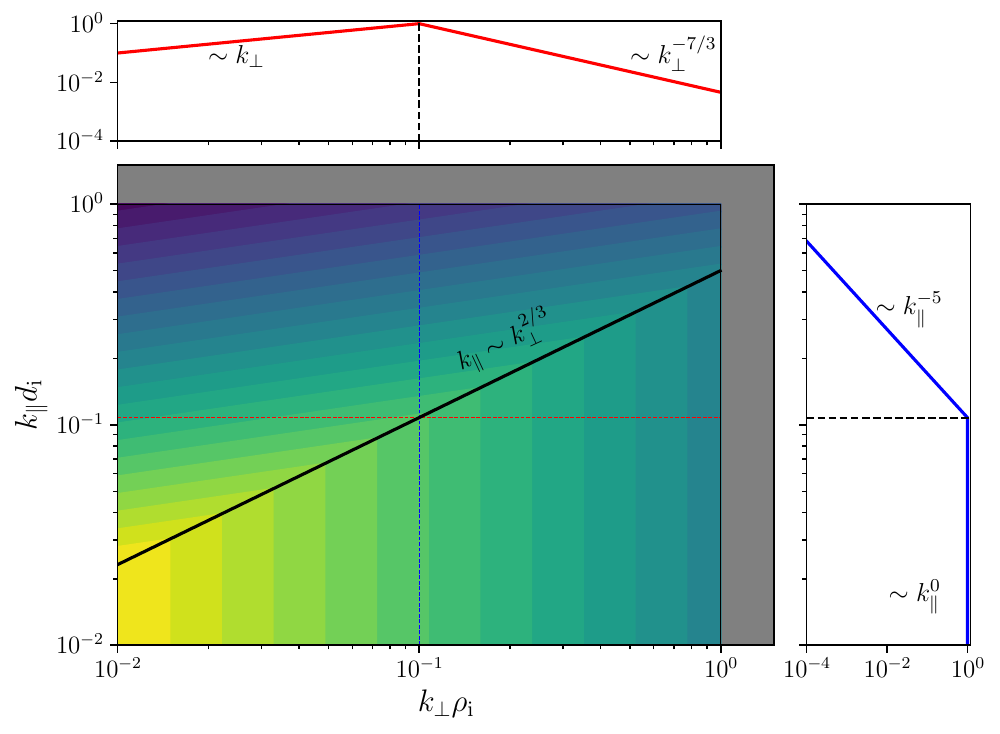}
    \caption{The 2D spectrum model \eqref{eq:RMHDSpecNorm} with the \citet{Goldreich1995-fv} critical balance scaling $\sCB = 2/3$, $C=1$, and $\stocfracth=0.5$ (with critical balance line $\kprl d_{\rm i}=0.5(\kprp\rhoi)^{2/3}$).
    The grey boundaries represent the spectrum cutoff at $|\kprl d_{\rm i}| > 1$ or $\kprp\rhoi > 1$ assumed in \eqref{eq:RMHDSpecNorm}.
    Side panels show slices through the spectra (normalised to the maximum value along the slice), showing the individual scalings of $\kprp$ and $\kprl$ above and below the CB line.}
    \label{fig:2DRMHDSpec}
\end{figure}

With this, \eqref{eq:RMHD2DSpec} can be written entirely in terms of the CB exponent $\sCB$:
\begin{equation}\label{eq:RMHD2DSpecUpdatedExponenets}
    \energyspecE_{\rm 2D}(\kprp,\kprl)\sim 
    \begin{cases}
        |\kprl\Lprl|^{-(2\sCB+2)/\sCB}(\kprp\lambda_{\rm CB}), & |\kprl\Lprl| > (\kprp\lambda_{\rm CB})^{\sCB},\\
        (\kprl\Lprl)^0(\kprp\lambda_{\rm CB})^{-2\sCB - 1}, & |\kprl\Lprl| \leq (\kprp\lambda_{\rm CB})^{\sCB}.
    \end{cases}
\end{equation}
For the \citet{Goldreich1995-fv} CB argument with $\sCB=2/3$, the spectrum follows $|\kprl\Lprl|^{-5}\kprp\lambda_{\rm CB}$ for $|\kprl\Lprl| > (\kprp\lambda_{\rm CB})^{2/3}$ and $(\kprl\Lprl)^0(\kprp\lambda_{\rm CB})^{-7/3}$ for $|\kprl\Lprl| \leq (\kprp\lambda_{\rm CB})^{2/3}$ (see figure~\ref{fig:2DRMHDSpec}).

The derivation of \eqref{eq:RMHD2DSpecUpdatedExponenets} assumes that the turbulence is balanced, and is identical for both the $\zp$ and $\zm$ fields.
In Appendix~\ref{app:RMHDModel}, we argue that this model can be extended to describe imbalanced turbulence by utilising RMHD turbulence simulations with different values of $\crosshel$ and comparing their spectra with \eqref{eq:RMHD2DSpecUpdatedExponenets}.
There, we show that the model of \citet{Schekochihin2022-nn} works well for describing the two-dimensional energy spectra at all levels of imbalance of the $\zp$ and $\zm$ fluctuations (with discrepancies above the CB cone for the $\zm$ spectra), with the measured scalings showing $\stwo\approx 1$ and $1/2 \lesssim\sCB\lesssim 2/3$.

\textit{Normalisation}---Assuming a collection of ions with thermal speed $\vthi$, thermal gyroradius $\rhoi$, and $\betai = \vthi^2/\vA^2$, we normalise \eqref{eq:RMHD2DSpecUpdatedExponenets} such that
\begin{equation}\label{eq:RMHDSpecNormDef}
    \frac{1}{\vthi^2}\intsingle{\kprp}{e^{-1/2}/\rhoi}{e^{1/2}/\rhoi}\intsingle{\kprl}{-\kprl^{\rm CB}(\kprp)}{\kprl^{\rm CB}(\kprp)}\energyspecE_{\rm 2D}(\kprp,\kprl)\approx\stocfracth^2,
\end{equation}
where $\kprl^{\rm CB}(\kprp) \propto \kprp^{\sCB}$ is the CB cone boundary.
This normalisation ensures that the energy in $\rhoi$-scale modes is given by the standard stochastic heating parameter \citep{Chandran2010-ow}
\begin{equation}
    \stocfracth = \frac{\delta u_{\rho_{\rm i}}}{\vthi}.
\end{equation}
Normalising the wavenumbers as $\kprptilde\equiv \kprp\rhoi$ and $\kprltilde\equiv \kprl\vA/\Omegai = \kprl\ion{d}$ (where $\ion{d}\equiv\vA/\Omegai$ is the ion inertial length), the spectrum becomes
\begin{align}\label{eq:RMHDSpecNorm}
    \energyspecE_{\rm 2D}(\kprp,\kprl)&=\frac{1}{2}V\vA^2\rhoi^2\betai^{1/2}\begin{cases}
        C^{-1}\stocfracth\kprptilde^{-\sthree}, & |\kprltilde|\leq \kprltildeCB\\ 
        C^{\sone-1}\stocfracth^{\sone+1}|\kprltilde|^{-\sone}\kprptilde^{\stwo}, & |\kprltilde|\geq \kprltildeCB\\
        0, & \kprptilde > 1 \text{ or } |\kprltilde| > 1,
    \end{cases}\nonumber\\
    &\equiv \frac{1}{2}V\vA^2\rhoi^2\betai^{1/2}\energyspecEtildetwoD(\kprp,\kprl),
\end{align}
where $\energyspecEtildetwoD(\kprp,\kprl)$ is the dimensionless part of the spectrum, $V$ is the volume of the plasma, $C$ is an arbitrary Kolmogorov-like constant to account for order-unity factors, and the exponents are $\sthree = 2\sCB+1$, and $\sone = (2\sCB+\stwo+1)/\sCB$.
With this normalisation, the critical balance line scales as
\begin{equation}
    \kprl\vA\sim\kprp\deltaz_\lambda\implies \kprltildeCB \equiv C \stocfracth \kprptilde^{\sCB}.\label{eq:KprlCB}
\end{equation}
For the remainder of this paper, we will assume the \citet{Goldreich1995-fv} CB scaling $\sCB=2/3$ and $\stwo=1$ (which follows from assuming the spectrum above the CB cone has reached a thermodynamic limit, as discussed above).

This spectrum is shown in figure~\ref{fig:2DRMHDSpec} with $C=1$ and $\stocfracth = 0.5$.
A common misunderstanding of the CB argument is that energy must be concentrated at the CB boundary $\kprl\sim\kprp^{2/3}$.
This is not true, with energy contained in all modes below the CB cone, as is clearly seen in figure~\ref{fig:2DRMHDSpec}.
The steep drop in the spectrum in $\kprl$ for fluctuations above the CB cone means that there is very little energy in these fluctuations.

The cutoff at $\kprptilde > 1$ stems from the assumption that $\kprp\rhoi\gtrsim 1$ fluctuations do not contribute strongly to heating, because they are suppressed by particles sampling many small-scale eddies over an orbit \citep{Chandran2000-pq}.
However, these fluctuations may still have some effect \citep{Arzamasskiy2019-qv,Isenberg2019-oy}; we quantify this contribution in Appendix~\ref{app:SubRhoHeating} where we use a model spectrum of strong balanced CB turbulence that includes $\kprp\rhoi\geq 1$ kinetic-Alfv\'en wave fluctuations, arguing their effect is often small when $\kprl/\kprp \ll 1$ at $\kprp\rhoi \sim 1$.
Additionally, by including a cutoff at $|\kprltilde| > 1$ this model also neglects the transition of Alfv\'en waves to dispersive ion-cyclotron waves (ICWs) as $\kprl$ approaches $d^{-1}_{\rm i}$.
Hybrid-kinetic simulations show the details of ICWs near $\kprl d_{\rm i}\sim 1$ are important for some aspects of ion heating \citep{Squire2022-dm,Squire2023-jn,Zhang2025-yd}; additionally, due to the difference in their dispersion relations, the parallel resonant wavenumber of the Alfv\'en waves and ICWs differ, which may bias the heating rate (details on this discrepancy are presented in Appendix~\ref{app:ICWHeating}).
Regardless, we neglect the effect of the Alfv\'en-wave dispersion (i.e. the change to ICWs at $\kprl d_{\rm i} \sim 1$) on ion heating in our model, and assume the heating is dominated by large-scale Alfv\'enic fluctuations.

\subsubsection{Temporal correlation function}

For a turbulent system with nonlinear frequency $\omeganl(\kvec)$, a general temporal correlation function $f[\tau\omeganl(\kvec)]$ encodes how correlated fluctuations are expected to be after a normalised lag time $\tau\omeganl(\kvec)$.
This function should satisfy the properties $f(0)=1$, $\lim_{\tau\to\infty}f[\tau\omeganl(\kvec)]=0$, and $\int_{0}^{\infty} \romand\tau f[\tau\omeganl(\kvec)] \approx \omega^{-1}_{\rm nl}(\kvec)$.
For our model of RMHD turbulence with variable imbalance, we take the nonlinear frequency to be that of the dominant $\zp$ fluctuations:
\begin{equation}
    \omeganl(\kprp) = \kprp\deltazm_\lambda = C\Omegai\forceratio^{-1/2}\stocfracth(\kprp\rhoi)^{\sCB},\label{eq:OmegaNL}
\end{equation}
where we have assumed the CB scaling $\omeganl\sim\omegaA = \kprl\vA$ and utilised the normalisation for the spectrum above.
To capture the reduced efficiency of the $\zp$ cascade and the corresponding decrease in their nonlinear frequency as imbalance increases, $\omeganl$ is scaled by $\forceratio^{-1/2}$; this scaling arises from a simple estimate using CB for the dominant $\zp$ fluctuations:
\begin{equation}
    \omeganl = \kprp\deltazm_\lambda \sim \frac{\deltazm_\lambda}{\deltazp_\lambda}\kprp\deltazp_\lambda \propto \forceratio^{-1/2}.
\end{equation}

A commonly-used choice for the temporal correlation function that satisfies the above properties is \citep[e.g.,][]{Schlickeiser1993-if,Chandran2000-pq}
\begin{equation}
    f[\tau\omeganl(\kprp)]=e^{-|\tau\omeganl(\kprp)|};\label{eq:ExponentialTimeCor}
\end{equation}
however, the discontinuity in slope of the function at $\tau=0$ causes an unphysically slow drop-off in the Fourier transform of $f$ at large frequencies, which can affect the calculated heating rates (see Appendix~\ref{app:TimeCorrCompare} for more details).
In contrast, a formal renormalisation group calculation in hydrodynamic turbulence shows that $f(\tau\omeganl)$ reduces to \eqref{eq:ExponentialTimeCor} for $\tau\omeganl \gg 1$ and to a Gaussian $e^{-(\tau\omeganl)^2}$ for $\tau\omeganl \ll 1$ \citep{Gorbunova2021-bh}, the latter of which is smooth at $\tau=0$.
Based on this, a convenient choice for our model spectrum is
\begin{equation}
f[\tau\omeganl(\kprp)]=\sech[\tau\omeganl(\kprp)],\label{eq:SechTimeCorrelationFunction}
\end{equation}
which has the asymptotic limits
\begin{equation}
    \sech(x)\approx\begin{cases}
        e^{-x^2/2}, & |x|\ll 1,\\
        2 e^{-|x|}, & |x|\gg 1,
    \end{cases}\label{eq:Sech}
\end{equation}
and satisfies $\intsingle{\tau}{0}{\infty} f(\tau\omeganl) = \frac{\upi}{2}\omeganl^{-1}$.
Taking the Fourier transform in time of \eqref{eq:SechTimeCorrelationFunction} with the convention 
\begin{equation}
    F(\omega) = \mathcal{F}\left[f(\tau)\right]= \frac{1}{2\upi}\int_{-\infty}^{\infty}\romand\tau e^{\textrm{i}\omega\tau}f(\tau)\label{eq:TemporalFTConvention}
\end{equation}
gives
\begin{equation}
    F(\omega;\kprp,\crosshel) = \frac{1}{2\omeganl(\kprp)}\sech\left(\frac{\upi\omega}{2\omeganl(\kprp)}\right).\label{eq:SechFrequencyCorrelationFunction}
\end{equation}

The choice of function used for $f(\tau\omeganl)$ can affect the results of the following sections; however, qualitatively similar results are obtained so long as the tails of the function's temporal Fourier transform decay sufficiently fast for large frequencies.
The effect of the choice on heating is studied in Appendix~\ref{app:TimeCorrCompare}.

\subsubsection{Full model}\label{sec:FullModel}

With the above choices for the energy spectrum and temporal correlation of the turbulence, the following functional form is set forward as a simple model to capture the general dependence of RMHD turbulence on imbalance.
Defining\footnote{The cylindrical Jacobian in $\kvec$-space within $\energyspecE_{\rm 2D}$ has been removed, as these quantities will be integrated over all $\kvec$ below.}
\begin{equation}
    \energyspecE^\pm\left(\kprp,\kprl,\omega;\crosshel\right)\equiv\frac{\energyspecE_{\rm 2D}(\kprp,\kprl)}{2\upi\kprp\vA^2}F\left(\omega\pm\kprl\vA;\kprp,\crosshel\right),\label{eq:EpmFourier}
\end{equation}
where $F(\omega)$ is the temporal Fourier transform of the correlation function $f(\tau)$, the total wavevector-frequency spectrum is
\begin{align}
\energyspecE_{\rm tot}\left(\kprp, \kprl, \omega;\crosshel\right) &\equiv \frac{1}{2}\left[(1+\crosshel)\energyspecE^+\left(\kprp,\kprl,\omega;\crosshel\right) +(1-\crosshel)\energyspecE^-\left(\kprp,\kprl,\omega;\crosshel\right)\right]\nonumber\\
&= \frac{V\rhoi^2\betai^{1/2}}{4}\frac{\energyspecEtildetwoD(\kprp,\kprl)}{2\upi\kprp} \sum_{\nu=\pm 1}(1+\nu\crosshel)F(\omega+\nu\kprl\vA;\kprp,\crosshel),\label{eq:RMHDFunctionalForm}
\end{align}
where \eqref{eq:RMHDSpecNorm} has been used for $\energyspecE_{\rm 2D}(\kprp,\kprl)$.
The amplitudes of the $\zp$ and $\zm$ spectra are weighted such that their ratio is $\forceratio$.

\begin{figure}
    \centering
    \includegraphics[width=0.95\textwidth]{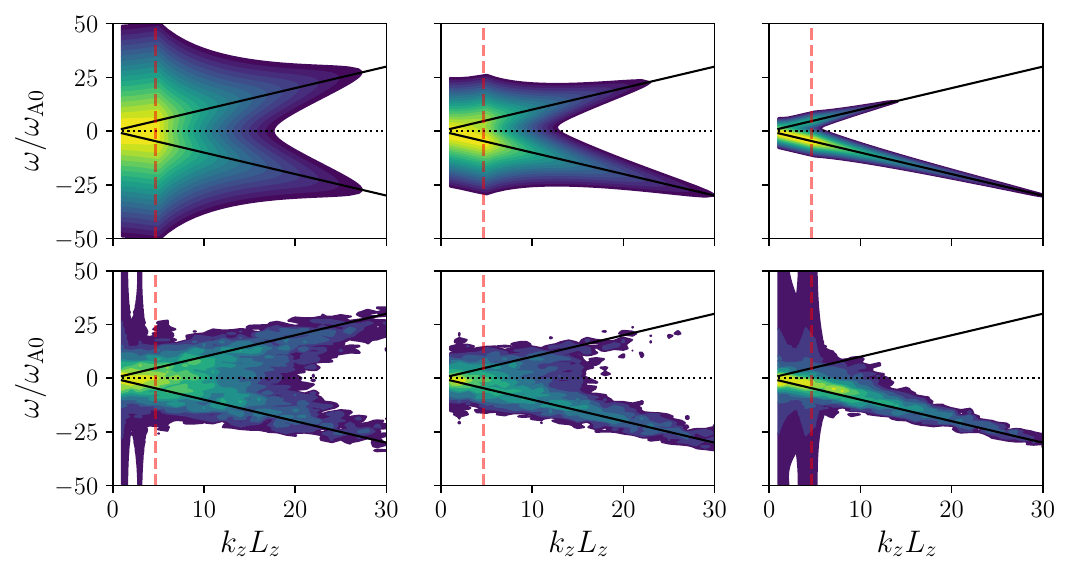}
    \caption{Slices at constant $\kprp\Lprp = 10$ through the model wavevector-frequency spectrum of RMHD turbulence (\ref{eq:RMHDFunctionalForm}, top), showing it qualitatively reproduces features of the RMHD simulations presented in Appendix~\ref{app:RMHDModel} (bottom).
    The turbulence has imbalance $\crosshel = 0,\ 0.59$, and $0.96$ (left, middle and right columns respectively), and the model sets $\stocfracth = 1$.
    The solid black lines correspond to zero frequency and the Alfv\'en dispersion relation $\omegaA = \pm k_z\vA$; the red dashed line is the critical balance scaling $\kprl^{\rm CB}$, such that everything with $\kprl < \kprl^{\rm CB}$ is within the CB cone.
    The frequencies are normalised to the outer-scale Alfv\'en frequency of the simulations, $\omega_{\rm A0}=\vA/L_z$.
    Note that the appearance of energy in high frequencies at small $k_z$ in the bottom left and right figures is an artefact of the method used to calculate the spectrum.}
    \label{fig:SpectroTheoryKprpSlice}
\end{figure}

\begin{figure}
    \centering
    \includegraphics[width=0.95\textwidth]{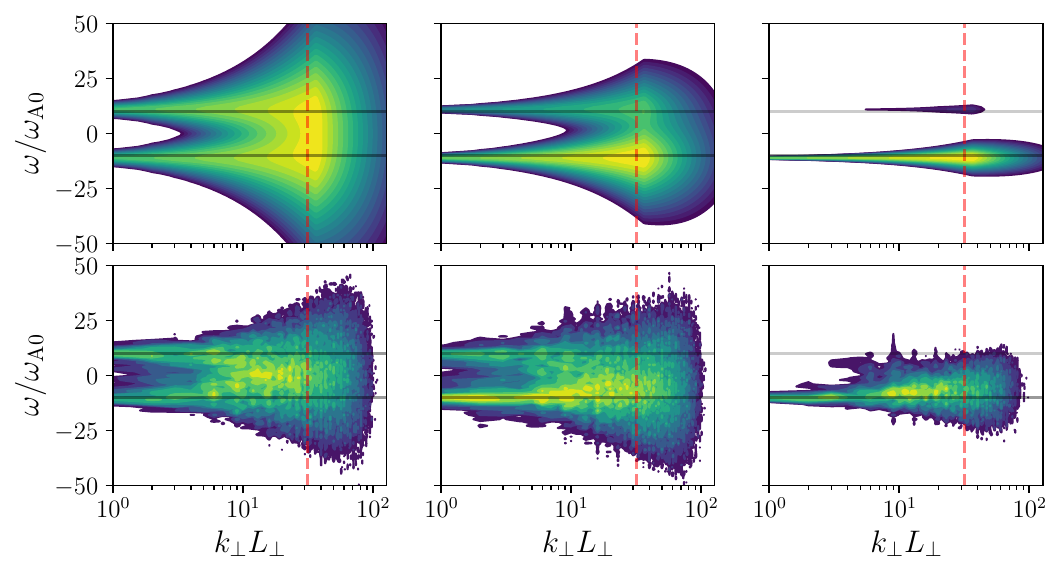}
    \caption{Slices at constant $k_z L_z = 10$ through the model wavevector-frequency spectrum of RMHD turbulence (\ref{eq:RMHDFunctionalForm}, top), showing it qualitatively reproduces features of the RMHD simulations presented in Appendix~\ref{app:RMHDModel} (bottom).
    The turbulence has imbalance $\crosshel = 0,\ 0.59$, and $0.96$ (left, middle and right columns respectively), and the model sets $\stocfracth = 1$.
    Solid black lines represent the centre of the Alfv\'en dispersion relation $\omegaA=\pm k_z\vA$ at the given value of $k_z$, and the red dashed line shows where $\kprp = (\kprl^{\rm CB})^{3/2}$.
    Frequencies are normalised to the outer-scale Alfv\'en frequency, $\omega_{\rm A0}=\vA/L_z$.}
    \label{fig:SpectroTheoryKzSlice}
\end{figure}

Figures~\ref{fig:SpectroTheoryKprpSlice} and \ref{fig:SpectroTheoryKzSlice} compare the model in \eqref{eq:RMHDFunctionalForm} to the wavevector-frequency spectra of the simulations in Appendix~\ref{app:RMHDModel} for $\crosshel=0$, 0.59, and 0.96.
Slices are taken through the spectra at $\kprp\Lprp=10$ and $k_z L_z=10$, respectively, with $\sCB=2/3$ and $\stocfracth=1$.
For $\kprl L_z > (\kprp \Lprp)^{2/3}$ (to the right or left of the red dashed line in figures~\ref{fig:SpectroTheoryKprpSlice} or \ref{fig:SpectroTheoryKzSlice}, respectively), the spectrum consists of two bands of fluctuations centred on the $\zpm$ dispersion relations $\omegaA^\pm = \mp\kprl\vA$, corresponding to the wave-like nature of turbulence in this region.
For modes below the CB cone with $\kprl L_z \lesssim (\kprp\Lprp)^{2/3}$ (to the left or right of the red dashed line in figures~\ref{fig:SpectroTheoryKprpSlice} or \ref{fig:SpectroTheoryKzSlice}, respectively) the turbulence becomes strong, with nonlinear interactions causing the spectrum to broaden; the width of this broadening increases with $\kprp$ (figure~\ref{fig:SpectroTheoryKzSlice}) as the nonlinear frequency $\omeganl(\kprp)$ is an increasing function of $\kprp$.
Additionally, the width of the spectrum decreases as the imbalance increases, corresponding to the reduced efficiency of the cascade of $\zp$ fluctuations as the $\zm$ fluctuations become subdominant.
In the limit of fully imbalanced turbulence ($\crosshel\to 1$), the spectrum would become a delta function along the $\zp$ dispersion relation $\omegaA = -\kprl\vA$.

Figures~\ref{fig:SpectroTheoryKprpSlice} and \ref{fig:SpectroTheoryKzSlice} show that the model \eqref{eq:RMHDFunctionalForm} qualitatively captures the features of the wavevector-frequency spectra of RMHD turbulence, as well as its dependence on the turbulence's imbalance.
The ability to express the spectrum in this form makes it useful for connecting theoretical models of RMHD turbulence to numerical or observational data across a range of plasma conditions.

\section{Calculation of heating rates in RMHD turbulence with varying imbalance}\label{sec:HeatingRates}

We now combine our quasi-linear theory results (Section~\ref{sec:QLTheory}) and turbulent spectrum (Section~\ref{sec:RMHDTurbulence}) to predict heating rates.
We will find that the heating rate has a similar form regardless of the value of $\crosshel$, illustrating the connection between the mechanisms of stochastic heating in balanced turbulence and cyclotron-resonant heating in imbalanced turbulence.

In Section~\ref{sec:generaldiffusioncoeff}, we present the form of the diffusion coefficients used, where additional assumptions are introduced to simplify the RMHD diffusion coefficients given in \eqref{eq:QLDiffusionCoefficientsAlfvenicPol}.
The corresponding expression for the perpendicular heating rate, $\Qprp$, is also provided under the simplifying assumption of an initially Maxwellian ion distribution.
We then examine ion heating in the fully imbalanced and balanced limits of these coefficients in Section~\ref{sec:limitingcases} to build intuition for the general dependence of $\Qprp$ on $\crosshel$, discussed in Section~\ref{sec:generalcase}.
Finally, in Section~\ref{sec:f0_Evolution} we show the general time evolution of a Maxwellian distribution due to the diffusion coefficients calculated in Section~\ref{sec:generaldiffusioncoeff}, and show how the turbulence's imbalance affects the structure of the distribution function.

\subsection{Further simplification of diffusion coefficients}\label{sec:generaldiffusioncoeff}

The RMHD framework introduced in Section~\ref{sec:RMHDLimitCoeffs} considerably reduces the complexity of the general quasi-linear diffusion coefficients derived in Appendix~\ref{app:QLTheoryDerivation}.
This section applies additional physical assumptions and symmetry considerations to further simplify the expressions in \eqref{eq:QLDiffusionCoefficientsAlfvenicPol}.

We first assume that there is equipartition between the normalised kinetic and magnetic energies of the turbulent fluctuations for all levels of imbalance (as observed in the RMHD simulations presented in Appendix~\ref{app:RMHDModel}; see figure~\ref{fig:RMHDSimKprpEnergySpec}), and set $\mathcal{K}(\kvec,\tau)=\mathcal{M}(\kvec,\tau)$ for the normalised power spectra in \eqref{eq:QLDiffusionCoefficientsAlfvenicPol}.
Additionally, we take
\begin{equation}
    \frac{\mathcal{C}(\kvec,\tau)}{\mathcal{K}(\kvec,\tau)} = \frac{4\mathcal{C}(\kvec,\tau)}{2(\mathcal{K}(\kvec,\tau)+\mathcal{M}(\kvec,\tau))} = \frac{\energyspecE^+(\kvec,\tau)-\energyspecE^-(\kvec,\tau)}{\energyspecE^+(\kvec,\tau) +\energyspecE^-(\kvec,\tau)} = \crosshel,
\end{equation}
where the normalised power spectra of the Elsasser fields are defined as
\begin{equation}
    \energyspecE^\pm(\kvec,\tau) \equiv \frac{\langle \zpm(\kvec, t)\dotprod(\zpm)^*(\kvec, t+\tau) \rangle}{\vA^2}=\mathcal{K}\pm2\mathcal{C}+\mathcal{M},\label{eq:ElsasserCorrelator}
\end{equation}
and $\crosshel$ is assumed to be independent of $\kvec$.
Using these assumptions in \eqref{eq:ElsasserCorrelator} and substituting them into the model form of $\energyspecE_{\rm tot}$, \eqref{eq:RMHDFunctionalForm}, we find that
\begin{equation}
    \mathcal{K}(\kvec,\tau) = \frac{1}{2(1+\crosshel^2)}\energyspecE_{\rm tot}(\kvec,\tau).
\end{equation}
Taking the inverse temporal Fourier transform of \eqref{eq:RMHDFunctionalForm} and using the property $\hat{\mathcal{F}}^{-1}\left[f(\omega \pm \omega_0)\right] = e^{\mp\Icplx\omega_0\tau}\hat{\mathcal{F}}^{-1}\left[f(\omega)\right]$ allows the diffusion coefficients in \eqref{eq:QLDiffusionCoefficientsAlfvenicPol} to be written as
\begin{equation}
    \frac{1}{\Omegai\vthi^2}\begin{pmatrix} \Dprpprp\\ \Dprlprp\\ \Dprlprl\end{pmatrix} = \begin{pmatrix} 1+2\crosshel\vprltilde+\vprltilde^2\\ -\vprptilde(\crosshel+\vprltilde)\\ \vprptilde^2\end{pmatrix} \frac{\mathcal{D}}{\Omegai\vthi^2},\label{eq:GeneralDiffusionCoefficients}
\end{equation}
with the \emph{generalised diffusion coefficient} defined as
\begin{align}
    \frac{\mathcal{D}}{\Omegai\vthi^2}&\equiv\frac{\Omegai\rhoi^2}{\sqrt{\betai}}\sum_{n=-\infty}^{\infty}\sum_{\nu=\pm 1}\times\nonumber\\
    &\frac{1+\nu\crosshel}{8(1+\crosshel^2)}\intsingle{\kvec}{}{}\frac{n^2\BesselJ{n}^2(\kappa)}{\kappa^2}\frac{\energyspecEtildetwoD(\kprp,\kprl)}{2\upi\kprp}\intsingle{\tau}{0}{\infty}e^{-\Icplx\left[\kprl\left(\vprl+\nu\vA\right)+n\Omegai\right]\tau}f(\tau\omeganl(\kprp)),
\end{align}
where \eqref{eq:RMHDFunctionalForm} has been inserted for $\energyspecE_{\rm tot}(\kprp,\kprl,\omega)$.

Changing variables to $\kprptilde = \kprp\rhoi$, $\kprltilde = \kprl\vA/\Omegai$, $\vprltilde = \vprl/\vA$, $\tautilde\equiv \Omegai\tau$, and $\omeganltilde(\kprptilde)\equiv \omeganl/\Omegai$ gives
\begin{align}
    \frac{\mathcal{D}}{\Omegai\vthi^2}&\approx\sum_{n,\nu=\pm 1}\frac{1+\nu\crosshel}{32(1+\crosshel^2)}\times\nonumber\\
    &\intsingle{\kprptilde}{0}{1}\intsingle{\kprltilde}{-1}{1}\energyspecEtildetwoD(\kprptilde,\kprltilde)\intsingle{\tautilde}{0}{\infty}e^{-\Icplx\left[\kprltilde\left(\vprltilde+\nu\right)+n\right]\tautilde}f(\tautilde\omeganltilde(\kprptilde)).\label{eq:RMHDDiffusionCoeffsSimplified}
\end{align}
The approximate equality above arises from the RMHD $\kprp\rhoi\ll 1$ limit: using the fact that the argument of the Bessel functions is $\kappa = \kprp\vprp/\Omegai = (\kprp\rhoi)\vprp/\vthi$ and assuming $\vprp\sim\vthi$, we have
\begin{equation}
    \frac{n^2 \BesselJ{n}^2(\kappa)}{\kappa^2} \approx \left(\frac{1}{2(n-1)!}\right)^2\left(\frac{\kprp\rhoi}{2}\right)^{2|n|-2}\to\begin{cases}
        1/4,& n=\pm 1,\\
        0,& |n| > 1,
    \end{cases}
\end{equation}
in the limit $\kprp\rhoi\ll 1$, allowing all but two terms in the Bessel function sum to be neglected.
While this approximation allows for the calculation to be simplified enormously, removing an extra dependence on $\kprp$ and $\vprp$, it is not fully justified near $\kprp\rhoi = 1$, where the Bessel functions begin to vary.\footnote{In Appendix~\ref{app:SubRhoHeating} we show that when the full Bessel functions are retained in the calculation of $\mathcal{D}$, the $n = \pm 1$ terms contribute significantly more than higher-order terms, justifying this approximation.}

The sum over $n=\pm 1$ in \eqref{eq:RMHDDiffusionCoeffsSimplified} can be simplified by taking advantage of the symmetries of $\energyspecEtildetwoD$ and $f(\tau)$.
Expanding the sum in $n$ and changing variables $\tautilde\to-\tautilde$ and $\kprltilde\to-\kprltilde$ in the $n=-1$ term gives a term of the form
\begin{equation}
    \intsingle{\kprltilde}{-1}{1}\energyspecEtildetwoD(\kprptilde,-\kprltilde)\intsingle{\tautilde}{-\infty}{0}e^{-\Icplx\left[\kprltilde\left(\vprltilde+\nu\right)+1\right]\tautilde}f(-\tautilde\omeganltilde(\kprptilde)).\label{eq:blah}
\end{equation}
As $f(\tau)$ and $\energyspecEtildetwoD$ are even in $\tau$ and $\kprltilde$, respectively, \eqref{eq:blah} is now equivalent to the $n=1$ term with flipped integration limits in $\tautilde$, allowing \eqref{eq:RMHDDiffusionCoeffsSimplified} to be written as
\begin{equation}
    \frac{\mathcal{D}}{\Omegai\vthi^2} = \sum_{\nu=\pm 1}\frac{1+\nu\crosshel}{32(1+\crosshel^2)}\intsingle{\kprptilde}{0}{1}\intsingle{\kprltilde}{-1}{1}\energyspecEtildetwoD\intsingle{\tautilde}{-\infty}{\infty}e^{-\Icplx\left[\kprltilde\left(\vprltilde+\nu\right)+1\right]\tautilde}f(\tautilde\omeganltilde(\kprptilde)).
\end{equation}
Finally, we note that the integral over $\tautilde$ is of a similar form to a temporal Fourier transform; namely, it is $2\upi\mathcal{F}\left[f(\tau)\right]$, \eqref{eq:TemporalFTConvention}, with $\omega\to \kprltilde\left(\vprltilde+\nu\right) + 1$.
With the Fourier transform of $f(\tautilde\omeganltilde)=\sech(\tautilde\omeganltilde)$ given in \eqref{eq:SechFrequencyCorrelationFunction}, we obtain the form of $\mathcal{D}$ to be used in this section:
\begin{align}
    \frac{\mathcal{D}}{\Omegai\vthi^2} &= \sum_{\nu=\pm 1}\frac{1+\nu\crosshel}{16(1+\crosshel^2)}\times\nonumber\\
    &\intsingle{\kprptilde}{0}{1}\frac{\upi}{2\omeganltilde(\kprptilde)}\intsingle{\kprltilde}{-1}{1}\energyspecEtildetwoD(\kprptilde,\kprltilde)\sech\left(\frac{\upi}{2\omeganltilde(\kprptilde)}\left[\kprltilde\left(\vprltilde+\nu\right) + 1\right]\right).\label{eq:DppGeneralIntegrand}
\end{align}

\subsubsection{Perpendicular heating rate of a Maxwellian distribution}\label{sec:MaxwellianHeatingRate}

The velocity-space diffusion of the distribution function results in particle heating.
In this section we derive the form of the perpendicular heating rate $\Qprp$ arising from this diffusion.
For simplicity, we assume the initial ion distribution is Maxwellian,
\begin{equation}\label{eq:MaxwellianDist}
    f_0(\vprptilde, \vprltilde)=n_0\left(\upi\betai\right)^{-3/2}\exp{\left(-\frac{\vprptilde^2+\vprltilde^2}{\betai}\right)},
\end{equation}
where $\tilde{v}_{\perp,\|} \equiv v_{\perp,\|}/\vA$ and $f_0$ is normalised such that $\intsingle{\vprltilde}{-\infty}{\infty}\intsingle{\vprptilde}{0}{\infty}2\upi\vprptilde f_0 = \vA^{3}n_0$.

The perpendicular heating rate per unit mass is given by
\begin{equation}
    \Qprp \equiv \totfrac{\langle \vprp^2 /2 \rangle}{t}= \intsingle{\vprl}{-\infty}{\infty}\intsingle{\vprp}{0}{\infty} 2\upi\vprp\left(\frac{\vprp^2}{2}\right) \frac{1}{n_0}\parfrac{f_0}{t}.\label{eq:PerpendicularHeatingRate}
\end{equation} 
Assuming that ions undergo a diffusion in velocity space in the quasi-linear form \eqref{eq:QLDiffusionForm}, \eqref{eq:PerpendicularHeatingRate} can be written as
\begin{align}
    \Qprp &= \intsingle{\vprl}{-\infty}{\infty}\intsingle{\vprp}{0}{\infty}2\upi\vprp \frac{\vprp^2}{2}\frac{1}{\vprp}\parfrac{}{\vprp}\left[\vprp\frac{1}{n_0}\left(\Dprpprp\parfrac{f_0}{\vprp}+\Dprlprp\parfrac{f_0}{\vprl}\right)\right]\nonumber\\
    &= -2\upi\intsingle{\vprl}{-\infty}{\infty}\intsingle{\vprp}{0}{\infty}\vprp^2\frac{1}{n_0}\left(\Dprpprp\parfrac{f_0}{\vprp}+\Dprlprp\parfrac{f_0}{\vprl}\right),\label{eq:PerpendicularHeatingRate2}
\end{align}
by assuming that the velocity gradients of a physically-reasonable distribution function should go to zero as $\norm{\pvel}\to\infty$ in order to remove $\parfracil{}{\vprl}$ terms via integration by parts; the second equality is obtained by further integration by parts over $\vprp$.
Changing variables to $\tilde{v}_{\perp,\|}$ and using the derivatives of the Maxwellian, \eqref{eq:PerpendicularHeatingRate2} becomes
\begin{equation}
    \Qprp = \frac{4}{\sqrt{\upi\betai^5}}\intsingle{\vprltilde}{-\infty}{\infty}e^{-\vprltilde^2/\betai}\intsingle{\vprptilde}{0}{\infty} \vprptilde^2 e^{-\vprptilde^2/\betai} \left(\vprptilde\Dprpprp+\vprltilde\Dprlprp\right).\label{eq:QprpDprp}
\end{equation}
Finally, inserting the diffusion coefficients \eqref{eq:GeneralDiffusionCoefficients} and noting that $\mathcal{D}$ in \eqref{eq:DppGeneralIntegrand} is independent of $\vprptilde$, we obtain
\begin{equation}
    \frac{\Qprp}{\Omegai\vthi^2} = \sqrt{\frac{4}{\upi\betai}}\intsingle{\vprltilde}{-\infty}{\infty}e^{-\vprltilde^2/\betai}(1+\crosshel\vprltilde)\frac{\mathcal{D}(\vprltilde)}{\Omegai\vthi^2}. \label{eq:QprpGeneralCase}
\end{equation}
This form will be used to calculate $\Qprp$ analytically and numerically in Sections~\ref{sec:limitingcases} and \ref{sec:generalcase} below.
We note that the form of $\Qprp$ in \eqref{eq:QprpGeneralCase} only gives the instantaneous heating rate of a Maxwellian distribution, and does not describe the heating once the diffusion causes $f_0$ to evolve away from a Maxwellian.
This evolution is shown in more detail in Section~\ref{sec:f0_Evolution}, where we numerically solve the quasi-linear diffusion equation \eqref{eq:QLDiffusionForm}.

\subsection{Limiting cases}\label{sec:limitingcases}

Before presenting the general calculation of the diffusion coefficients for arbitrary $\crosshel$, we first examine the limiting cases of fully imbalanced and balanced RMHD turbulence.
These cases represent well-understood physical regimes and provide intuition for interpreting the more general results discussed in Section~\ref{sec:generalcase}.

\subsubsection{Fully imbalanced turbulence: $\crosshel=1$}\label{sec:imbalancedlimit}

The fully imbalanced limit may be viewed as the case $\crosshel \to 1$, corresponding to a collection of $\zpm$ fluctuations undergoing a cascade that is infinitely slow compared to their propagation rate.
In this limit, the quasi-linear framework reduces to studying the effect of strong interactions between ions and waves following a linear dispersion relation $\omega(\kvec)$, with the evolution of the distribution function controlled by the resonance condition \eqref{eq:ResonanceCondition}.
When this condition is satisfied, the distribution function diffuses in velocity space along constant energy contours in the wave frame (\citealp{Kennel1966-rx}; we prove this explicitly in Appendix~\ref{app:KandEReduction}).
For the Alfv\'enic $\zp$ fluctuations, with $\omega(\kvec) = -\kprl\vA$, the resonance condition becomes
\begin{equation}
    \kprl(\vprl+\vA)+n\Omegai = 0.\label{eq:ZpResonance}
\end{equation}

Taking the limit $\crosshel\to 1$ in the model of the wavevector-frequency spectrum in Section~\ref{sec:FullModel}, the nonlinear frequency $\omeganltilde(\kprptilde)$ goes to zero ($\alpha = (1+\crosshel)/(1-\crosshel) \to \infty$ in \ref{eq:OmegaNL}) and the temporal correlation function approaches a delta function \citep{Wheeler2002-wk},
\begin{equation}
    \frac{1}{2\omeganltilde(\kprptilde)}\sech\left(\frac{\upi}{2\omeganltilde(\kprptilde)}\left[\kprltilde\left(\vprltilde+\nu\right) + 1\right]\right)\to \delta\left(\kprltilde\left(\vprltilde+\nu\right) + 1\right),
\end{equation}
which is exactly the resonance condition \eqref{eq:ZpResonance} with $n=1$.
With this, in the limit $\crosshel\to 1$ the generalised diffusion coefficient in \eqref{eq:DppGeneralIntegrand} becomes
\begin{equation}
    \frac{\mathcal{D}}{\Omegai\vthi^2} = \frac{\upi}{16}\intsingle{\kprptilde}{0}{1}\intsingle{\kprltilde}{-1}{1}\energyspecEtildetwoD(\kprptilde,\kprltilde)\delta\left[\kprltilde(\vprltilde +1)+1\right].
\end{equation}
The delta function enforces that ions with parallel speed $\vprl$ interact with fluctuations with resonant wavenumber
\begin{equation}
\kprltilde^{(1)}\equiv-(1+\vprltilde)^{-1}.\label{eq:KprlResonance}
\end{equation}
Note that, because $|\kprltilde|\leq 1$ in this model (as for $|\kprltilde| = \kprl\ion{d}\gtrsim 1$ the fluctuations are no longer Alfv\'enic, transitioning to ICWs), resonance only occurs for ions with $\vprltilde>0$, counter-propagating along magnetic-field lines with respect to the waves.

The resonance condition \eqref{eq:KprlResonance} shows that the speed of the ions controls the part of the turbulent spectrum with which they are able to interact.
If ions are too slow,
\begin{equation}
\stocfracth \leq |\kprltilde^{(1)}|\leq 1 \implies 0 \leq\vprltilde\leq\stocfracth^{-1}-1,
\end{equation}
they can only interact with the portion of the spectrum above the CB cone; however, if $|\kprltilde^{(1)}| <\stocfracth$ they are also able to interact with that below the cone.
Performing the integrals, setting $\kprltilde = \kprltilde^{(1)}$ using the delta function and allowing for $\stocfracth\gtrsim 1$, we obtain (setting $C=1$ in $\energyspecEtildetwoD$, \ref{eq:RMHDSpecNorm})
\begin{equation}\label{eq:DppImbalancedTurb}
    \frac{\mathcal{D}}{\Omegai\vthi^2}=\frac{\upi}{16(\stwo+1)}\begin{cases}
    0, & \vprltilde < 0,\\
    \stocfracth^{\sone+1}(1+\vprltilde)^{\sone}, & 0\leq\vprltilde\leq\max(0,\stocfracth^{-1}-1),\\ 
    \frac{1}{2}\stocfracth^3\left[\sone(1+\vprltilde)^2-\frac{\stwo+1}{\sCB}\stocfracth^{-2}\right], & \vprltilde > \max(0,\stocfracth^{-1}-1);
    \end{cases}
\end{equation}
using the scalings for the \citet{Goldreich1995-fv} spectrum ($\sCB = 2/3$ and $\stwo=1$) gives
\begin{equation}\label{eq:DppImbalancedTurbGS95Spec}
    \frac{\mathcal{D}}{\Omegai\vthi^2}=\frac{\upi}{32}\begin{cases}
    0, & \vprltilde < 0,\\
    \stocfracth^6(1+\vprltilde)^5, & 0\leq\vprltilde\leq\max(0,\stocfracth^{-1}-1),\\ 
    \frac{1}{2}\stocfracth^3\left[5(1+\vprltilde)^2-3\stocfracth^{-2}\right], & \vprltilde > \max(0,\stocfracth^{-1}-1).\\
    \end{cases}
\end{equation}
This form of $\mathcal{D}$ transitions smoothly from a $\stocfracth^3$ to a $\stocfracth^6$ scaling with increasing $\vprltilde$.\footnote{The $\stocfracth^6$ scaling arises from the choice $\sCB = 2/3$ used in this section for $\energyspecEtildetwoD$, \eqref{eq:RMHDSpecNorm}; generally, $\Qprp \propto \stocfracth^{\sone+1}$ for $\stocfracth\ll 1$, where $\sone = (2\sCB + 2)/\sCB$.}

\begin{figure}
  \centering
  \includegraphics[width=0.8\textwidth]{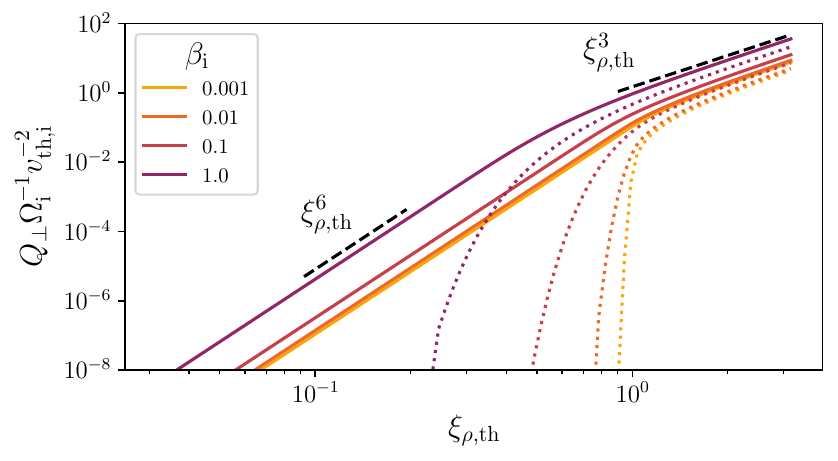}
  \caption{The perpendicular heating rate $\Qprp$ in the fully imbalanced limit, where the general quasi-linear theory of velocity-space diffusion reduces to diffusion along contours of constant energy in the wave frame \citep{Kennel1966-rx}.
  $\Qprp$ is calculated numerically from \eqref{eq:QprpImbalancedCase} using the imbalanced limit of the generalised diffusion coefficient $\mathcal{D}$ \eqref{eq:DppImbalancedTurbGS95Spec} for different values of $\betai$ (with $C=1$ in $\energyspecEtildetwoD$, \ref{eq:RMHDSpecNorm}).
  Dotted lines show only the contribution of fluctuations below the CB cone to $\Qprp$.}
\label{fig:QprpImbalanced}
\end{figure}

In the limit $\crosshel\to 1$, the general heating rate in \eqref{eq:QprpGeneralCase} reduces to
\begin{equation}
    \frac{\Qprp}{\Omegai\vthi^2} = \sqrt{\frac{4}{\upi\betai}}\intsingle{\vprltilde}{-\infty}{\infty}e^{-\vprltilde^2/\betai}(1+\vprltilde)\frac{\mathcal{D}(\vprltilde)}{\Omegai\vthi^2}. \label{eq:QprpImbalancedCase}
\end{equation}
Figure~\ref{fig:QprpImbalanced} shows $\Qprp$ as a function of $\stocfracth$ and $\betai$, calculated numerically using the integral in \eqref{eq:QprpImbalancedCase}.
The solid lines are calculated using the form of $\mathcal{D}$ in \eqref{eq:DppImbalancedTurbGS95Spec}, which shows the contribution of modes in $\kvec$-space both above and below the CB cone to the heating rate.
These inherit the scaling with $\stocfracth$ present in $\mathcal{D}$, transitioning from a $\stocfracth^6$ scaling to a $\stocfracth^3$ scaling with increasing turbulent amplitude.
The presence of the $\stocfracth^6$ scaling at small $\stocfracth$ (a feature also seen in the balanced limit and general calculation below) suggests that ions can be heated by weak fluctuations above the CB cone (although the assumptions used in deriving this model may mean this is not quite correct; see Section~\ref{sec:Conclusion} for a discussion of the caveats of this model).
The heating rates in figure \ref{fig:QprpImbalanced} show little to no variation with $\betai$ for $\betai\ll 1$.
This is because only ions with $\vprltilde\ll 1$ contribute to the integral; ions with large $\vprltilde$ will contribute less to the total heating rate $\Qprp$ as there are comparatively fewer of them.

If we consider ions only interacting with strong fluctuations below the CB cone (dotted lines in figure~\ref{fig:QprpImbalanced}), we instead see that the heating rate is strongly suppressed at small $\stocfracth$.
This follows from an analogous calculation for $\mathcal{D}$ in \eqref{eq:DppImbalancedTurbGS95Spec} where only the contribution of modes below the CB cone ($|\kprltilde| \leq \kprltildeCB$) are taken into account, which gives
\begin{equation}
    \frac{\mathcal{D}}{\Omegai\vthi^2} = \frac{3\upi}{64}\stocfracth^3\left[(1+\vprltilde)^2 - \stocfracth^{-2}\right]\label{eq:DppImbalancedStrong}
\end{equation}
for $\vprltilde > \max(0,\stocfracth^{-1}-1)$ and $0$ otherwise.
For ions that are fast enough to interact with modes below the CB cone, $\mathcal{D}$ scales as $\stocfracth^3$ and drops off rapidly as $\stocfracth$ approaches $1/(1+\vprltilde)$.
This is quite restrictive compared to the $\stocfracth\gtrsim 0.1$ requirement of stochastic heating for $\vprltilde\ll1$, although the exact threshold depends on the choice of $C$. It is a result of the resonance condition \eqref{eq:KprlResonance} allowing ions to interact with a single frequency.

\subsubsection{Balanced turbulence: $\crosshel=0$}\label{sec:balancedlimit}

In contrast to the imbalanced limit, balanced turbulence consists of a collection of $\zp$ and $\zm$ fluctuations with equal energies.
The nonlinear interactions between
the two populations lead to broadening of the wavevector-frequency spectrum, which is
captured by the temporal correlation function \eqref{eq:SechTimeCorrelationFunction}.
Using \eqref{eq:DppGeneralIntegrand} for the generalised diffusion coefficient with $\crosshel=0$ (where $\alpha=1$ in $\omeganltilde$), we have
\begin{align}
    \frac{\mathcal{D}}{\Omegai\vthi^2} &= \frac{1}{16}\sum_{\nu=\pm 1}\times\nonumber\\
    &\intsingle{\kprptilde}{0}{1}\frac{\upi}{2\omeganltilde(\kprptilde)}\intsingle{\kprltilde}{-1}{1}\energyspecEtildetwoD(\kprptilde,\kprltilde)\sech\left(\frac{\upi}{2\omeganltilde(\kprptilde)}\left[\kprltilde\left(\vprltilde+\nu\right)+1\right]\right).\label{eq:BalancedDpp3}
\end{align}
The integral \eqref{eq:BalancedDpp3} is now symmetric in $\vprltilde$ because there are fluctuations travelling in both directions along the magnetic field.
The general heating rate \eqref{eq:QprpGeneralCase} in this limit reduces to
\begin{equation}
    \frac{\Qprp}{\Omegai\vthi^2} = \sqrt{\frac{4}{\upi\betai}}\intsingle{\vprltilde}{-\infty}{\infty}e^{-\vprltilde^2/\betai}\frac{\mathcal{D}(\vprltilde)}{\Omegai\vthi^2}. \label{eq:QprpBalancedCase}
\end{equation}

The full integral of \eqref{eq:BalancedDpp3} is not able to be completed analytically over all $\kvec$-space, and can only be calculated numerically.
To investigate its limiting behaviour, we first use the assumption of low-$\betai$ ions such that $\vprltilde = \vprl/\vA\ll 1$, removing the dependence of $\mathcal{D}$ on $\vprltilde$ in \eqref{eq:BalancedDpp3}.
This approximation essentially ignores the equivalent resonant condition in the imbalanced limit, where ions are only able to interact with a single frequency at a given $\kvec$ (see \ref{eq:KprlResonance}).
However, because the turbulent fluctuations in the balanced case have a greater spread in frequency due to their nonlinear broadening, ions are still able to diffuse in velocity space as they interact with a variety of fluctuations with differing frequencies at a given $\kvec$.

\begin{figure}
  \centering
  \includegraphics[width=0.8\textwidth]{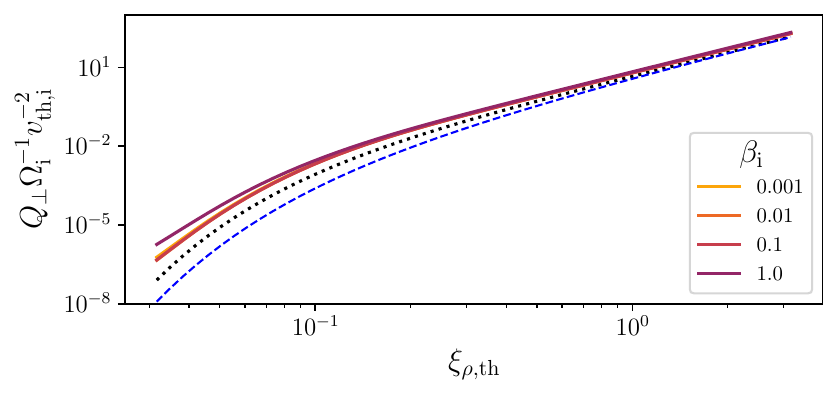}
  \caption{The perpendicular heating rate $\Qprp$ in the balanced limit for different values of $\betai$, calculated numerically using \eqref{eq:QprpBalancedCase} with $C=5$ in $\energyspecEtildetwoD$ \eqref{eq:RMHDSpecNorm}.
  To better highlight the suppression, only the contribution from modes below the CB cone are considered.
  The black dotted line shows the $\betai$-independent analytic expression for $\Qprp$ \eqref{eq:QprpBalancedLowBeta} (obtained in the limit $|\vprltilde| \ll 1$ and $\kprltilde\ll \kprltildeCB$) with $\hat{c}_1\approx 5$ and $\hat{c}_2 \approx 0.3$.
  These show qualitatively similar behaviour to the empirical stochastic heating formula \eqref{eq:SHRate} \citep{Chandran2010-ow}, shown by the blue dashed line with the same values for $c_1$ and $c_2$.}
\label{fig:QprpBalanced}
\end{figure}

With the $\vprltilde\ll 1$ limit, the contributions of the integral \eqref{eq:BalancedDpp3} to ion heating from below the CB cone can be investigated by looking at fluctuations such that the ratio
\begin{equation}
\frac{\kprltilde}{\omeganltilde(\kprptilde)} = \frac{\kprltilde}{C\stocfracth \kprptilde^{\sCB}}\lesssim 1.\label{eq:BelowCBCone}
\end{equation}
In this region $\energyspecEtildetwoD = C^{-1}\stocfracth\kprptilde^{-2\sCB-1}$ and, after taking the $x\gg 1$ limit of $\sech(x)$ in \eqref{eq:Sech} as $1/ \omeganltilde = \Omegai/ \omeganl \gtrsim 1$, \eqref{eq:BalancedDpp3} becomes\footnote{The integral of the $\sech$ function over the CB cone ($|\kprltilde|\leq\kprltildeCB$) can be computed exactly in terms of special functions because $\energyspecEtildetwoD$ is independent of $\kprltilde$ in this region; however, the resulting expression is cumbersome and does not offer further physical insight, so it is not shown here.}
\begin{align}
    \frac{\mathcal{D}}{\Omegai\vthi^2} &\approx \frac{1}{4}\intsingle{\kprptilde}{0}{1}C^{-1}\stocfracth\kprptilde^{-2\sCB-1}\frac{\upi}{2\omeganltilde(\kprptilde)}\exp\left(-\frac{\upi}{2\omeganltilde(\kprptilde)}\right)\intsingle{\kprltilde}{-C\stocfracth\kprptilde^{\sCB}}{C\stocfracth\kprptilde^{\sCB}}\nonumber\\
    &= \frac{C}{\upi\sCB}\stocfracth^3\intsingle{u}{\upi/(2C\stocfracth)}{\infty}u e^{-u}\nonumber\\
    &= \frac{C}{\upi\sCB}\stocfracth^3 \left(1 + \frac{\upi}{2C\stocfracth}\right)e^{-\upi / (2C\stocfracth)},\label{eq:BalancedDppStocFrac}
\end{align}
where a change of variables to $u = \upi / (2\omeganltilde(\kprptilde)) = (\upi/2)(C\stocfracth)^{-1}\kprptilde^{-\sCB}$ was made in the second equality.
Finally, as $\mathcal{D}$ is independent of $\vprltilde$ for low-$\betai$ ions, the heating rate \eqref{eq:QprpBalancedCase} becomes
\begin{equation}
    \frac{\Qprp}{\Omegai\vthi^2} = \hat{c}_1\stocfracth^3 \left(1 + \frac{\hat{c}_2}{\stocfracth}\right)e^{-\hat{c}_2/ \stocfracth},\label{eq:QprpBalancedLowBeta}
\end{equation}
where $\hat{c}_1 \equiv 2C/(\upi\sCB) \approx C$ (for the \citet{Goldreich1995-fv} CB scaling with $\sCB=2/3$) and $\hat{c}_2 \equiv \upi / (2C) \approx 1.5C^{-1}$.

Remarkably, \eqref{eq:QprpBalancedLowBeta} has an almost identical form to the empirical stochastic heating formula \eqref{eq:SHRate} \citep{Chandran2010-ow}.
The coefficient $c_2$ in \eqref{eq:SHRate} is generally found to be $\lesssim 0.5$ in numerical simulations of stochastic heating \citep{Chandran2010-ow,Xia2013-ob,Cerri2021-xo,Johnston2025-ss}; using this to constrain $C$ we find $C\gtrsim 3$ and $\hat{c}_1\gtrsim 3$.
This may be larger than results reported in past simulations, although there is no agreement on exact values for $c_1$; for example, \citet{Xia2013-ob} find $0.5\lesssim c_1 \lesssim 1.5$ in low-$\betai$ balanced RMHD turbulence and $c_1\approx 4$ at $\betai=1$, but \citet{Johnston2025-ss} find $c_1\approx 4$ in balanced turbulence at $\betai = 0.05$.
Given the drastic assumptions of quasi-linear theory and the likely dependence on other properties such as the choice of forcing or inertial-range length (for instance, as a consequence of intermittency; \citealp{Mallet2019-ks,Bowen2025-tv}), the general agreement of the functional form and order-unity agreement of the coefficients is encouraging.

Figure~\ref{fig:QprpBalanced} shows the integral \eqref{eq:QprpBalancedCase} used to calculate $\Qprp$ for different values of $\betai$.
To compare to previous simulation results discussed in the previous paragraph, we determine $C$ from a representative value of the coefficient $\hat{c}_2\approx 0.3$ using the relation $\hat{c}_2\approx 1.5C^{-1}$ from \eqref{eq:QprpBalancedLowBeta}, giving $C=5$; the value $\hat{c}_1\approx 5$ then follows from the same relations.
Additionally, we only take into account the contribution from modes below the CB cone to better show the form of the heating-rate suppression.
The calculated curves show little to no variation with $\betai$ when $\betai\ll1$, with small changes at $\stocfracth \ll 1$ when $\betai=1$.
This behaviour is exactly what is expected from stochastic heating, where $\Qprp/(\Omegai\vthi^2)$ is independent of $\betai$ for $\betai\ll 1$ and heating in small-amplitude turbulence is suppressed due to the conservation of the magnetic moment of the ions \citep{Chandran2010-ow} (however, due to the term $1+\hat{c}_2/\stocfracth$ in \eqref{eq:QprpBalancedLowBeta}, the suppression of $\Qprp$ calculated by this model decreases more slowly than that of the empirical form for the choice of parameters used).
If the contribution of modes above the CB cone are also taken into account, the model predicts the heating of ions by weak fluctuations as in the imbalanced case; this is explicitly shown by the $\stocfracth^6$ scaling for small $\stocfracth$ in figure~\ref{fig:DppGeneral} below.

Finally, we note that the suppression in \eqref{eq:BalancedDppStocFrac} and \eqref{eq:QprpBalancedLowBeta} arises in part from the form of the temporal correlation function, rather than just from the exponential dependence of the Maxwellian distribution function assumed.
This indicates that the effect is fundamentally determined by the dynamical properties of the turbulence and its coupling to the ions, as noted previously \citep{Cerri2021-xo,Mallet2026-ei}.
This perspective contrasts with the original interpretation in \eqref{eq:SHRate}, where the suppression is attributed to the conservation of the magnetic moment of ions in the presence of small-amplitude fluctuations \citep{Chandran2010-ow}.

\subsection{Heating-rate properties: arbitrary imbalance}\label{sec:generalcase}

After gaining intuition for the heating of ions in the limiting cases of balanced and imbalanced turbulence in Section~\ref{sec:limitingcases}, we now study the properties of the heating rate for general values of $\crosshel$.
In particular, we show how the scalings of $\Qprp$ with $\stocfracth$ arise from the properties of the wavevector-frequency spectrum in the calculation of $\mathcal{D}$.

\begin{figure}
    \centering
    \includegraphics[width=\textwidth]{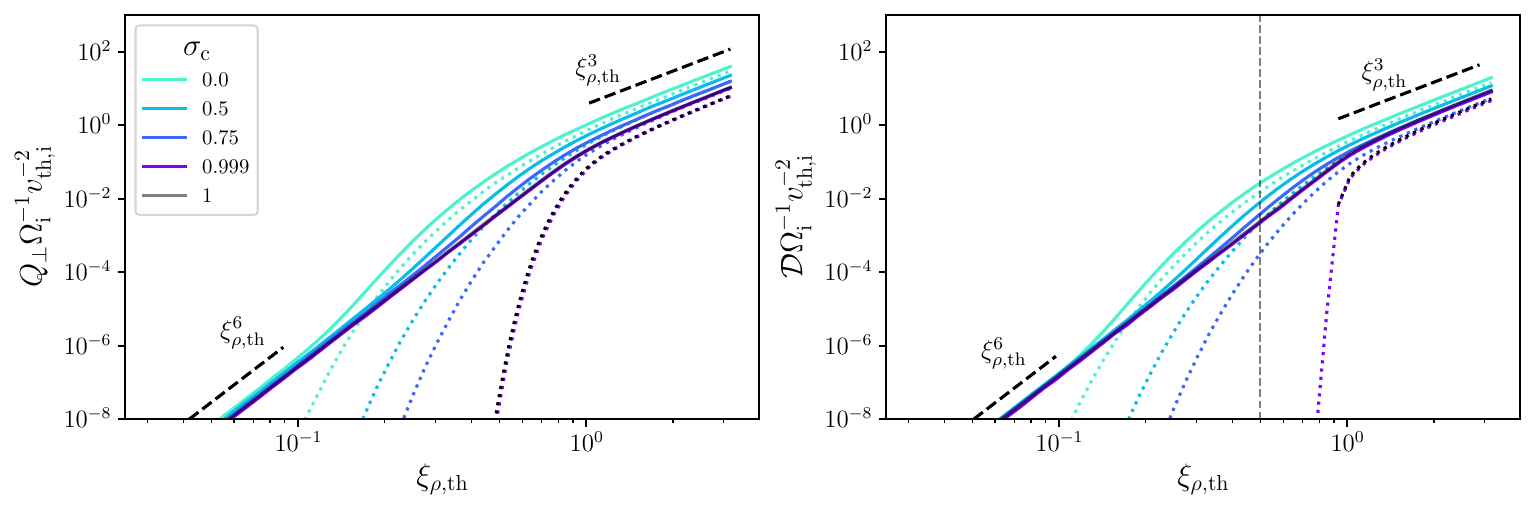}
    \caption{Left panel: The perpendicular heating rate $\Qprp$, \eqref{eq:QprpGeneralCase}, as a function of $\stocfracth$, with the integral in \eqref{eq:QprpGeneralCase} calculated numerically with $\betai=0.1$ and for different values of $\crosshel$.
    Solid lines show the contribution to $\Qprp$ from all modes, and dotted lines the contribution from only modes below the CB cone to highlight the suppression in heating for small $\stocfracth$ (with $C=1$ in $\energyspecEtildetwoD$, \ref{eq:RMHDSpecNorm}).
    The form of $\Qprp$ in the $\crosshel=1$ limit \eqref{eq:QprpImbalancedCase} is also shown in grey; this overlaps $\Qprp$ calculated with $\crosshel=0.999$.
    Right panel: The generalised diffusion coefficient $\mathcal{D}$ \eqref{eq:DppGeneralIntegrand} as a function of $\stocfracth$, showing the similarity in scaling to $\Qprp$. The integral in \eqref{eq:DppGeneralIntegrand} is calculated numerically with $\vprltilde=0.1$ and with the same values of $\crosshel$ in the left panel; dotted lines show the contribution from only modes below the CB cone.
    The dashed vertical line at $\stocfracth=0.5$ corresponds to the value of $\stocfracth$ used in figure~\ref{fig:DppGeneralIntegrand}.
    The form of $\mathcal{D}$ in the $\crosshel=1$ limit, \eqref{eq:DppImbalancedTurbGS95Spec}, is also shown in black; as above, this overlaps $\mathcal{D}$ calculated with $\crosshel=0.999$.}
    \label{fig:DppGeneral}
\end{figure}

The left panel of figure~\ref{fig:DppGeneral} shows how $\Qprp$ varies with $\stocfracth$, calculated numerically from \eqref{eq:QprpGeneralCase} (using the general form of $\mathcal{D}$ in \ref{eq:DppGeneralIntegrand} and setting $C=1$ in $\energyspecEtildetwoD$, \ref{eq:RMHDSpecNorm}) with $\betai=0.1$ and $\crosshel = 0.0, 0.75,$ and $0.999$.
As observed in both the balanced and imbalanced limits, the heating rate scales as $\stocfracth^3$ at large $\stocfracth$; additionally, the model predicts the heating of ions from interactions with weak wave-like fluctuations lying above the CB cone, as shown by the $\stocfracth^6$ scaling at small $\stocfracth$.
The dotted lines show the heating rate when only considering fluctuations below the CB cone; as in the balanced and imbalanced limits, the heating rate is suppressed at small values of $\stocfracth$.
Thus, the heating rate has the general form
\begin{equation}
    \frac{\Qprp}{\Omegai\vthi^2}\propto \stocfracth^3 F(\stocfracth;\crosshel),
\end{equation}
where $F(\stocfracth;\crosshel)$ is an imbalance-dependent suppression factor that goes to $0$ as $\stocfracth\to 0$.
This suppression is stronger when only considering modes beneath the CB cone, with the increase in suppression with imbalance arising from the decreasing width of the wavevector-frequency spectrum as $\crosshel\to 1$ (which will be discussed in more detail below).

At large $\stocfracth$, the heating becomes moderately less efficient as $\crosshel$ increases.
This is due to the Gaussian-weighted peak in the integral of $\Qprp$, \eqref{eq:QprpGeneralCase}, arising from the choice of a Maxwellian distribution.
For the imbalanced case, the sharp peak at $\kprltilde^{(1)}$ above the CB cone contributes strongly to $\Qprp$, as this is near the peak of the Gaussian for small $\vprltilde$.
In contrast, due to the broadening of the wavevector-frequency spectrum, the balanced and moderately imbalanced cases peak along the CB cone (which lies at parallel scales $|\kprltilde| \leq \stocfracth \ll 1$ for small $\stocfracth$) rather than at $\kprltilde^{(1)}$, so $\Qprp$ is smaller.
The numerical simulations of \citet{Johnston2025-ss} recently found similar values of $\Qprp$ at large $\stocfracth$ in both balanced and imbalanced turbulence, in contrast to what is seen in figure~\ref{fig:DppGeneral}.
This may be due to a difference in the definition of $\stocfracth$ (which in this work is defined in terms of the Elsasser variable spectrum \eqref{eq:RMHDSpecNormDef} compared to the r.m.s. $\ExB$ velocity $\deltau{\rhoi}$ in \citet{Johnston2025-ss}, which may lead to an imbalance-dependent factor between the heating rates), or additional factors such as intermittency or $\kprp\rhoi \gtrsim 1$ effects within their simulations. 

The behaviour of $\mathcal{D}$ as $\crosshel$ transitions between these limits is shown in the right panel of figure~\ref{fig:DppGeneral}, where the integral \eqref{eq:DppGeneralIntegrand} is calculated numerically with $\vprltilde = 0.1$ and the same values of $\crosshel$ as used in the left panel.
In all cases, $\mathcal{D}$ scales as $\stocfracth^6$ at amplitudes $\stocfracth\ll 1$, and transitions to a $\stocfracth^3$ scaling as $\stocfracth$ increases.
Since $\mathcal{D}$ does not vary much for ions with velocities $\vprltilde\ll 1$, $\Qprp$ follows the same scalings as $\mathcal{D}$ at small and large $\stocfracth$.

\subsubsection{Physical origin of the heating-rate scalings}

\begin{figure}
  \centering
  \includegraphics[width=\textwidth]{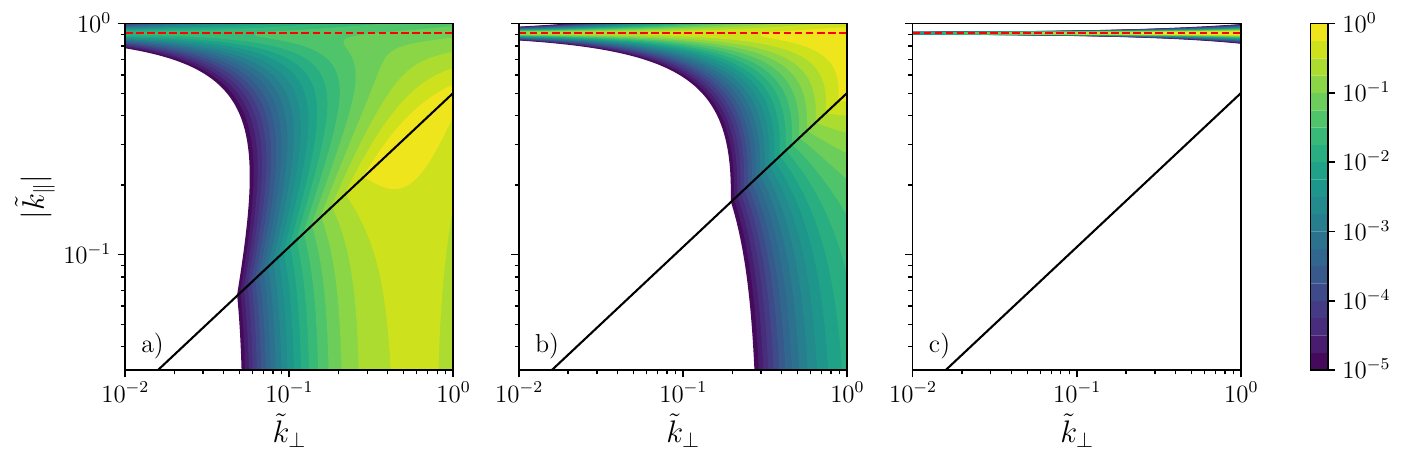}
  \caption{The integrand of the generalised diffusion coefficient $\mathcal{D}$, \eqref{eq:DppGeneralIntegrand}, which therefore shows the contribution of fluctuations at a given $(\kprptilde,\kprltilde)$ to the overall heating.
  Each integrand is normalised to its maximum value and is plotted with $\stocfracth = 0.5$, $\vprltilde=0.1$, and $\crosshel = 0.0$ (a), $0.75$ (b), and $0.999$ (c). The black line corresponds to the CB cone $\kprltilde^{\rm CB} = \stocfracth\kprptilde^{2/3}$. The horizontal red dashed line is $\kprltilde^{(1)}\approx 0.9$, the resonant $\kprl$ with which ions interact in the imbalanced limit defined in \eqref{eq:KprlResonance}. Note that we are looking at the $\kprltilde < 0$ modes.}
\label{fig:DppGeneralIntegrand}
\end{figure}

The reason for this scaling behaviour can be understood by studying the properties of the integrand of the generalised diffusion coefficient $\mathcal{D}$ \eqref{eq:DppGeneralIntegrand}. 
This is shown in figure~\ref{fig:DppGeneralIntegrand} for the same values of $\crosshel$ as in figure~\ref{fig:DppGeneral} with $\stocfracth=0.5$ and $\vprltilde=0.1$; we show only $\kprl < 0$ modes, which correspond to the sign of the dispersion relation of the $\zp$ fluctuations ($-\omegaA=-\kprl\vA$).
The integrand of $\mathcal{D}$ represents the contribution that fluctuations with a given $(\kprp,\kprl)$ provide to the heating of ions with parallel velocity $\vprltilde$.
The integrand consists of a peak at $|\kprltilde^{(1)}| = (1+\vprltilde)^{-1}$ (approximately $0.9$ for $\vprltilde=0.1$, red dashed line in figure~\ref{fig:DppGeneralIntegrand}), the resonant parallel wavenumber in the imbalanced limit \eqref{eq:KprlResonance}, as well as a region of $\kvec$-space both above and below the CB cone that is dependent on the broadening of the wavevector-frequency spectrum arising from nonlinear interactions.
Because these interactions become less dominant as $\crosshel$ is increased, the width of the broadening decreases and the integrand becomes increasingly sharply peaked around $\kprltilde^{(1)}$.

The scaling of $\Qprp$ and $\mathcal{D}$ with $\stocfracth$ in figure~\ref{fig:DppGeneral} can be intuitively understood by noting that varying $\stocfracth$ changes the position of the CB cone, $\kprltildeCB = \stocfracth\kprptilde^{2/3}$, in figure~\ref{fig:DppGeneralIntegrand}.
As shown above in the balanced limit (Section~\ref{sec:limitingcases}), the region below the CB cone contributes a term to $\mathcal{D}$ proportional to $\stocfracth^3$; similarly, the region above the cone contributes to the $\stocfracth^6$ scaling in both the $\crosshel=0$ and $1$ limits.
As $\stocfracth$ is increased the cone shifts upwards in figure~\ref{fig:DppGeneralIntegrand}, causing fluctuations with $\kprltilde<\kprltildeCB$ to have a greater contribution to the integral and leading to the $\stocfracth^3$ scaling in \eqref{eq:BalancedDppStocFrac} and \eqref{eq:DppImbalancedTurbGS95Spec}.
Similarly, the region above the cone becomes the dominant contribution as $\stocfracth$ is decreased and the cone shifts downwards, leading to the $\stocfracth^6$ scaling.
Because the width of the broadening function decreases with increasing $\crosshel$, a larger value of $\stocfracth$ is required so that the region below the cone can contribute to the integral. 
This leads to the transition to a $\stocfracth^3$ scaling at larger values of $\stocfracth$ for highly imbalanced turbulence, as seen in figure~\ref{fig:DppGeneral}.

Figure~\ref{fig:DppGeneralIntegrand} also illustrates the suppression of $\Qprp$ that arises when only modes below the CB cone are included.
This suppression depends on the contribution from the tails of the Fourier transform of the temporal correlation function, which---at fixed imbalance---becomes increasingly small as $\stocfracth$ decreases.
The imbalance-dependent width of these tails also plays an important role in determining the strength of the suppression: as $\crosshel$ increases, the integrand below the cone contributes progressively less to the total heating, since the spectrum becomes more sharply peaked around $\kprltilde^{(1)}$. This leads to the increased suppression seen in both $\mathcal{D}$ and $\Qprp$ in figure~\ref{fig:DppGeneral}.
It should be emphasised that, although the integral for $\Qprp$ in \eqref{eq:QprpGeneralCase} includes a Gaussian weighting $\exp(-\vprltilde^2/\betai)$ arising from the assumed Maxwellian distribution, the observed suppression is primarily a consequence of the broadening of the wavevector–frequency spectrum discussed above, rather than the form of the distribution (see Section~\ref{sec:balancedlimit}).

\subsubsection{The differential heating rate $\romand\Qprp/\romand \vprl$}

\begin{figure}
  \centering
  \includegraphics[width=\textwidth]{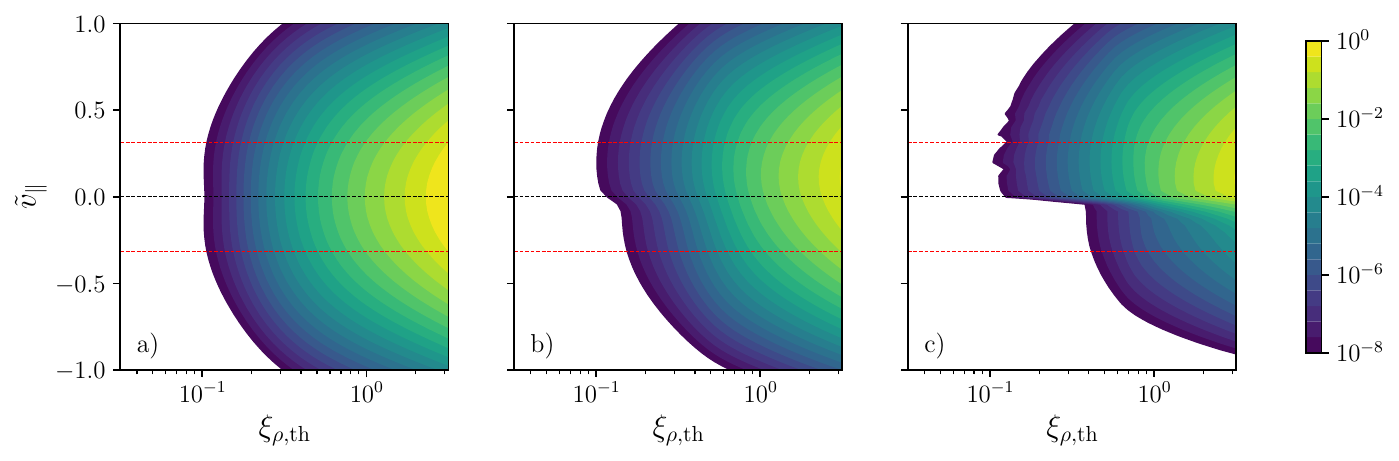}
  \caption{The differential heating rate $\romand \Qprp/\romand \vprl$ \eqref{eq:dQprpdvprl}, which quantifies the heat given to ions with parallel velocities between $\vprl$ and $\vprl+\romand\vprl$, calculated using $\betai=0.1$ and $\crosshel=0.0$ (a), $0.75$ (b), and $0.999$ (c).
  The red dashed lines correspond to where $|\vprltilde|=\vthi/\vA=\sqrt{\betai}$.
  These plots are normalised to the maximum value in the balanced case in (a).
  }
\label{fig:dQprpdvprl}
\end{figure}

An additional metric of ion heating is the differential heating rate $\romand \Qprp/\romand \vprl$, which quantifies the heat given to ions with parallel velocities between $\vprl$ and $\vprl+\romand\vprl$.
This can be calculated from \eqref{eq:QprpGeneralCase} where, after setting the upper integral limit to $\vprltilde$, we have
\begin{equation}
    \frac{1}{\Omegai\vthi^3}\frac{\romand \Qprp}{\romand \vprl}(\vprltilde) = \sqrt{\frac{4}{\upi}}e^{-\vprltilde^2/\betai}(1+\crosshel\vprltilde)\mathcal{D}(\vprltilde). \label{eq:dQprpdvprl}
\end{equation}
Figure~\ref{fig:dQprpdvprl} plots \eqref{eq:dQprpdvprl} as a function of $\vprltilde$ and $\stocfracth$ for $\betai=0.1$ and $\crosshel=0.0$, $0.75$, and $0.999$.
All cases show increased heating at larger $\stocfracth$ for ions with $|\vprl|\lesssim \vthi$ (corresponding to a $\stocfracth^3$ scaling). 
In the balanced case (figure~\ref{fig:dQprpdvprl}a) $\romand \Qprp/\romand \vprl$ is symmetric around $\vprl=0$ due to the presence of equal populations of $\zpm$ fluctuations, with a moderate peak at $|\vprl|\sim\vthi$ for small $\stocfracth$ and at $|\vprl|\sim 0$ for large $\stocfracth$.
This symmetry is broken as $\crosshel$ increases, with the peak at large $\stocfracth$ shifting slightly to positive $\vprl$; as the amplitude of $\zm$ fluctuations decrease, ions with $\vprl < 0$ are heated less.
This transition occurs slowly with increasing imbalance and only becomes starkly apparent when $\crosshel\gtrsim0.9$, as seen in figure~\ref{fig:dQprpdvprl}c).
The tail at $\stocfracth\ll 1$ is sharply cut off for particles with $\vprltilde < 0$; the remaining heating for $\vprltilde < 0$ particles at $\stocfracth\sim 1$ arises from the fact that, although the peak of the correlation function at $|\kprltilde^{(1)}| >1$ is outside the spectrum domain in \eqref{eq:RMHDSpecNorm}, the finite width of the temporal correlation function allows for a contribution from modes with $|\kprltilde| < 1$. 
This contribution disappears completely in the limit $\crosshel\to 1$ as the temporal correlation function approaches a delta function (cf.~\ref{eq:DppImbalancedTurbGS95Spec}).

\subsection{Numerical evolution of the distribution function}\label{sec:f0_Evolution}

\begin{figure}
  \centering
  \includegraphics[width=\textwidth]{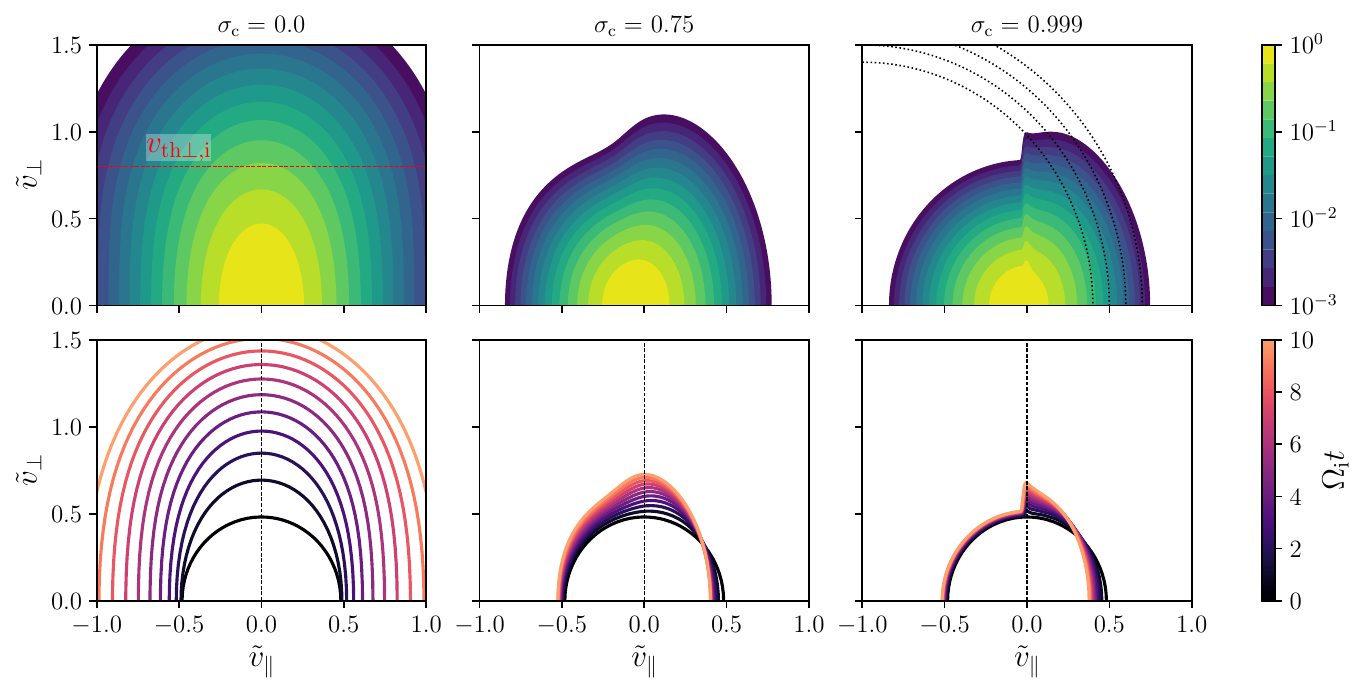}
  \caption{The distribution function $f_0$ numerically evolved from an initial Maxwellian using the diffusion coefficients \eqref{eq:GeneralDiffusionCoefficients} in the diffusion equation \eqref{eq:QLDiffusionForm}, with $\betai=0.1$, $\stocfracth=0.5$, and $\crosshel=0.0$ (left column), $0.75$ (middle column) and $0.999$ (right column).
  Top row: Snapshots of $f_0$ at time $\Omegai t = 4$, showing their evolved structure.
  The distribution functions are normalised to their maximum value, with contours of the initial Maxwellian shown in grey. The red dashed line in the $\crosshel=0.0$ case represents the current perpendicular thermal velocity $v_{\rm th\perp, i}$ of $f_0$, and the dotted lines in the $\crosshel=0.999$ case represent contours of constant energy in the frame of the $\zp$ fluctuations, which are circles centred on $\vprltilde=-1$.
  Bottom row: Evolution of $f_0$ up to time $\Omegai t = 10$, with contours enclosing 90\% of the distribution shown at intervals of $\Omegai t=1$.
  As the imbalance increases, $f_0$ becomes asymmetric around $\vprltilde=0$ (dashed lines).}
\label{fig:f0_evo}
\end{figure}

The expression \eqref{eq:QprpGeneralCase} for the heating rate $\Qprp$ used above calculates the instantaneous heating of a Maxwellian distribution, as noted in Section~\ref{sec:MaxwellianHeatingRate}.
Because of this, it does not directly reveal the velocity-space structure of the distribution function as it evolves away from its initial state.
To illustrate the physical mechanisms embodied in the diffusion coefficients \eqref{eq:GeneralDiffusionCoefficients}, we now numerically evolve $f_0$ according to the quasi-linear diffusion equation \eqref{eq:QLDiffusionForm}, showing that these coefficients reproduce velocity-space signatures of stochastic heating in the balanced case and cyclotron-resonant heating in the imbalanced case.

The distribution function is evolved in time using the method presented in \citet{Isenberg2019-oy}, where $f_0$ is discretised on a grid in ($\vprltilde,\vprptilde$)-space and \eqref{eq:QLDiffusionForm} is advanced using a fully implicit flux-conservative scheme. The grid in $\vprptilde$ is staggered by half a gridcell to avoid division by zero at $\vprptilde=0$, with reflecting boundary conditions used; absorbing boundaries are used at all other grid edges.
Figure~\ref{fig:f0_evo} shows the evolution of $f_0$ using this method, with $\betai = 0.1$, $\stocfracth = 0.5$, and $\crosshel=0$, $0.75$, and $0.999$.
A grid of resolution $1000\times 500$, sufficient to resolve the diffusion in all cases, is used in $(\vprltilde,\vprptilde)$-space.
To ensure the grid boundaries minimally affect the evolution of $f_0$, the grid extends between $-15\vthi/\vA \leq \vprltilde \leq 15 \vthi/\vA$ and $0 < \vprptilde \leq 15\vthi/\vA$ (where $\vthi/\vA = \sqrt{\betai} \approx 0.316$); this choice conserves the number of particles in the distribution for all cases.

The top row of figure~\ref{fig:f0_evo} shows the structure of $f_0$ after a (gyrofrequency-normalised) time $\Omegai t= 4$.
The balanced case shows that the distribution has been preferentially heated in the perpendicular direction, although some parallel heating has also occurred.
Additionally, the distribution is approximately flat for $|\vprl|\approx 0$ and $\vprp < v_{\rm th\perp, i}$ (the perpendicular thermal velocity of the distribution, shown by the red dashed line in figure~\ref{fig:f0_evo}).
This strong perpendicular heating and ``flat-topped'' distribution is a hallmark signature of stochastic heating \citep{Klein2016-qo,Cerri2021-xo,Zhang2025-yd,Johnston2025-ss}.
This further demonstrates that the diffusion coefficients derived in this work can reproduce stochastic-heating-like behaviour in the balanced case, connecting the quasi-linear framework to these prior results.

The symmetric evolution of $f_0$ about $\vprltilde=0$ in the balanced case follows from the symmetry of the diffusion coefficients \eqref{eq:GeneralDiffusionCoefficients} in $\vprltilde$ when $\crosshel = 0$.
As the turbulence's imbalance increases, the structure of the diffusion coefficients becomes increasingly asymmetric, as does the evolution of $f_0$, shown in the highly imbalanced ($\crosshel = 0.999$) case in figure~\ref{fig:f0_evo}.
Due to the sharp form of the generalised diffusion coefficient $\mathcal{D}$ (as seen in figure~\ref{fig:DppGeneralIntegrand}c), the evolution of $f_0$ resembles that due to the wave-particle resonance predicted by quasi-linear theory: the distribution flattens along contours of constant energy in the frame of the $\zp$ fluctuations, which are circles centred on $\vprl = -\vA$ ($\vprltilde=-1$).
Only ions with $\vprltilde \gtrsim 0$ are able to diffuse, as discussed in Section~\ref{sec:imbalancedlimit}.
In contrast to the $\crosshel=0.999$ case, the sharp features in $f_0$ are smoothed out in modest ($\crosshel=0.75$) imbalance, which follows from the broadening of the sharp resonant structures in the diffusion coefficients.\footnote{The smooth structure of $f_0$ in the moderately imbalanced case is similar to what is seen in imbalanced hybrid-kinetic simulations \citep{Squire2022-dm,Squire2023-jn,Zhang2025-yd}, suggesting that the smooth distribution functions observed there may, at least in part, result from the broadening in frequency of the turbulent fluctuations with which ions interact.}

The full evolution of $f_0$ at intervals of $\Omegai t = 1$ up to $\Omegai t = 10$ is shown in the bottom row of figure~\ref{fig:f0_evo}, with contours enclosing 90\% of the distribution at each time; the imbalance-dependent features and heating discussed above at $\Omegai t = 4$ remain qualitatively unchanged throughout.
While the numerical method is capable of evolving $f_0$ to arbitrarily long times, the restriction to $\Omegai t = 10$ here is sufficient to demonstrate the general evolution of $f_0$ using the diffusion coefficients \eqref{eq:GeneralDiffusionCoefficients}.

\section{Summary and discussion}\label{sec:Conclusion}

In this work, we analytically calculate the perpendicular heating rate $\Qprp$ of ions interacting with critically balanced RMHD turbulence with different levels of imbalance.
This is achieved using the framework of quasi-linear theory to describe the ions' interactions with small-amplitude electromagnetic fluctuations as a diffusion of the ion distribution function in velocity-space.
This diffusion depends on the wavevector-frequency spectrum of the turbulence, whose nonlinear broadening (due to interactions between Alfv\'enic fluctuations) depends on the imbalance.
Generalising previous arguments in \citet{Schekochihin2022-nn}, supplemented with a suite of numerical simulations (presented in Appendix~\ref{app:RMHDModel}), we develop a model wavevector-frequency spectrum of RMHD turbulence as a function of imbalance.
This spectrum is used in the quasi-linear diffusion coefficients, after various simplifications, to predict the turbulent heating rate as a function of turbulence and imbalance.

We show that the heating rate has the general form
\begin{equation}
    \frac{\Qprp}{\Omegai\vthi^2}\propto \stocfracth^3 F(\stocfracth;\crosshel),\label{eq:HeatingSuppressionConc}
\end{equation}
with an imbalance-dependent suppression factor $F(\stocfracth;\crosshel)$ that goes to zero as $\stocfracth\to 0$.
Using an analytical calculation, taking into account only the interaction of ions with strong fluctuations below the CB cone, the heating rate in balanced turbulence is shown to have the form
\begin{equation}
    F(\stocfracth;\crosshel=0)\approx \hat{c}_1\left(1+\frac{\hat{c}_2}{\stocfracth}\right)e^{-\hat{c}_2/\stocfracth},\label{eq:QprpSuppBal}
\end{equation}
where $\hat{c}_1 = 3C/\upi$ and $\hat{c}_2 = \upi / (2 C)$ with $C$ an order-unity turbulence constant used in the normalisation of the spectrum (cf. Sections~\ref{sec:SchekochihinSpectrum} and \ref{sec:balancedlimit}).
The form of \eqref{eq:QprpSuppBal} is similar to the exponential suppression commonly used in the phenomenology of stochastic heating \eqref{eq:SHRate}; a similar suppression is observed in the imbalanced limit, with a sharper cutoff due to the requirement of ions resonating with a single mode.
The fact that we see this general form for $\Qprp$ in a quasi-linear calculation, and that stochastic-heating-like behaviour is obtained in the balanced case, illustrates the connection between stochastic heating and cyclotron-resonant heating through the general form of the wavevector-frequency spectrum of the turbulence.

Similar ideas were discussed and studied numerically in recent work by \citet{Johnston2025-ss}, who used high-resolution numerical simulations to study the heating of a distribution of ions interacting with strong low-$\betai$ turbulence and showed that their measured ion heating rate is of the general form \eqref{eq:HeatingSuppressionConc} regardless of whether the turbulence is balanced or imbalanced.
The results of the work in this paper provide an analytical complement to these simulations by demonstrating in detail how this suppression arises through the properties of the wavevector-frequency spectrum with which the ions interact, as well as showing that standard quasi-linear theory is able to recover a stochastic-heating-like process.
We note that the results are not exactly consistent, with the analytic calculations of $\Qprp$ in figure~\ref{fig:DppGeneral} showing a modest dependence on imbalance, whereas the measurements of \citet{Johnston2025-ss} show a more universal form independent of imbalance.
This may be because the dependence of $\Qprp$ on imbalance is relatively minor and therefore difficult to measure well in simulations.
Despite this distinction, the qualitative behaviour of the heating remains broadly consistent across both the analytic calculations and simulation measurements.

The ideas presented in this paper are also complementary to recent work by \citet{Mallet2026-ei}, who studied the interactions of ions with coherent electromagnetic fluctuations and also showed that the predicted heating rate exhibits an exponential suppression factor of the form $e^{-1/\eta}$, where $\eta\sim\tau\Omegai$ and $\tau$ is a characteristic timescale of the fluctuation.
The suppression \eqref{eq:HeatingSuppressionConc} seen in this work arises similarly from the temporal correlations of the turbulence fluctuations rather than the details of the ion velocity distribution (as shown by the presence of the suppression in the general diffusion coefficient $\mathcal{D}$; figure~\ref{fig:DppGeneral}), showing that it is a result of the properties of the turbulence and its interaction with ions.

\subsection{Approximations and outlook}\label{sec:Approximations}

To ensure that we model the interaction of ions with purely Alfv\'enic fluctuations, the two-dimensional RMHD energy spectrum $\energyspecE_{\rm 2D}$ \eqref{eq:RMHDSpecNorm} incorporates two key simplifying assumptions.
The first is that the spectrum is cut off at $\kprp\rhoi > 1$, which follows from the assumption that heating is dominated by fluctuations with $\kprp\rhoi \lesssim 1$, which is justified because $\kprp\rhoi\gtrsim 1$ fluctuations are averaged out over particle orbits \citep{Chandran2000-pq}.
Although kinetic-Alfvén wave fluctuations at $\kprp\rhoi \geq 1$ may play a role in heating \citep{Arzamasskiy2019-qv,Isenberg2019-oy}, their overall contribution is small in the context of our model, as explored in Appendix~\ref{app:SubRhoHeating}.

The second is a cutoff at $|\kprl d_{\rm i}| > 1$, where $d_{\rm i} = \vA/\Omegai$ is the ion inertial length.
By assuming that ion heating is dominated by Alfv\'enic fluctuations, this choice neglects the transition from Alfvén to ion-cyclotron waves (ICWs) as $\kprl$ approaches $d_{\rm i}^{-1}$ and their frequency approaches $\Omegai$, effectively meaning that the dispersion relation of the fluctuations remains Alfv\'enic ($\omegaA = \kprl\vA < \Omegai$) up to this limit.
In the imbalanced limit, where the wavevector-frequency spectrum is sharply peaked around the linear dispersion relation of the fluctuations, this approximation is less well justified for protons because, due to the difference in their dispersion relation, the resonant wavenumbers of ICWs and Alfvénic fluctuations with which ions interact differ; in Appendix~\ref{app:ICWHeating}, we argue that the effect of this approximation is less than might naively be expected due to critical balance.
Despite this, this model is likely accurate for describing the heating of minor ions with mass $A m_{\rm p} > m_{\rm p}$, charge $Z e$, and gyrofrequency $\Omegai = (Z/A)\Omegap < \Omegap$ (and thus $d_{\rm i} > d_{\rm p}$, where $d_{\rm p}$ is the proton inertial length), as their smaller gyrofrequency allows them to interact with the Alfv\'enic branch of the ICW dispersion relation.

An interesting result arising from the heating rate calculations above is the prediction that, in addition to heating from fluctuations below the CB cone, ions are also heated through interactions with weak (wave-like) fluctuations above it at $|\kprl d_{\rm i}| \lesssim 1$. Such fluctuations naturally occur in CB turbulence due to the continuity of the spectrum across the CB cone and the thermal spectrum at low~$k_\perp$.
In figure~\ref{fig:DppGeneral}, this heating manifests as a steep $\stocfracth^6$ scaling at small $\stocfracth$ in the heating rate across all levels of imbalance (although the exact exponent of $\stocfracth$ depends on the $\kprl$ spectrum; see \ref{eq:RMHDSpecNorm}).\footnote{Details on a tentative simulation result that appears to match this prediction are given in the supplementary material of \citet{Johnston2025-ss}, where test particles interacting with small-$\stocfracth$ balanced turbulence appear to show a similar power-law form of the heating rate.}
Because our model neglects the transition to ICWs---which modifies both the spectrum and resonance structure near $\kprl d_{\rm p}\sim 1$ \citep{Squire2022-dm,Squire2023-jn,Zhang2025-yd}---this scaling with $\stocfracth$ may be modified for protons; however, it may be valid for minor ions interacting with lower-frequency Alfv\'enic fluctuations due to their smaller gyrofrequency.
Regardless, the presence of heating above the CB cone may be a generic feature even if the detailed properties of the spectrum differ; hybrid-kinetic simulations that include $d_{\rm i}$-scale physics are required to verify this effect.

We note that the form of the heating rate in this work, \eqref{eq:QprpDprp}, assumes that the ions are in an initially isotropic Maxwellian distribution.
This was chosen for simplicity, but does not reflect more complicated velocity distributions that are observed.
However, given a functional form of the distribution, the heating rate can be similarly calculated (either analytically or numerically as done in Section~\ref{sec:f0_Evolution}) using the framework in this paper; further work would be needed to investigate how this would modify these results.

\subsubsection{Imbalanced turbulence and the helicity barrier}

RMHD is well-suited for describing large-amplitude Alfv\'enic turbulent fluctuations relevant to ion heating as discussed in this work.
However, it excludes key sub-$\rhoi$ physics and, in particular, the helicity barrier: a mechanism 
that plays a key role in imbalanced, low-$\betai$ turbulence \citep{Meyrand2021-ix,Squire2022-dm,Squire2023-jn,Adkins2024-iw,Adkins2025-by,Johnston2025-ss}.
The helicity barrier arises due to the failure of the conservation of a gyrokinetic ``generalised helicity'' at perpendicular scales above and below $\rhoi$ in imbalanced turbulence \citep{Schekochihin2019-al,Meyrand2021-ix}, which prevents turbulent energy from cascading to small perpendicular scales through $\rhoi$, trapping it at scales such that $\kprp\rhoi\lesssim 1$, resulting in a decrease in the energy of $\rhoi$-scale fluctuations.
The helicity barrier is well-supported observationally within the solar wind, manifesting as a steep ``transition range'' drop in measured energy spectra \citep{Leamon1998-kl,Bowen2020-wa} as well as in other features \citep{Bowen2022-qt,Bowen2024-fy,McIntyre2025-oc}.

Using simulations of a reduced gyrokinetic model able to capture the effects of the helicity barrier, \citet{Johnston2025-ss} showed that the heating of ions is unaffected by the barrier, despite the drop in energy of turbulent fluctuations at  $\kprp\rhoi\sim 1$ scales.
This is because ions are most efficiently heated by fluctuations whose frequencies approach $\Omegai$, which occurs at scales $\lambda \sim \rhoi$ in balanced turbulence and shifts to $\lambda > \rhoi$ in the presence of a helicity barrier \citep{Johnston2025-ss}.
Since the blocked cascade accumulates more power at these larger scales than at $\rhoi$-scales, the ion heating rate is maintained.

The effect of the helicity barrier on ion heating can be captured in the quasi-linear framework used in this paper by shifting the cutoff in the 2D wavevector energy spectrum $\energyspecE_{\rm 2D}$ to perpendicular scales $\lambda > \rhoi$ (i.e. by changing the cutoff from $\kprptilde > 1$ to $\kprptilde >\rhoi/\lambda$ in \ref{eq:RMHDSpecNorm}), where the value of $\lambda$ can be measured from simulations of the helicity barrier.
Because the fluctuations that ions interact with at this scale contain more power than those at $\rhoi$, the heating rate will be increased compared to the predictions in Section~\ref{sec:generalcase} of imbalanced RMHD without a helicity barrier.
Although this cutoff provides a good approximation to the helicity barrier for $\kprp\rhoi \lesssim 1$ fluctuations, the corresponding modification for the spectrum at $\kprp\rhoi\gtrsim 1$ (as would be needed in the calculations of Appendix~\ref{app:SubRhoHeating}) is not clear.

\subsubsection{Future work}

Recent work has shown that intermittency---the heavy-tailed distribution of fluctuation amplitudes intrinsic to turbulence---can significantly affect the efficiency of perpendicular ion heating \citep{Mallet2019-ks,Cerri2021-xo,Bowen2025-tv}.
Since our formalism relies on the wavevector-frequency spectrum of the turbulence, it does not directly capture such effects; extending it to account for intermittency remains an important direction for future work.

We note that the quasi-linear framework employed in this work assumes that cyclotron-resonant heating dominates in imbalanced turbulence, with ions directly resonating with the sharply-peaked wavevector-frequency spectrum of the fluctuations.
However, non-resonant heating of ions from wave-like fluctuations is also possible, given sufficiently large wave amplitude \citep{Chen2001-wd,Johnson2001-ui}.
This mechanism cannot be captured using quasi-linear theory as it involves higher-order nonlinear interactions beyond the quadratic products retained in quasi-linear theory (see Section~\ref{sec:QLTheory} and Appendix~\ref{app:QLTheoryDerivation}).
Detailed comparisons between quasi-linear theory predictions and idealised single-wave simulations could help quantify these effects.

An interesting direction for future work is to investigate how ion heating is modified in the presence of purely weak turbulence, rather than in addition to strong turbulence as in this work.
In weak turbulence, the wavevector–frequency spectrum is sharply peaked around the linear dispersion relation, even in the balanced limit (in contrast to the broadened spectrum of strong turbulence), reflecting the wave-like nature of its fluctuations \citep{Meyrand2016-nc}.
Our results predict cyclotron-resonant heating would dominate in this case for all $\crosshel$.
Exploring this regime by, for example, comparing heating in weak and strong turbulence with similar $\stocfracth$, could therefore help to further disentangle the difference between stochastic heating and cyclotron-resonant heating.

While this work has focused primarily on the theoretical aspects of ion heating in turbulent plasmas, its predictions can be connected to space and astrophysical observations.
Observations indicate that turbulent ion heating in the solar corona and wind is likely driven by a combination of stochastic heating \citep{Bourouaine2013-gh,Klein2016-qo,Vech2017-jq,Martinovic2019-xf,Martinovic2020-do} and cyclotron-resonant heating by ICWs \citep{Kasper2013-ri,Bale2019-sd,Bowen2020-kd,Bowen2022-qt,Bowen2024-fy}.
The heating rates associated with these mechanisms can be directly inferred from spacecraft measurements---for example, \citet{Bowen2025-tv} explicitly quantify both the ICW- and stochastic-heating rates of ions in \emph{Parker Solar Probe} data.
Extending such analyses across a wider range of plasma conditions (varying imbalance $\crosshel$, plasma beta $\betai$, turbulent amplitude $\stocfracth$, and minor-ion composition) could allow direct tests of the general phenomenological ion-heating formula \eqref{eq:HeatingSuppressionConc}.

\section*{Acknowledgements}
The authors thank T.~Bowen, B.~D.~G.~Chandran, M.~Kunz, A.~Mallet, R.~Meyrand, and M.~Zhang for interesting discussions over the course of this work.
High-performance computing resources were provided by the New Zealand eScience Infrastructure (NeSI) under project grant uoo02637.

\section*{Funding}
Support for Z.J. was provided by a postgraduate publishing bursary from the University of Otago.
Support for J.S. was provided by the Royal Society Te Ap\=arangi, through Marsden Fund grant MFP-UOO2221.

\section*{Declaration of interests}

The authors report no conflict of interest.

\appendix

\section{Numerical test of 2D RMHD turbulent spectrum model}\label{app:RMHDModel}

\begin{figure}
    \centering
    \includegraphics[width=0.75\textwidth]{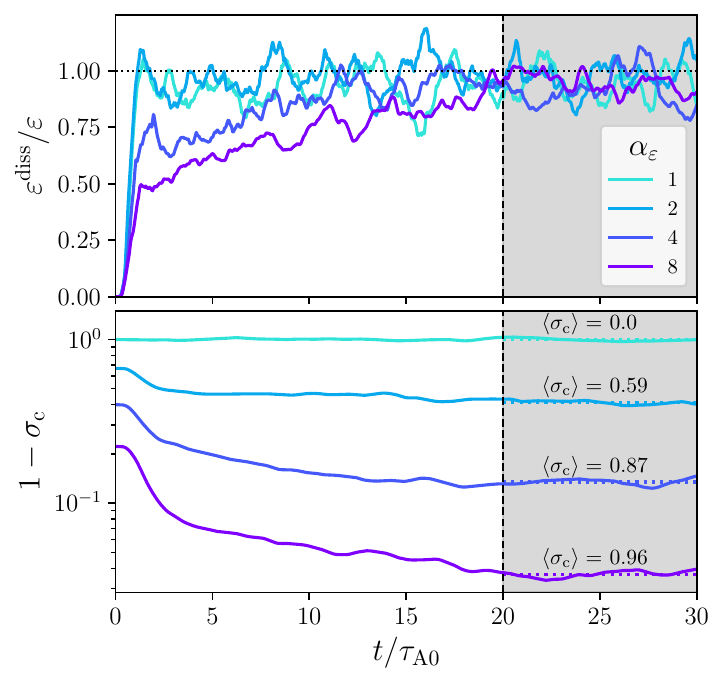}
    \caption{Dissipation and normalised cross-helicity versus time from all simulations, with different levels of injection imbalance $\injectratio$ as listed in the legend. The turbulence reaches approximate steady state when dissipation balances the injection energy, with all simulations reaching this point after $20\tauAouter$. The grey region represents the period of time over which all statistics are calculated and averaged (denoted by $\langle\cdot\rangle$).}
    \label{fig:RMHDSimDissipationCrossHel}
\end{figure} 

In Section~\ref{sec:RMHDModel}, we propose a model for the wavevector-frequency spectrum of RMHD turbulence, which includes the effects of imbalance on the spectrum. This model is then used in the quasi-linear diffusion coefficients to calculate the heating rate of ions in RMHD turbulence (Section~\ref{sec:HeatingRates}).
In this Appendix, we test its validity using a series of four RMHD simulations with varying levels of imbalance. The wavevector–frequency spectra from each simulation are examined in detail to assess how well they reproduce the predicted properties.
While the results herein provide basic model validation to the level needed for our heating studies, we caution that the diagnostics are not sufficiently stringent to constitute a careful check of different imbalanced turbulence phenomenologies.

We use the pseudospectral code AsteriX \citep{Meyrand2016-nc,Johnston2025-ss}, a modified version of the pseudo-spectral code Turbo \citep{Teaca2009-go}, to solve the RMHD equations.
These equations are advanced in time with a third-order modified Williamson algorithm (a four-step, low-order Runge-Kutta method; \citealp{Williamson1980-hz}).
To prevent aliasing of information at small-scales, where modes with wavelengths smaller than the grid size are incorrectly interpreted as larger-scale fluctuations, a grid-shifting method is used \citep{Teaca2009-go}.

The simulations have box lengths $\Lprp=L_z=2\upi$\footnote{The RMHD equations admit a rescaling symmetry \citep[arising from the ordering $\kprl/\kprp \ll 1$;][]{Beresnyak2011-or,Beresnyak2012-eb,Schekochihin2022-nn}, which allows us to use a cubic box without loss of generality; the turbulent fluctuations will develop the correct anisotropy regardless of the aspect ratio of the box.
All wavenumbers are measured in units of $2\upi/L_{\perp,z} = 1$; to reduce notation, we suppress these normalisation factors in this section.} and resolution $N_\perp = N_z = 256$, and are run for $30\tauAouter$, where $\tauAouter=L_z/\vA$ is the outer-scale Alfv\'en time.
These simulations are forced with different values of injection imbalance
\begin{equation}
    \injectimb = \frac{|\forcep - \forcem|}{\forcep + \forcem}
\end{equation}
in order to generate varying levels of turbulence imbalance.
The injection ratio
\begin{equation}
    \injectratio = \frac{\forcep}{\forcem} = \frac{1+\injectimb}{1-\injectimb},
\end{equation}
was chosen to take the values $1, 2, 4$, and 8, corresponding to $\injectimb = 0.0, 0.33, 0.6,$ and 0.77.
Energy is injected into the Elsasser fields at outer scales $1 < \sqrt{\kprp^2 + k^2_z} < 3$ at a rate $\cascadeE \equiv \forcep+\forcem = 0.1 E_0 / \tauAouter$, where $E_0 = 0.1 \vA^2$ is the initial energy of the fields.
The forcing is in the form of negative damping, allowing $\injectimb$ to be controlled exactly \citep{Meyrand2021-ix}.
To ensure all simulations have approximately the same energy in the dominant Elsasser field $\zp$ at steady-state, the forcing in the imbalanced cases is reduced by a factor of $\injectratio^{2/3}$ compared to the balanced simulation 
\citep{Schekochihin2022-nn}.

\begin{figure}
    \centering
    \includegraphics[scale=0.7]{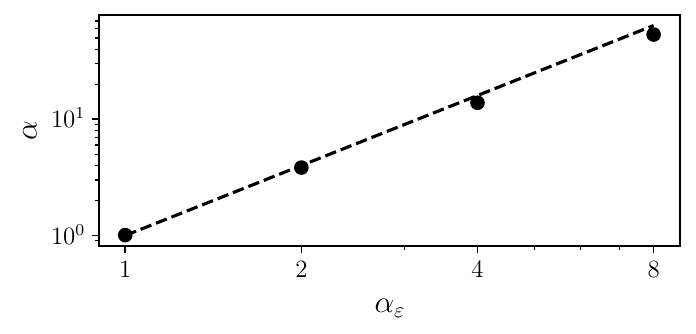}
    \caption{The Elsasser energy ratio calculated from the time-averaged cross helicity using \eqref{eq:ZpmEnergyRatioApp}. Most phenomenological theories of imbalanced turbulence predict that this should scale as $\propto\injectratio^2$ (dashed line), which agrees well with our results.}
    \label{fig:RMHDSimEnergyRatio}
\end{figure}

Imbalanced turbulence takes longer to reach steady state due to the reduced efficiency of the cascade for the dominant Elsasser field; this is illustrated in figure~\ref{fig:RMHDSimDissipationCrossHel}, showing the ratio of energy dissipated near grid scales to that injected at large scales.
All simulations reach approximate steady state at $20\tauAouter$, and all statistics are averaged over the remaining $10\tauAouter$.
The evolution of the normalised cross-helicity $\crosshel$ is also shown in figure~\ref{fig:RMHDSimDissipationCrossHel}, with the simulations reaching steady-state values of $0.0, 0.59, 0.87$, and 0.96, respectively.
Most phenomenological theories of imbalanced turbulence predict that the ratio of Elsasser energies should scale as \citep{Schekochihin2022-nn}
\begin{equation}
    \forceratio \equiv\frac{E^+}{E^-}= \frac{1+\crosshel}{1-\crosshel} \sim \left(\frac{\forcep}{\forcem}\right)^2 = \injectratio^2. \label{eq:ZpmEnergyRatioApp}
\end{equation}
For the RMHD simulations in this Appendix, this scaling is confirmed in figure~\ref{fig:RMHDSimEnergyRatio} using the time-averaged values of the normalised cross-helicity to calculate $\forceratio$ in \eqref{eq:ZpmEnergyRatioApp}.

\begin{figure}
    \centering
    \includegraphics[width=\textwidth]{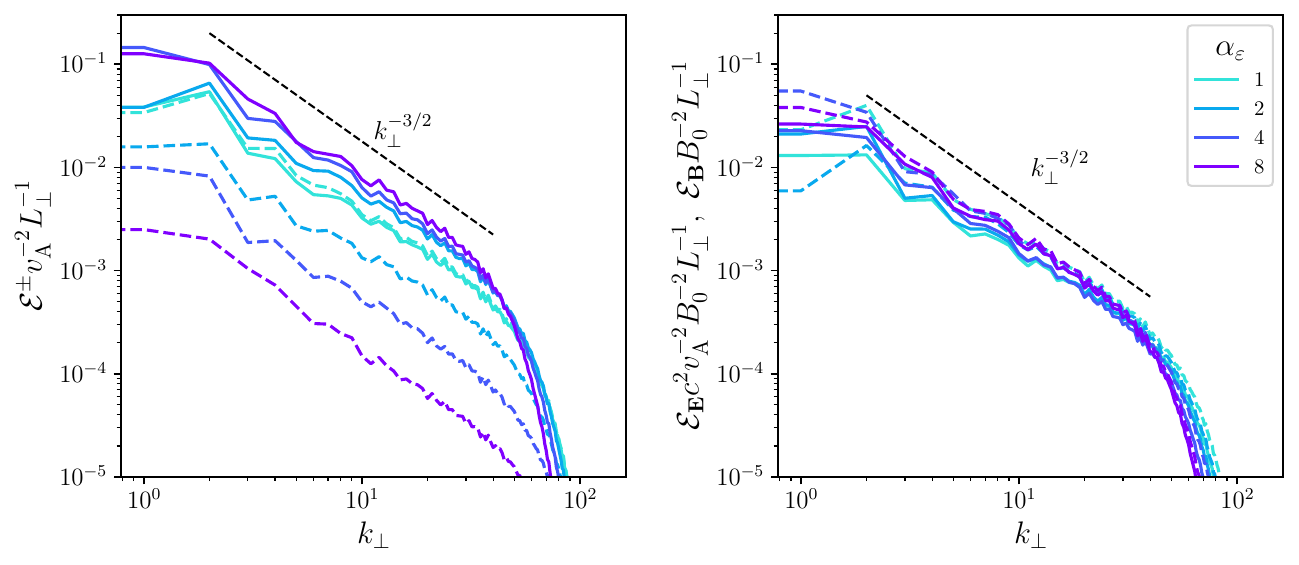}
    \caption{Left: One-dimensional energy spectra of the $\zp$ ($\energyspecE^+(\kprp)$, solid) and $\zm$ ($\energyspecE^-(\kprp)$, dashed) fluctuations, respectively. Right: One-dimensional energy spectra of the electric- ($\energyspecE_{\Evec}(\kprp)$, solid) and magnetic-field fluctuations ($\energyspecE_{\B}(\kprp)$, dashed).}
    \label{fig:RMHDSimKprpEnergySpec}
\end{figure}

\subsection{Energy spectra}\label{app:RMHDModelEnergySpec}

As described above in Section~\ref{sec:SchekochihinSpectrum}, the two-dimensional energy spectrum $\energyspecE(\kprp,\kprl)$ of a turbulent system measures the energy contained in fluctuations around a given $\kprp$ and $\kprl$.
We calculate the two-dimensional energy spectrum of fluctuations in a manner analogous to this, with the energy of fluctuations summed in two-dimensional $(\kprp,k_z)$ bins.\footnote{For simplicity, spectra are measured along the background magnetic field ($z$-axis) rather than parallel to the exact magnetic field, which may cause discrepancies due to the wandering of field lines on small enough scales \citep{Cho2000-bi,Schekochihin2022-nn}. To emphasise this difference, we use $k_z$ rather than $\kprl$ in this section.
However, we note that spectra calculated using the method of \citet{Squire2022-dm} (which calculates spectra along the local magnetic field) gives similar results for this resolution.}
One-dimensional spectra, such as $\energyspecE(\kprp)$, are then obtained via $\energyspecE(\kprp)=\intsingle{k_z}{}{} \energyspecE_{\rm 2D}(\kprp,k_z)$.

\begin{figure}
    \centering
    \includegraphics[width=0.8\textwidth]{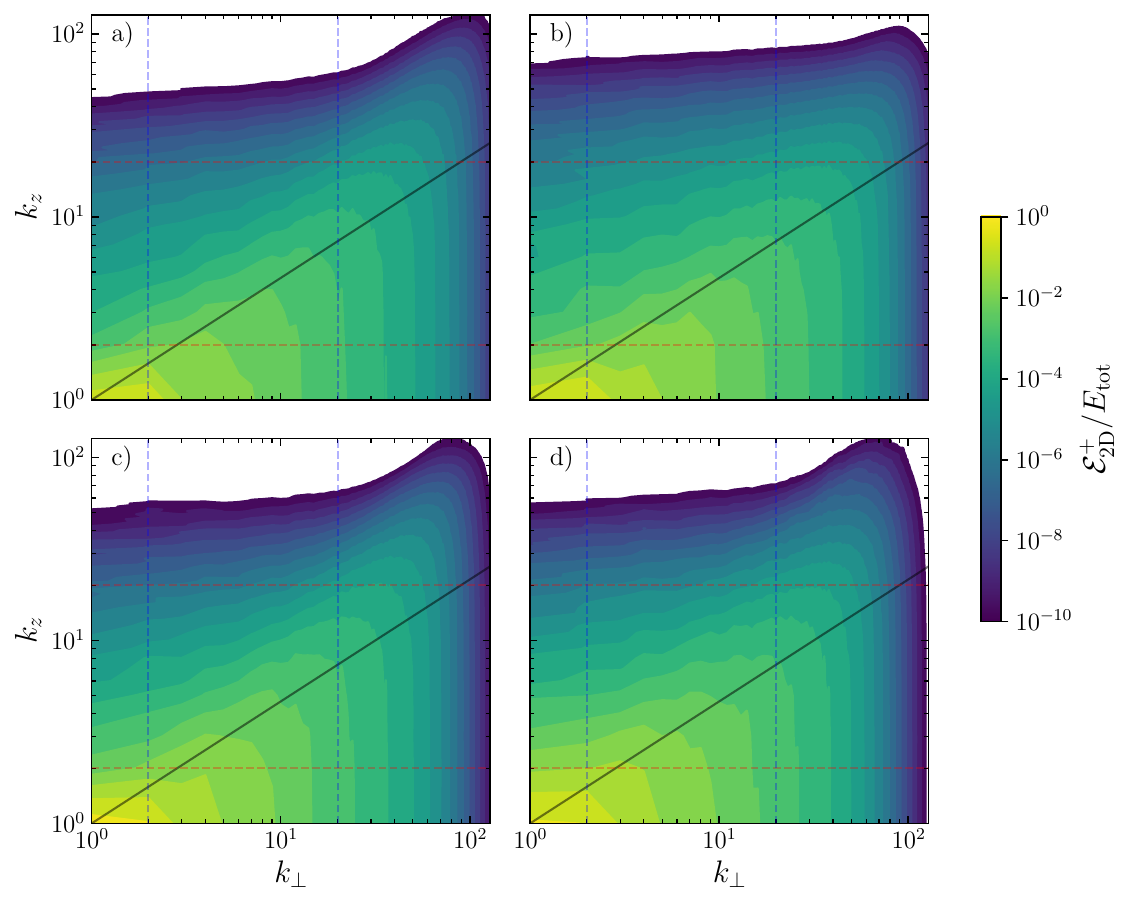}
    \caption{Two-dimensional spectra of $\zp$ from the $\injectratio = 1$ (a), 2 (b), 4 (c), and 8 (d) simulations. Blue and red dashed lines correspond to the slices at constant $\kprp$ and $k_z$ used in figure~\ref{fig:RMHDSimZp2DEnergySpecSlice}, and the black line shows $k_z \propto \kprp^{2/3}$.}
    \label{fig:RMHDSimZp2DEnergySpec}

    \bigskip

    \includegraphics[width=0.8\textwidth]{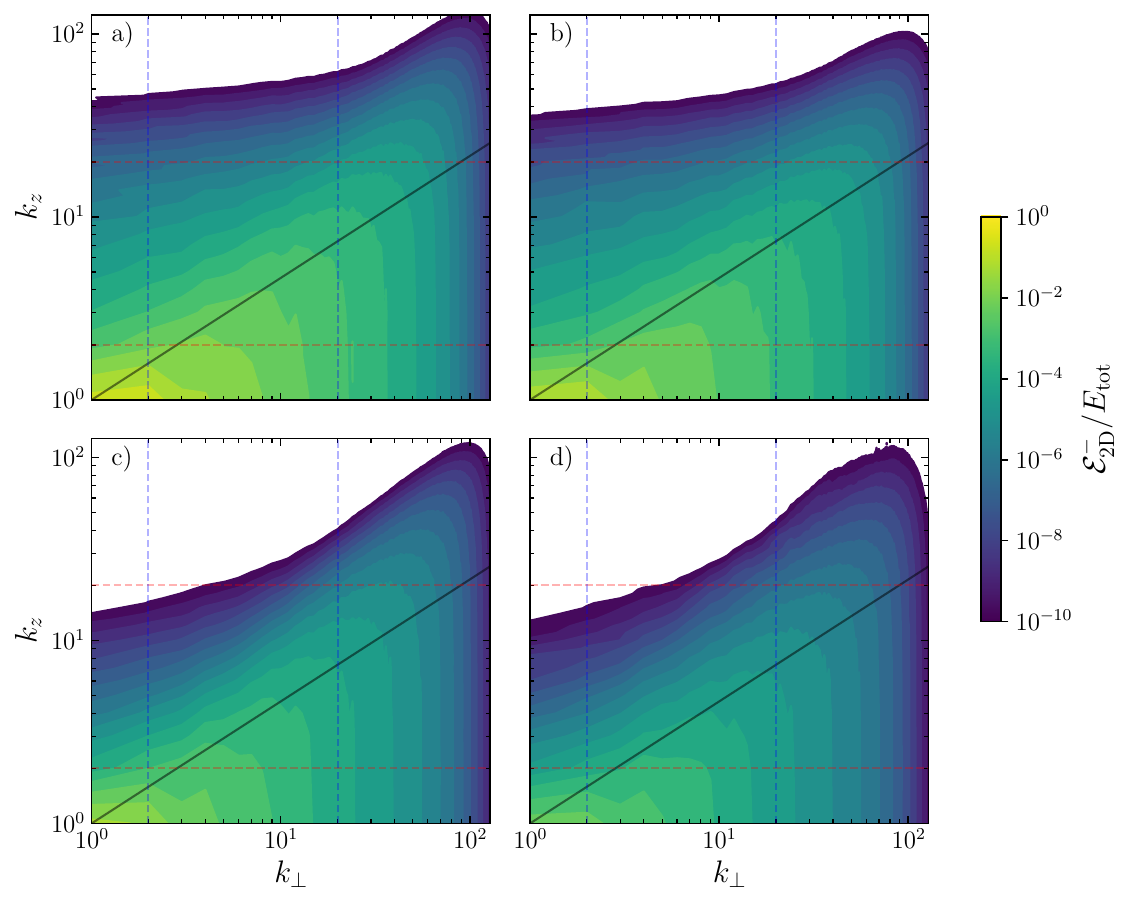}
    \caption{Two-dimensional spectra of $\zm$ from the $\injectratio = 1$ (a), 2 (b), 4 (c), and 8 (d) simulations. As in figure~\ref{fig:RMHDSimZp2DEnergySpec}, blue and red dashed lines correspond to the slices used in figure~\ref{fig:RMHDSimZm2DEnergySpecSlice}.}
    \label{fig:RMHDSimZm2DEnergySpec}
\end{figure}

The one-dimensional spectra of the $\zpm$ fields, as well as that of the electric- and magnetic-field fluctuations, in each of the simulations are shown in figure~\ref{fig:RMHDSimKprpEnergySpec}.
The amplitudes of the $\zp$ fluctuations are approximately equal, showing that, in line with figure~\ref{fig:RMHDSimEnergyRatio}, our scaling of forcing strength with $\injectratio^{2/3}$ works approximately as intended.
As they are being forced with increasingly less energy, the amplitudes of the $\zm$ fluctuations decrease with increasing imbalance.
Both fields follow similar scaling with $\kprp$, perhaps closer to a $\kprp^{-3/2}$ dynamic alignment scaling \citep{Boldyrev2006-xd,Beresnyak2011-or,Beresnyak2012-eb} than the \citet{Goldreich1995-fv} scaling of $\kprp^{-5/3}$, although our resolution is insufficient to make any strong claim.
Additionally, because $\vel_\perp = (c/\Bmeanmag)\unitvector{z}\crossprod\gradprp\scapot = (c/\Bmeanmag)\Evec\crossprod\unitvector{z}$ in RMHD, the kinetic energy spectrum is identical to that of the electric-field fluctuations up to a factor $(c/\Bmeanmag)^2$; they are equivalent in the code units used in AsteriX.
Figure~\ref{fig:RMHDSimKprpEnergySpec} thus shows that there is an approximate equipartition in the normalised kinetic and magnetic energies, regardless of the imbalance of the turbulence.

The model of the two-dimensional spectrum of RMHD turbulence is expected to describe spatiotemporal statistics of $\zp$ and $\zm$ fluctuations in balanced turbulence.
This has not been well tested, especially the temporal correlations, and theory is even less well understood for imbalanced turbulence.
To study how well this model can be applied to describe measured spectra, the two-dimensional energy spectra $\energyspecE^+(\kprp,k_z)$ and $\energyspecE^-(\kprp,k_z)$ of the $\zpm$ fluctuations from all simulations are shown in figures~\ref{fig:RMHDSimZp2DEnergySpec} and~\ref{fig:RMHDSimZm2DEnergySpec}, each normalised to the total energy of the system $E_{\rm tot}= E^+ + E^-$.
Up to dissipation scales at around $\kprp\approx 50$, the 2D spectra of both fluctuations qualitatively agree with the model, with a clear demarcation between turbulence below the CB cone and above.
Importantly, the $\zp$ spectra appear nearly identical for all values of imbalance, which agrees with the propagation CB model \citep[see Section~\ref{sec:ImbalancedTurbulence};][]{Lithwick2007-ao}: the linear and nonlinear frequencies scale as $\vA/\lprl^+\sim\delta z_\lambda / \lambda$ without a strong dependence of $\lprl^+$ on $\injectratio$, even though the true nonlinear timescale for the $\zp$ fluctuations arises from the $\zm$ fluctuations.

\begin{figure}
    \centering
    \includegraphics[width=0.9\textwidth]{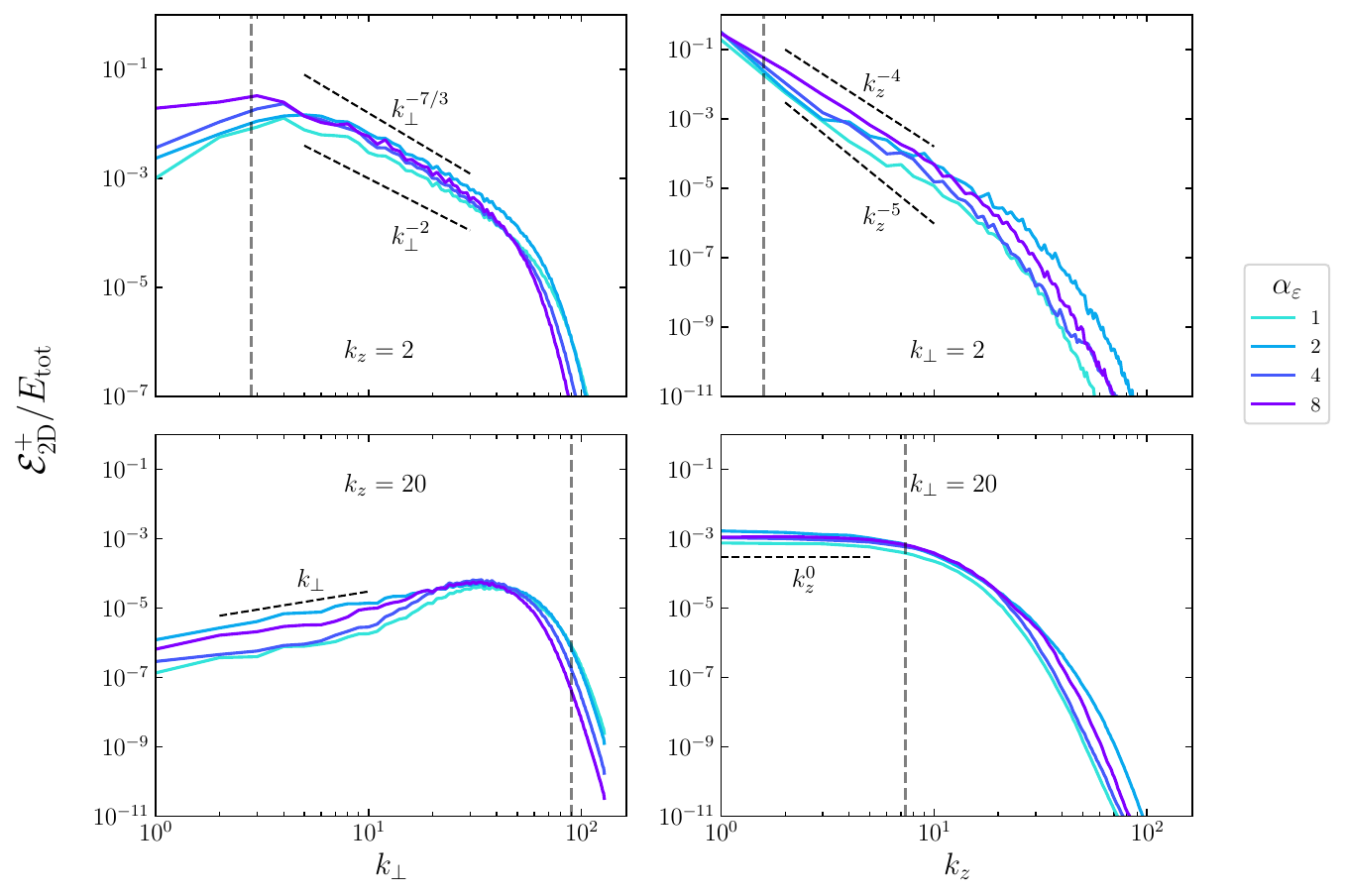}
    \caption{Slices taken through the two-dimensional spectra of $\zp$ at constant $k_z$ (left panels) and $k_\perp$ (right panels), and at low (top panels) and high (bottom panels) values of $k_z$ and $\kprp$, respectively. Vertical dashed lines correspond to the approximate position of the CB cone, $\approx \kprp^{2/3}$ for the constant-$\kprp$ slices and $\approx k^{3/2}_z$ for the constant-$k_z$ slices.
    Note the difference in $y$-axis scaling for the $k_z=2$ slice.}
    \label{fig:RMHDSimZp2DEnergySpecSlice}

    \bigskip

    \includegraphics[width=0.9\textwidth]{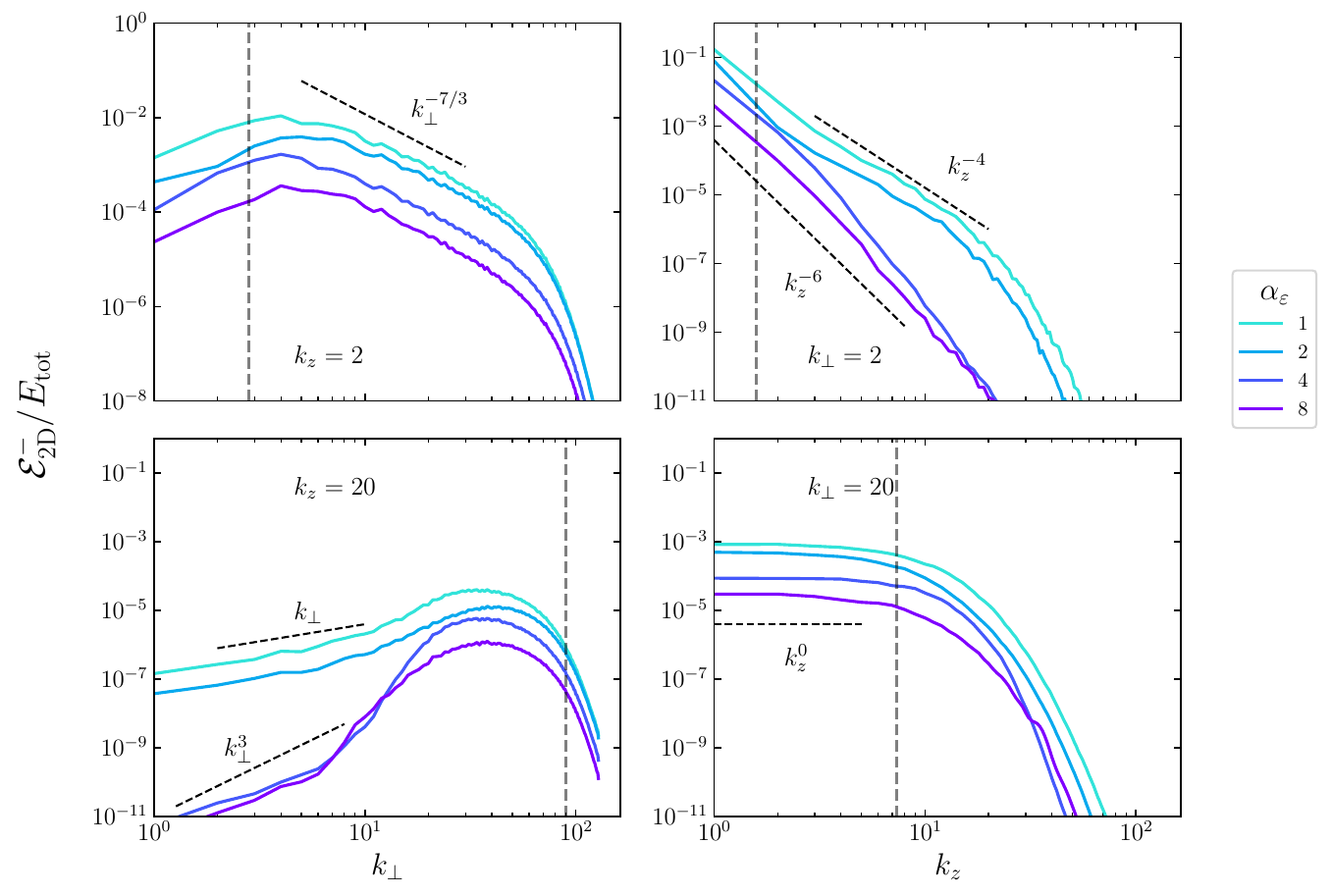}
    \caption{As in figure~\ref{fig:RMHDSimZp2DEnergySpecSlice} but for the $\zm$ fluctuations.}
    \label{fig:RMHDSimZm2DEnergySpecSlice}
\end{figure}

To compare the slopes of these two-dimensional spectra to the model in Section~\ref{sec:SchekochihinSpectrum}, we take slices at $\kprp=2,\ 20$ (blue lines in figures~\ref{fig:RMHDSimZp2DEnergySpec} and~\ref{fig:RMHDSimZm2DEnergySpec}) and $k_z=2,\ 20$ (red lines) in all spectra.
These values are chosen to sample as much of the regions above and below the CB cone as possible before dissipative scales (at $\kprp\gtrsim 50$) take over.
Slopes for the $\zp$ fluctuations (figure~\ref{fig:RMHDSimZp2DEnergySpecSlice}) are nearly consistent across all values of imbalance.
Turbulence below the CB cone qualitatively matches the predictions of Section~\ref{sec:SchekochihinSpectrum}, lying somewhere between a $k_z^0\kprp^{-7/3}$ power law (predicted with $\sCB=2/3$) and a $k_z^0\kprp^{-2}$ power law ($\sCB=1/2$).
Above the cone, the turbulence appears to follow the $\kprp^1$ thermal spectrum with energy equally partitioned among all $\kprp$ modes; however, it does not quite agree with the predicted $k_z$ scaling, showing an approximate $k^{-4}_z$ scaling compared to $k_z^{-5}$ for $\sCB=2/3$ (or $k_z^{-6}$ for $\sCB=1/2$).
The discrepancy in power-law scalings presented here may be a result of the method by which the spectrum is calculated: the present calculation uses scales parallel to the mean magnetic field rather than to the local magnetic field at similar scales.
Additionally, the low resolution of the simulations could also affect the measured results; future work with higher-resolution simulations would be needed to verify these scalings.\footnote{We note that \citet{Maron2001-vz} find $\energyspecE_{\rm 2D}(k_\perp,k_z) \propto k_z^{-6}$ above the CB cone in balanced MHD simulations that were in a highly anisotropic RMHD-like regime as well as a $\kprp^{-3/2}$ one-dimensional perpendicular spectrum, both consistent with the prediction of \citet{Schekochihin2022-nn} for $\sCB = 1/2$.}

The model in Section~\ref{sec:SchekochihinSpectrum} also appears to describe the $\zm$ fluctuations below the CB cone reasonably well, with all simulations showing a similar slope slightly shallower than $-7/3$ in $\kprp$ (figure~\ref{fig:RMHDSimZm2DEnergySpecSlice}).
As the imbalance increases, the parallel correlation length of the $\zm$ fluctuations becomes greater than that of the $\zp$ fluctuations, consistent with previous numerical and observational evidence \citep{Schekochihin2022-nn}; this effect is not included in our model spectrum but is unimportant for ion heating.
Additionally, the fact that both the $\zp$ and $\zm$ fluctuations show a $k^0_z$ scaling below the CB cone in all simulations shows that a CB description of the turbulence is valid, regardless of the level of imbalance. 
However, as seen in figure~\ref{fig:RMHDSimZm2DEnergySpec} there is a clear change in the spectrum of the $\zm$ fluctuations above the CB cone compared to that of the $\zp$ fluctuations, with the spectrum dropping off dramatically at small $\kprp$ or large $k_z$ for large imbalance, appearing to follow a $k_z^{-6}\kprp^2$ power law.
The reason for this change in slope is not yet clear; one possible reason may be that the $\zm$ fields have not been able to ``thermalise'' to a $\kprp^{1}$ spectrum for some reason, instead displaying a spectrum closer to the $\kprp^3$ kinematic one (Section~\ref{sec:SchekochihinSpectrum}; \citealp{Schekochihin2022-nn}).
This steepening would also lead to the corresponding steeper slope in $k_z$, due to the constraint of continuity along the CB line. 
Further investigation with high-resolution simulations is needed to verify whether this is a physical or numerical result.

\subsection{Wavevector-frequency spectrum}

\begin{figure}
    \centering
    \includegraphics[width=\textwidth]{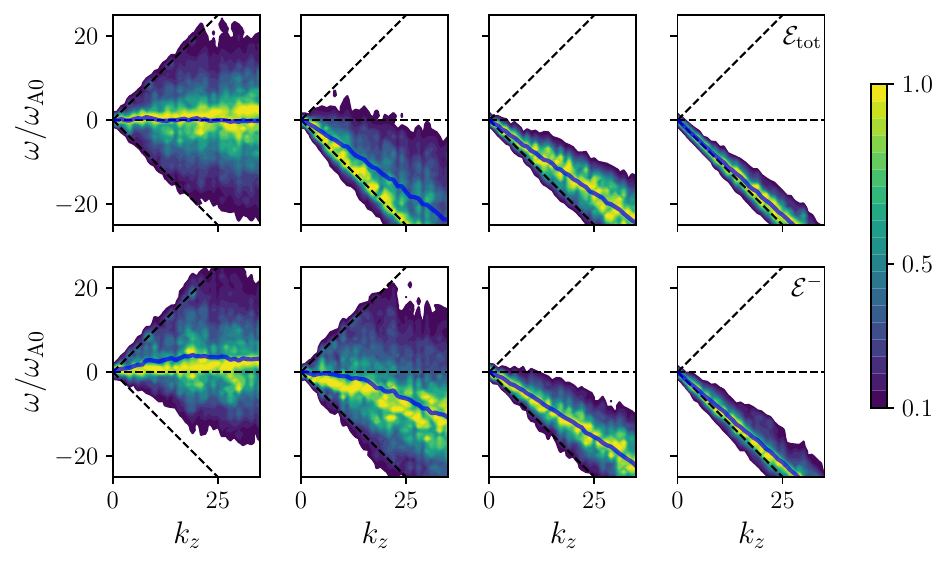}
    \caption{Top row: The wavevector-frequency spectrum $\energyspecE_{\rm tot}$ \eqref{eq:kfreqEtot} normalised to its maximum value at each value of $k_z$, with $\injectratio$ increasing from left to right. The dashed lines correspond to zero frequency and the Alfv\'en dispersion relation $\omegaA = \pm k_z\vA$; the blue line is the mean value of $\omega$ at each value of $k_z$.
    Bottom row: The wavevector-frequency spectrum $\energyspecE^-$ of the $\zm$ fluctuations normalised to its maximum value at each value of $k_z$, with $\injectratio$ increasing from left to right.
    Frequencies in all plots are normalised to the outer scale Alfv\'en frequency $\omega_{\rm A0}=\vA/L_z$.}
    \label{fig:KprpNormSpectroZpZm}
\end{figure}

The wavevector-frequency spectrum of the turbulence can be calculated by taking the space-time Fourier transform of the fluctuations.
The top row of figure~\ref{fig:KprpNormSpectroZpZm} shows the wavevector-frequency spectrum,
\begin{equation}
    \energyspecE_{\rm tot}(k_z,\omega)=\energyspecE^+(k_z, \omega) + \energyspecE^-(k_z, \omega),\label{eq:kfreqEtot}
\end{equation}
for all simulations, where $\energyspecE^{\pm}(k_z, \omega)$ is the $\kprp$-averaged wavevector-frequency spectrum
\begin{equation}
    \energyspecE^{\pm}(\kprp, k_z, \omega) \equiv \frac{1}{2}|\zpm(k_\perp, k_z, \omega)|^2.
\end{equation}
These wavevector-frequency spectra are calculated in AsteriX by specifying an output cadence ($\tauAouter/50$ for these simulations) and outputting the complex Fourier amplitudes of the $\zpm$ fluctuations at every point in the plane containing the $k_z$ axis and the ray along the direction $(\unitvector{x}+\unitvector{y})/\sqrt{2}$ in the $\kprp$-plane.\footnote{The spectrum is assumed to be isotropic around the $k_z$ axis, so this direction is arbitrary.}
The temporal Fourier transform of these coefficients is taken to obtain the wavevector-frequency spectrum, with a Hamming window used to improve resolution.

As the imbalance increases, the centre of the band moves increasingly closer to the dispersion relation of the $\zp$ fluctuations ($\omegaA=-k_z\vA$) as they become energetically dominant.
The centre of the band (blue line), calculated from the mean value of $\omega$ at each value of $k_z$, follows $\omega \approx -\langle\crosshel\rangle k_z\vA$ for each simulation.
This result is not obvious, and follows from the behaviour of the $\zm$ fluctuations as the imbalance grows.
The bottom row of figure~\ref{fig:KprpNormSpectroZpZm} shows the $\energyspecE^-$ component of the total wavevector-frequency spectrum which, instead of peaking around $\omega = k_z\vA$, shifts to near the peak of the $\zp$ fluctuations.
This effect, often termed anomalous coherence \citep{Lithwick2007-ao,Lugones2016-xz,Lugones2019-ur,Schekochihin2022-nn,Meyrand2025-bp}, is a result of the weaker $\zm$ fluctuations being swept up by the stronger $\zp$ fluctuations, allowing $\zm$ to stay coherent with $\zp$ (as described by propagation CB; \citealp{Beresnyak2008-aw}).
The origin of the peak in the total wavevector–frequency spectrum of the $\injectratio=2$ simulation around $\omega = -k_z\vA$ remains unclear; longer simulations are likely needed for better statistics.

\begin{figure}
    \centering
    \includegraphics[width=\textwidth]{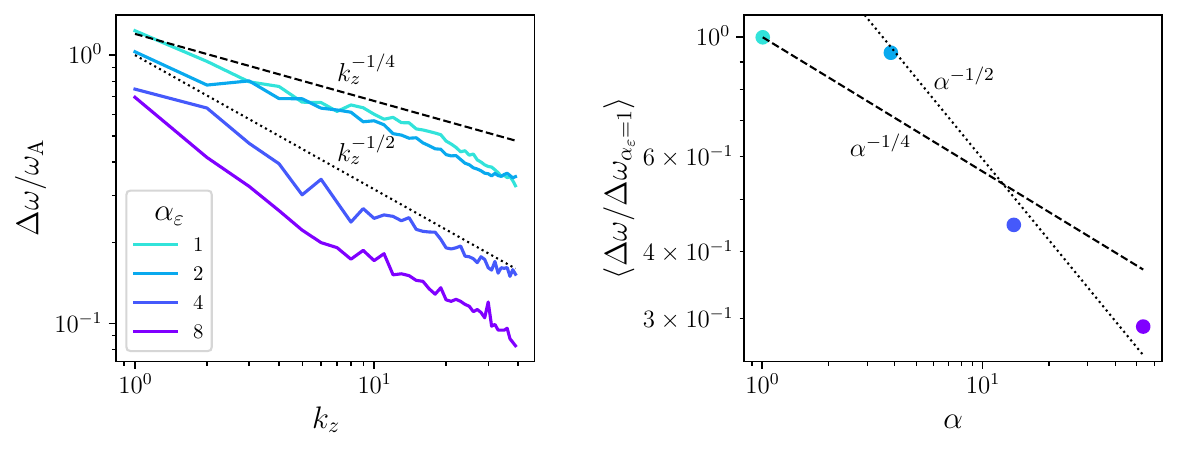}
    \caption{Left: Widths $\Delta\omega$ of the bands in the $\kprp$-averaged wavevector-frequency spectrum (figure~\ref{fig:KprpNormSpectroZpZm}) relative to the Alfv\'en frequency $\omegaA=k_z\vA$. These scale approximately between $k^{-1/4}_z$ and $k^{-1/2}_z$. Right: Average value of the ratio of $\omega_{\rm rms}$ for each simulation to that of the $\injectratio =1$ simulation.}
    \label{fig:ZpmEnergyRatioScaling}
\end{figure}

Additionally, we see that the nonlinear broadening (or width) of the wavevector-frequency spectrum decreases with increasing imbalance as the $\zp$ fluctuations become energetically dominant.
In the limit of complete imbalance ($\crosshel\to 1$), the wavevector-frequency spectrum would become a delta function $\delta(\omega + k_z\vA)$.
As discussed in Section~\ref{sec:HeatingMechanismImbalance}, this imbalance-dependent transition is argued to be important in controlling the heating mechanism of ions.
The width $\Delta\omega$ of the distribution $\energyspecE_{\rm tot}$ can be quantified by the r.m.s. value of $\omega$ relative to the mean at each value of $k_z$ in figure~\ref{fig:KprpNormSpectroZpZm}.
The dependence of $\Delta\omega$ on $k_z$, relative to the linear frequency $\omegaA = k_z\vA$, is shown in the left plot in figure~\ref{fig:ZpmEnergyRatioScaling}.
These scale between $\Delta\omega/\omegaA \propto k^{-1/4}_z$ and $\propto k^{-1/2}_z $, increasing in steepness with $\injectratio$.
This is surprising, as a simple estimate using CB predicts that $\Delta \omega / \omegaA$ is independent of $k_z$:
\begin{equation}
    \frac{\Delta\omega}{\omegaA} \sim \frac{\omeganl}{\omegaA} = \frac{\kprp\deltazm}{k_z \vA} \sim \frac{\deltazm}{\deltazp}\frac{\kprp\deltazp}{k_z\vA} \sim \forceratio^{-1/2},\label{eq:PredictedDeltaOmegaAlpha}
\end{equation}
with the ratio $\deltazp/\deltazm$ assumed to be independent of scale.
The right plot of figure~\ref{fig:ZpmEnergyRatioScaling} shows the ratio of $\Delta\omega$ relative to that of the $\injectratio=1$ simulation, averaged over $k_z$, suggesting a decrease in width with $\forceratio$ broadly consistent with a scaling between $\forceratio^{-1/4}$ and $\forceratio^{-1/2}$.
The differences between the measured scalings and expectations may be due to the fact that the amplitudes of the $\zp$ fluctuations are not truly constant across the simulations, with figure~\ref{fig:RMHDSimKprpEnergySpec} showing that the imbalanced simulations have a slightly higher amplitude compared to the balanced case.
A detailed verification of the scalings in \eqref{eq:PredictedDeltaOmegaAlpha} may require higher resolution simulations, with care taken to ensure the amplitudes of the $\zp$ fluctuations are as constant as possible when varying imbalance.

\begin{figure}
    \centering
    \includegraphics[width=\textwidth]{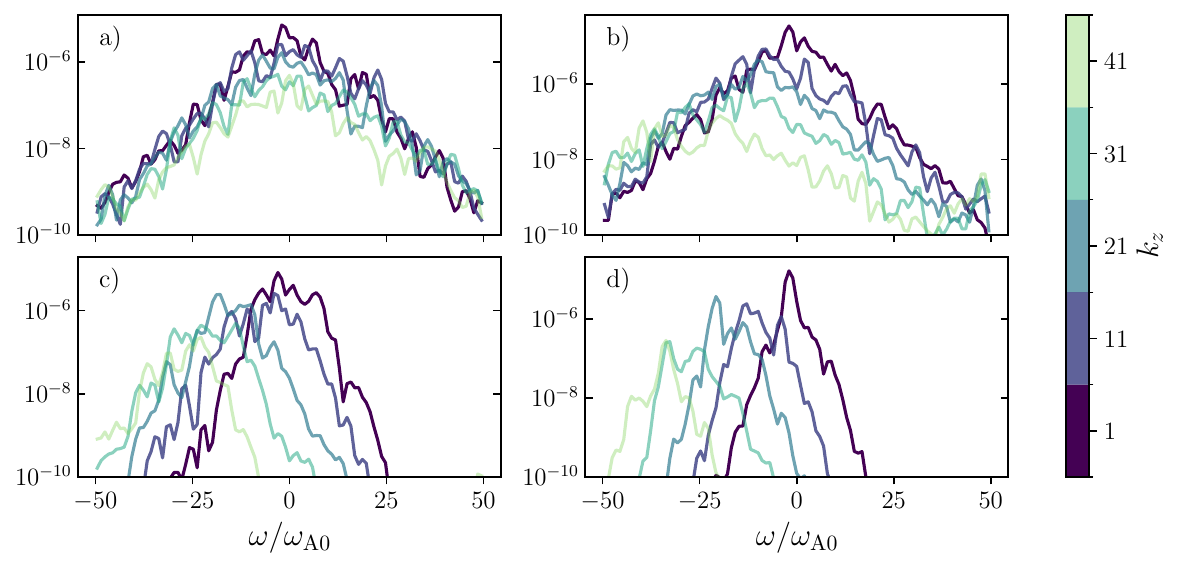}
    \caption{Slices at $\kprp=50$ and increasing values of $k_z$ (from 1 to 41 in steps of 10) from the $\injectratio = 1$ (a), 2 (b), 4 (c), and 8 (d) wavevector-frequency spectra $\energyspecE_{\rm tot}(\kprp,k_z,\omega)$. Frequencies are normalised to the outer scale Alfv\'en frequency $\omega_{\rm A0}=\vA/L_z$.}
    \label{fig:SpectroSliceConstKprp}
\end{figure}

\begin{figure}
    \centering
    \includegraphics[width=\textwidth]{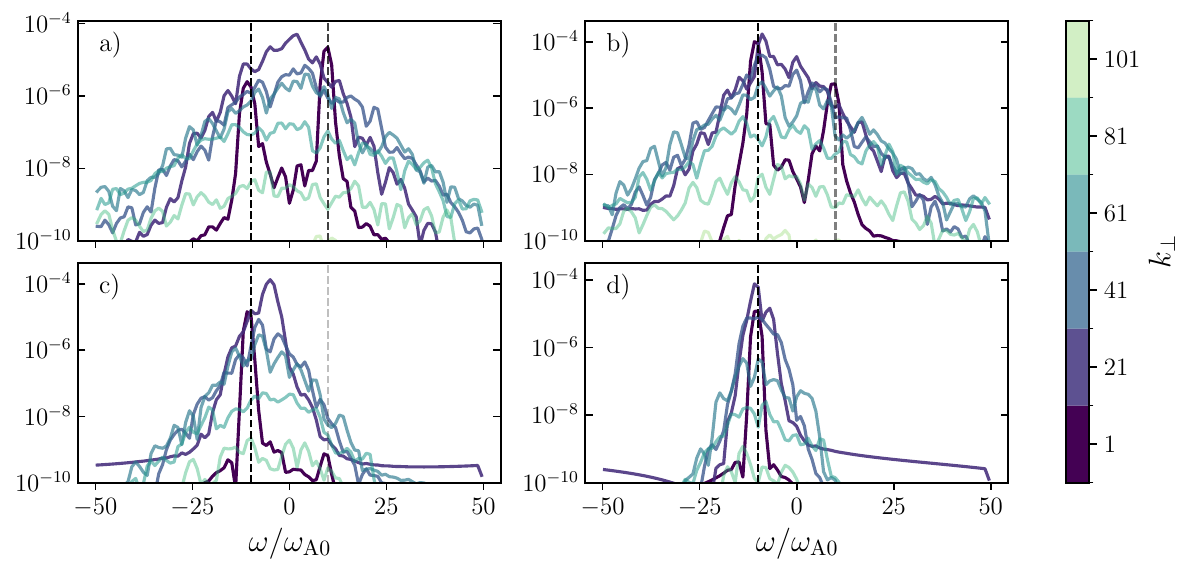}
    \caption{Slices at $k_z=10$ and increasing values of $\kprp$ (from 1 to 101 in steps of 20) from the $\injectratio = 1$ (a), 2 (b), 4 (c), and 8 (d) wavevector-frequency spectra $\energyspecE_{\rm tot}(\kprp,k_z,\omega)$.
    Dashed vertical lines at $\omegaA/\omega_{\rm A0}=\pm 10$ correspond to the Alfv\'en wave dispersion relation at this value of $k_z$.
    Frequencies are normalised to the outer scale Alfv\'en frequency $\omega_{\rm A0}=\vA/L_z$.}
    \label{fig:SpectroSliceConstKz}
\end{figure}

Figures~\ref{fig:SpectroSliceConstKprp} and~\ref{fig:SpectroSliceConstKz} show slices through the full $\energyspecE_{\rm tot}(\kprp,k_z,\omega)$ wavevector-frequency spectrum at $\kprp=50$ and $k_z = 10$, respectively.
The slices at $\kprp=50$ (figure~\ref{fig:SpectroSliceConstKprp}) reflect the reduction of the nonlinear broadening with increasing imbalance seen in figure~\ref{fig:KprpNormSpectroZpZm}, as well as the shift in the peak towards the $\zp$ dispersion relation ($\omegaA=-k_z\vA$) for each value of $k_z$.
The amplitude of the wavevector-frequency spectrum also stays approximately constant (reflecting the region of the turbulence spectrum $\propto k^0_z$, where $\tauA\gg\taunl$) until it rises above the CB threshold, where it begins to decrease.

The slices at $k_z=10$ (figure~\ref{fig:SpectroSliceConstKz}) explicitly show how the $\zp$ fluctuations dominate over the $\zm$ fluctuations with increasing imbalance.
In the region above the CB cone, where the width of the broadening is less than $\omeganl$, the $\zpm$ fluctuations show up in the wavevector-frequency spectrum as strongly-peaked regions centred on $\omegaA/\vA = \pm k_z = \pm 10$.
As the imbalance increases, the $\zm$ peak (at $\omegaA/\vA= 10$) becomes increasingly subdominant and disappears almost entirely at large values of $\crosshel$.
The amplitude of these peaks increases as we get closer to the CB threshold, beyond which it reaches the region inside the CB cone where $\tauA\gg\taunl$; here, the peaks broaden, and their amplitude decreases (in accord with the $\kprp^{-7/3}$ scaling of the energy spectrum in this regime); in this region of strong turbulence, there are also no signs of any peaks at the Alfv\'enic dispersion relation in $\zp$ or $\zm$.
As in figure~\ref{fig:SpectroSliceConstKprp}, the width of the nonlinear broadening decreases with increasing imbalance.

Overall, these diagnostics justify the core assumptions of our imbalanced turbulence model.
However, higher resolution simulations run for longer would be valuable to assess subtler features of the imbalanced phenomenology and to understand the minor discrepancies found above.

\section{Derivation of quasi-linear diffusion coefficients in $(\vprp,\vprl)$ coordinates for general electromagnetic fields}\label{app:QLTheoryDerivation}

In this Appendix, we derive the diffusion coefficients that form the basis for the calculations presented in Sections~\ref{sec:RMHDLimitCoeffs} and \ref{sec:HeatingRates}.
Several related formulations have appeared previously---most commonly in the context of pitch-angle scattering for cosmic ray applications \citep[e.g.,][]{Schlickeiser1993-if,Chandran2000-pq}---but a derivation of the general diffusion coefficients in $(\vprp,\vprl)$ coordinates appears to have been presented only recently by \citet{Schlickeiser2022-vz} (who give a more general treatment of the approach outlined below) and by \citet{Brizard2022-dz} (who develop a Hamiltonian formulation of quasi-linear theory).
The important difference compared to the standard treatments of \citet{Kennel1966-rx,Stix1992-bo} is that we do not assume waves with a linear dispersion relation, and make no assumptions about the space and time correlations of the fields.
For completeness, we provide a detailed derivation and discussion, although much of this is now standard.

We first start with the Vlasov equation describing the collisionless evolution of a particle species' distribution function $f(\ppos,\pvel,t)$:
\begin{equation}
    \parfrac{f}{t}+\pvel\dotprod\boldnabla f + \frac{q}{m}\left(\Evec+\frac{\pvel}{c}\crossprod\B\right)\dotprod\parfrac{f}{\pvel}=0.\label{eq:VlasovEq}
\end{equation}
Then, split the distribution function and fields into background and fluctuating components:
\begin{align}
        f(\ppos,\pvel,t)&=f_0(\pvel,t)+f_1(\ppos,\pvel,t),\\
        \Evec(\ppos,t)&=\Evec_1(\ppos,t),\\
        \B(\ppos,t)&=\Bmeanmag\unitvector{z}+\B_1(\ppos,t);
\end{align}
the background distribution function $f_0$ is assumed to be homogenous in space, and the background electric field is assumed to vanish.

We assume that the fluctuations are small in amplitude and vary on timescales faster than the evolution of the background distribution.\footnote{Namely, we assume that $\parfracil{f_0}{t}\to 0$ over the fast timescales of the fluctuations; this is more formally justified by an average over the gyroangle, which is used below when calculating the evolution of $f_0$.}
With this assumption, the zeroth-order terms $f_0$ and $\Bmean$ satisfy
\begin{equation}
    \Omegai(\pvel\crossprod\unitvector{z})\dotprod\parfrac{f_0}{\pvel}=0.\label{eq:zerothOrderVlasov}
\end{equation}
After defining cylindrical coordinates in velocity space aligned along $\Bmeanmag\unitvector{z}$,
\begin{align}
    \pvel &= \vprp (\unitvector{x}\cos\phi+\unitvector{y}\sin\phi)+\vprl\unitvector{z}\nonumber\\
    &= \vprp\unitvector{v}_\perp+\vprl\unitvector{z},\label{eq:VelocityCylindricalCoords}
\end{align}
where $\phi$ is the gyroangle and $\unitvector{v}_\perp = \unitvector{x}\cos\phi+\unitvector{y}\sin\phi$, $\unitvectorgreek{\phi}=-\unitvector{x}\sin\phi+\unitvector{y}\cos\phi$ and $\unitvector{z}$ are the cylindrical basis vectors in velocity space, \eqref{eq:zerothOrderVlasov} reduces to $\parfracil{f_0}{\phi}=0$. This shows that $f_0$ is gyrotropic, and can be written as $f_0(\pvel,t)=f_0(\vprp,\vprl,t)$.

Using the decomposition above, to first-order the Vlasov equation becomes
\begin{equation}
    \parfrac{f_1}{t}+\pvel\dotprod\boldnabla f_1 + \Omegai\left(\pvel\crossprod\unitvector{z}\right)\dotprod\parfrac{f_1}{\pvel} = -\frac{q}{m}\left(\Evec_1+\frac{\pvel}{c}\crossprod\B_1\right)\dotprod\parfrac{f_0}{\pvel},\label{eq:VlasovFirstOrder}
\end{equation}
with $\Omegai=\ion{q}\Bmeanmag/\ion{m}c$ the gyrofrequency.
The left-hand side of \eqref{eq:VlasovFirstOrder} is just $(\totfracil{f_1}{t})_0$, the total time derivative of $f_1$ along the zero-order trajectory of particles (i.e. unperturbed helices along the background magnetic field).
Integrating \eqref{eq:VlasovFirstOrder} from $t'=-\infty\to t$, we obtain
\begin{align}
    f_1(\ppos,\pvel,t)&=-\frac{q}{m}\intsingle{t'}{-\infty}{t}\left[\Evec_1(\ppos',t')+\frac{\pvel'}{c}\crossprod\B_1(\ppos',t')\right]\dotprod\parfrac{f_0(\pvel')}{\pvel'} \nonumber \\ 
    &=-\frac{q}{m}\intsingle{\tau}{0}{\infty} \left[\Evec_1(\ppos',t-\tau)+\frac{\pvel'}{c}\crossprod\B_1(\ppos',t-\tau)\right]\dotprod\parfrac{f_0(\pvel')}{\pvel'},\label{eq:f1EvoSpatial}
\end{align}
where in the second equality we change variables to the lag time $\tau\equiv t-t'\geq 0$.
Here, $\ppos'\equiv \ppos'(t')$ and $\pvel' \equiv \pvel'(t')$ are Lagrangian coordinates following the zero-order trajectory of particles through phase space satisfying
\begin{align}
    \totfrac{\ppos'}{t'}&=\pvel,\label{eq:LagrangianPosition}\\
    \totfrac{\pvel'}{t'}&=\Omegai(\pvel'\crossprod\unitvector{z})\label{eq:LagrangianVelocity}
\end{align}
such that $\ppos'(t'=t) = \ppos$ and $\pvel'(t'=t) = \pvel$ (the Eulerian coordinate $(\ppos,\pvel)$ in phase space at time $t$ at which the distribution function is evaluated).

At this point, if the fluctuations of interest follow a linear dispersion relation $\omega(\kvec)$, we can choose the ansatz $\Evec_1(\ppos',t')=\Eveck e^{\Icplx(\kvec\dotprod\ppos'-\omega(\kvec) t')}$ and use Faraday's law to write $\B_1$ in terms of $\Evec_1$.
This formulation allows the description of strong wave–particle interactions, where resonance between the particle gyrofrequency and the Doppler-shifted wave frequency causes diffusion along constant-energy contours in the wave frame \citep{Kennel1966-rx,Stix1992-bo}.
However, a spectrum of turbulent fluctuations is generally broadened due to nonlinear interactions compared to a collection of waves following a dispersion relation.
As we are interested in the effect of general turbulent fluctuations on ion heating, we do not use a dispersion relation in this calculation and make no assumptions about the space and time correlations of the fields.
The electric and magnetic field terms are also kept separate, as this allows us to see which field is responsible for a given effect on the distribution function if needed \citep{Hall1967-cg}.
However, if the wave ansatz introduced above is inserted at any stage of the following derivation, the general diffusion coefficients derived below reduce to the forms presented in \citet{Kennel1966-rx,Stix1992-bo}; a brief outline of the steps in this procedure is provided at the end of this Appendix.

\subsection{Spatial Fourier transform of the first-order distribution function}

We now calculate the spatial Fourier transform of \eqref{eq:f1EvoSpatial}, which is to be used in the calculation of its effect on the background distribution below.
The first-order perturbation can be expressed in terms of its Fourier components as\footnote{We use the spatial transform convention 
$\hat{\mathcal{F}}_{\kvec}[f(\ppos)] = (2\upi)^{-3}\intvol{e^{-\Icplx\kvec\dotprod\ppos}f(\ppos)}$,
where the integration is carried over a windowing volume $V$ that allows the Fourier transform to converge, and the inverse transform carries no factors of $2\upi$.}
\begin{equation}
    f_1(\ppos,\pvel,t) = \intsingle{\kvec}{}{} e^{\Icplx\kvec\dotprod\ppos}f_1(\kvec,\pvel,t);
\end{equation}
the electric and magnetic field perturbations can be expressed in a similar way.
Inserting this into \eqref{eq:f1EvoSpatial}, we obtain
\begin{align}
    \intsingle{\kvec}{}{} e^{\Icplx\kvec\dotprod\ppos} f_1(\kvec,\pvel,t) &= -\frac{q}{m}\intsingle{\tau}{0}{\infty}\times\nonumber\\
    &\intsingle{\kvec}{}{} e^{\Icplx\kvec\dotprod\ppos'}\left[\Evec_1(\kvec,t-\tau)+\frac{\pvel'}{c}\crossprod\B_1(\kvec,t-\tau)\right]\dotprod\parfrac{f_0(\pvel')}{\pvel'}.\label{eq:f1FTPartial}
\end{align}
We cannot yet identify $f_1(\kvec,\pvel,t)$ with the right-hand integrand as the exponential contains the Lagrangian trajectory $\ppos'$ evaluated at $t-\tau$, rather than the Eulerian coordinate at which the left-hand side is evaluated.
This can be fixed by solving \eqref{eq:LagrangianPosition} and \eqref{eq:LagrangianVelocity} to express the Lagrangian coordinate $\ppos'$ in terms of $\tau$ and the Eulerian coordinate $\ppos$.

First, we change variables from $t'$ to $\tau$ in \eqref{eq:LagrangianPosition} and \eqref{eq:LagrangianVelocity} (resulting in a minus sign in both equations).
Solving \eqref{eq:LagrangianVelocity} with initial condition $\pvel'(\tau=0)=\pvel$ gives
\begin{align}
    v'_x(\tau) &= \vprp\cos(\phi+\Omegai\tau),\\
    v'_y(\tau) &= \vprp\sin(\phi+\Omegai\tau),\\
    v'_z(\tau) &= \vprl.
\end{align}
Using this velocity in \eqref{eq:LagrangianPosition} and $\ppos'(\tau=0)=\ppos$ gives $\ppos'(\tau) = \ppos + \Bvectorgreek{\xi}(\tau)$, where
\begin{align}
     \Bvectorgreek{\xi}(\tau) &\equiv -\frac{\vprp}{\Omegai}\left[\left(\sin(\phi+\Omegai\tau)-\sin(\phi)\right)\unitvector{x}-\left(\cos(\phi+\Omegai\tau)-\cos(\phi)\right)\unitvector{y}\right] - \vprl\tau\unitvector{z}\nonumber\\
     &= -\frac{\vprp}{\Omegai}\left[\sin(\Omegai\tau)\unitvector{v}_\perp + (1-\cos(\Omegai\tau))\unitvectorgreek{\phi}\right] - \vprl\tau\unitvector{z},
\end{align}
using the cylindrical unit vectors in \eqref{eq:VelocityCylindricalCoords}.
Inserting this into \eqref{eq:f1FTPartial} allows us to now identify $f_1(\kvec,\pvel,t)$ as
\begin{align}
f_1(\kvec,\pvel,t)&=-\frac{q}{m}\intsingle{\tau}{0}{\infty} e^{\Icplx\kvec\dotprod\Bvectorgreek{\xi}(\tau)}\left[\Evec_1(\kvec,t-\tau)+\frac{\pvel'}{c}\crossprod\B_1(\kvec,t-\tau)\right]\dotprod\parfrac{f_0(\pvel')}{\pvel'}\nonumber\\
&= -\frac{q}{m}\intsingle{\tau}{0}{\infty} e^{\Icplx\beta_{\kvec}(\tau)}\left[\Eveck'+\frac{\pvel'}{c}\crossprod\Bveck'\right]\dotprod\parfrac{f_0(\pvel')}{\pvel'},\label{eq:f1EvoFourier}
\end{align}
defining $\beta_{\kvec}(\tau)\equiv \kvec\dotprod\Bvectorgreek{\xi}(\tau)$, $\Eveck\equiv \Evec_1(\kvec,t)$, and $\Bveck\equiv \B(\kvec,t)$ with the prime denoting evaluation at the time $t-\tau$.
Next, define cylindrical coordinates in wavevector space aligned along $\Bmeanmag\unitvector{z}$:
\begin{align}
    \kvec &= \kprp(\unitvector{x}\cos\psi+\unitvector{y}\sin\psi)+\kprl\unitvector{z}\nonumber\\
    &= \kprp\cos(\phi-\psi)\unitvector{v}_\perp - \kprp\sin(\phi-\psi)\unitvectorgreek{\phi}+\kprl\unitvector{z},
\end{align}
where we have expressed $\kvec$ in terms of the cylindrical unit vectors in \eqref{eq:VelocityCylindricalCoords}.
Applying trigonometric identities, it can then be shown that
\begin{equation}
    \beta_{\kvec}(\tau) = -\frac{\kprp\vprp}{\Omegai}\left[\sin(\tilde\phi+\Omegai\tau) - \sin(\tilde\phi)\right]-\kprl\vprl\tau,\label{eq:betaKvecDef}
\end{equation}
with $\tilde\phi\equiv\phi-\psi$.

Next, write $\Eveck'$ and $\Bveck'$ in cylindrical coordinates:
\begin{equation}
    \Eveck' = \Eveckcomp{\rm T}'\unitvector{v}_\perp + (\cdots)\unitvectorgreek{\phi} + \Eveckcomp{z}'\unitvector{z}\textrm{ and } 
    \Bveck' = (\cdots)\unitvector{v}_\perp -\Icplx\Bveckcomp{\rm P}'\unitvectorgreek{\phi} + (\cdots)\unitvector{z},
\end{equation}
where 
\begin{equation}
\Eveckcomp{\rm T}' \equiv \Eveckcomp{x}'\cos(\phi+\Omegai\tau)+\Eveckcomp{y}'\sin(\phi+\Omegai\tau) = \Eveckcomp{+}'e^{-\Icplx(\tilde\phi+\Omegai\tau)} + \Eveckcomp{-}'e^{\Icplx(\tilde\phi+\Omegai\tau)}\label{eq:EvecT}
\end{equation}
and
\begin{align}
-\Icplx\Bveckcomp{\rm P}' &\equiv -\Bveckcomp{x}'\sin(\phi+\Omegai\tau)+\Bveckcomp{y}'\cos(\phi+\Omegai\tau)\nonumber\\
&= -\Icplx\left[\Bveckcomp{+}'e^{-\Icplx(\tilde\phi+\Omegai\tau)} - \Bveckcomp{-}'e^{\Icplx(\tilde\phi+\Omegai\tau)}\right].\label{eq:BvecP}
\end{align}
Here, we use circularly polarised components of $\Eveck$ and $\Bveck$, defined as
\begin{equation}
    A_{\kvec\pm}\equiv\frac{1}{2}\left(A_{\kvec x}\pm \Icplx A_{\kvec y}\right)e^{\mp \Icplx\psi}
\end{equation}
for general vector $\Bvector{A}_{\kvec}$.
The subscripts T and P are labelled to denote the fact that these fields drive changes in transverse velocity (when $\Bveckcomp{\rm P}$ = 0) and pitch angle (when $\Eveckcomp{\rm T}=\Eveckcomp{z}=0$).
The $\unitvectorgreek{\phi}$ component of $\Eveck'$ and the $\unitvector{v}_\perp$ and $\unitvector{z}$ components of $\Bveck'$ are irrelevant due to the gyrotropy of $f_0$ (the last two give a term $\propto\unitvectorgreek{\phi}$ from $\pvel'\crossprod\Bveck$).
Defining $\hat{v}_{\perp,\|}\equiv v_{\perp,\|}/c$, we then have
\begin{equation}
    \Eveck'+\frac{\pvel'}{c}\crossprod\Bveck' = \left[\Eveckcomp{\rm T}'+\Icplx\vprlhat\Bveckcomp{\rm P}'\right]\unitvector{v}_\perp + \left[\Eveckcomp{z}'-\Icplx\vprphat\Bveckcomp{\rm P}'\right]\unitvector{z}\label{eq:LorentzForceLag}
\end{equation}
giving
\begin{equation}
    f_1(\kvec,\pvel,t) = -\frac{q}{m}\intsingle{\tau}{0}{\infty} e^{\Icplx\beta_{\kvec}(\tau)} \Biggl\{\left[\Eveckcomp{\rm T}'+\Icplx\vprlhat\Bveckcomp{\rm P}'\right]\parfrac{}{\vprp}+\left[\Eveckcomp{z}'-\Icplx\vprphat\Bveckcomp{\rm P}'\right]\parfrac{}{\vprl}\Biggr\}f_0.\label{eq:f1Fourier}
\end{equation}

\subsection{Evolution of background distribution in response to first-order perturbation}

With the first-order response calculated, we can determine its effect on the background distribution.
Before continuing, let us clarify the meaning of the timescales involved in the problem, as it may seem contradictory that we are integrating $f_0$ over an infinite time to obtain $f_1$, and then using this to calculate how $f_0$ varies in time.
The key point here is that $f_0$ and $f_1$ are assumed to vary on a slow and fast timescale, respectively.
Intuitively, the separation of these timescales lets us treat $f_0$ as fixed when calculating $f_1$, with the limit $\tau\to\infty$ taken to allow $f_1$ to reach its asymptotic form over the fast timescale in the presence of the background distribution.
The effect on the slow evolution of $f_0$ can then be calculated with this asymptotic form (see Chapter 10, Section 4 of \citealp{Stix1992-bo} for further discussion).

Start with the full Vlasov equation \eqref{eq:VlasovEq} and use the facts that $\Evec$ and $\B$ are independent of $\pvel$ and that $\parfrac{}{\pvel}\cdot(\pvel\crossprod\B)=0$ to write the velocity term as a divergence:
\begin{equation}
    \parfrac{f}{t}+\pvel\dotprod\bnabla f + \frac{q}{m}\parfrac{}{\pvel}\cdot\left[\left(\Evec+\frac{\pvel}{c}\crossprod\B\right)f\right]=0.\label{eq:VlasovDivergence}
\end{equation}
To get an evolution equation for $f_0$ we take the spatial average of the above equation, performed using a window volume $V$ in space and taking the limit of this volume to infinity.
This average annihilates $\pvel\cdot\bnabla f$, and we assume that the first-order averages $\langle\Evec_1\rangle=\langle\B_1\rangle=0$; higher order contributions to this average are ignored.
Additionally, we take the gyroaverage due to the assumed slow evolution of $f_0$ compared to fluctuations; this allows us to write $f_0(\pvel, t)=f_0(\vprp,\vprl,t)$.
This gives
\begin{align}
    \parfrac{f_0}{t}&=\lim_{V\to\infty}-\frac{q}{m}\intfrac{\ppos}{V}{}{}\intfrac{\phi}{2\upi}{0}{2\upi}\parfrac{}{\pvel}\cdot\left[\left(\Evec_1(\ppos,t)+\frac{\pvel}{c}\crossprod\B_1(\ppos,t)\right)f_1(\ppos,\pvel,t)\right]\nonumber\\
    &= \lim_{V\to\infty}-\frac{q}{m}\intfrac{\kvec}{V}{}{}\intfrac{\phi}{2\upi}{0}{2\upi}\parfrac{}{\pvel}\cdot\left[\left(\Eveck+\frac{\pvel}{c}\crossprod\Bveck\right)^*f_{\kvec}\right],\label{eq:f0Evolution}
\end{align}
where the equality follows from the product of the Fourier transforms of the electromagnetic and distribution function terms in the integrand (with $f_{\kvec}\equiv f_1(\kvec,\pvel,t)$), and using the fact that $f_0$ and $f_1$ are real (so $f^*_{-\kvec}=f_{\kvec}$).

Ignoring the $\parfracil{}{\phi}$ derivative in \eqref{eq:f0Evolution}, because it is annihilated by the gyroaverage as a total derivative, the $\Eveck+\pvel\crossprod\Bveck/c$ term can be expanded in a manner similar to \eqref{eq:LorentzForceLag}, with the only difference being the replacement $\tilde\phi+\Omegai\tau\to\tilde\phi$ as the fields are evaluated at the current time $t$ rather than the lag time $t-\tau$.
Using \eqref{eq:f1Fourier} and changing variables to $\tilde\phi$ in the integral\footnote{All intervals of width $2\upi$ in $\tilde\phi$ are equivalent, so we keep the bounds from $0\to2\upi$.} allows the gyroaveraged velocity divergence to be written as
\begin{equation}\label{eq:f0Gyroaverage}
    I\equiv-\frac{q}{m}\intfrac{\tilde\phi}{2\upi}{0}{2\upi}\parfrac{}{\pvel}\cdot\left[\left(\Eveck+\frac{\pvel}{c}\crossprod\Bveck\right)^*f_{\kvec}\right] =\frac{q^2}{m^2}\intfrac{\tilde\phi}{2\upi}{0}{2\upi}\intsingle{\tau}{0}{\infty}\mathcal{P}f_0,
\end{equation}
where
\begin{align*}
    \mathcal{P}&= \frac{1}{\vprp}\parfrac{}{\vprp}\left[\vprp(\Eveckcomp{\rm T}^*-\Icplx\vprlhat\Bveckcomp{\rm P}^*)\mathcal{Q}\right]+\parfrac{}{\vprl}\left[(\Eveckcomp{z}^*+\Icplx\vprlhat\Bveckcomp{\rm P}^*)\mathcal{Q}\right],\\
    \mathcal{Q} &= e^{\Icplx\beta_{\kvec}(\tau)}\left\{ \left[\Eveckcomp{\rm T}'+\Icplx\vprlhat\Bveckcomp{\rm P}'\right]\parfrac{}{\vprp}+\left[\Eveckcomp{z}'-\Icplx\vprphat\Bveckcomp{\rm P}'\right]\parfrac{}{\vprl}\right\}.
\end{align*}
Swapping the order of the integrals and moving them in between the differential operators in $\mathcal{P}$ and $\mathcal{Q}$ leads to terms of the form
\begin{equation}
    \frac{q^2}{m^2}\intsingle{\tau}{0}{\infty} e^{-\Icplx\kprl\vprl\tau}  \intfrac{\tilde\phi}{2\upi}{0}{2\upi} e^{-\Icplx \kappa(\sin(\tilde\phi+\Omegai\tau)-\sin(\tilde\phi))} R^*S',
    \label{eq:GyroaverageSimplifed}
\end{equation}
where $\kappa\equiv \kprp\vprp/\Omegai$ and $R$ and $S$ are any of $\Eveckcomp{\rm T},\ \Eveckcomp{z}$, or $\Icplx\hat{v}_{\perp,\|}\Bveckcomp{\rm P}$.
Using the definitions \eqref{eq:EvecT} and \eqref{eq:BvecP}, these multiplications lead to terms proportional to $e^{\pm\Icplx(l\tilde\phi+\Omegai\tau)}$ with $l=0,1,$ or $2$.
The gyroaverage of these products can then be written as an infinite sum of Bessel function products using the identities $e^{\Icplx \kappa\sin\phi}=\sum_{n=-\infty}^\infty e^{\Icplx n\phi}\BesselJ{n}(\kappa)$ and $\intfrac{\phi}{2\upi}{0}{2\upi}e^{\Icplx(m-n)\phi}=\delta_{m,n}$ \citep{Arfken2012-ra}:
\begin{align}
    \intfrac{\tilde\phi}{2\upi}{0}{2\upi} e^{-\Icplx \kappa(\sin(\tilde\phi+\Omegai\tau)-\sin(\tilde\phi))} R^*S'&\propto \sum_{n,m=-\infty}^{\infty} \BesselJ{n}\BesselJ{m}e^{-\Icplx(n\mp 1)\Omegai\tau}\intfrac{\tilde\phi}{2\upi}{0}{2\upi}e^{-\Icplx[m-(n\mp l)]\tilde\phi}\nonumber\\
    &= \sum_{n=-\infty}^{\infty} \BesselJ{n\mp(l-1)}\BesselJ{n\pm 1}e^{-\Icplx n\Omegai\tau} \ (l=0,1,2),
\end{align}
where the index was changed from $n\to n\pm 1$ in the last equality.
After the gyroaveraging, the terms can be factorized into products of terms of the form
\begin{subequations}
\label{eq:GyroAvgComponents}
\begin{align}
E^{(n)}_{\kvec \rm T}&\equiv \Eveckcomp{+}\BesselJ{n-1}+\Eveckcomp{-}\BesselJ{n+1},\\
E^{(n)}_{\kvec z}&\equiv\Eveckcomp{z}\BesselJ{n},\\
B^{(n)}_{\kvec \rm P}&\equiv \Bveckcomp{+}\BesselJ{n-1}-\Bveckcomp{-}\BesselJ{n+1}.
\end{align}
\end{subequations}
Finally, inserting this gyroaverage into \eqref{eq:f0Evolution} and taking an ensemble average shows that the background distribution function undergoes a diffusion in velocity space of the form
\begin{equation}
    \parfrac{f_0}{t} = \frac{1}{\vprp}\parfrac{}{\vprp}\left[\vprp\left(\Dprpprp\parfrac{f_0}{\vprp}+\Dprlprp\parfrac{f_0}{\vprl}\right)\right] + \parfrac{}{\vprl}\left[\Dprpprl\parfrac{f_0}{\vprp}+\Dprlprl\parfrac{f_0}{\vprl}\right],
\end{equation}
with the diffusion coefficients given by 
\begin{equation}
    D_{ab} = 
    \lim_{V\to\infty}\frac{q^2}{m^2}\sum_{n=-\infty}^\infty\intfrac{\kvec}{V}{}{}\intsingle{\tau}{0}{\infty} e^{-\Icplx(\kprl\vprl +n\Omegai)\tau} \mathcal{D}^{(n)}_{ab}(\kvec,\tau),\label{eq:OuterDiffusionCoeff}
\end{equation}
where $a,b\in \{\perp, \|\}$ and $V$ is a windowing volume used when taking the spatial averaging of the distribution function.
The quantities $\mathcal{D}^{(n)}_{ab}(\kvec,\tau)$ are given by
\begin{subequations}
    \label{eq:InnerDiffusionCoeffs}
\begin{align}
    \mathcal{D}^{(n)}_{\perp\perp}(\kvec,\tau) &= \left\langle \left[E^{(n)}_{\rm T}(\kvec, t)+\Icplx\vprlhat B^{(n)}_{\rm P}(\kvec, t)\right]\left[E^{(n)}_{\rm T}(\kvec, t+\tau)+\Icplx\vprlhat B^{(n)}_{\rm P}(\kvec, t+\tau)\right]^*\right\rangle,\\
    \mathcal{D}^{(n)}_{\|\perp}(\kvec,\tau) &= \left\langle \left[E^{(n)}_{z}(\kvec, t)-\Icplx\vprphat B^{(n)}_{\rm P}(\kvec, t)\right]\left[E^{(n)}_{\rm T}(\kvec, t+\tau)+\Icplx\vprlhat B^{(n)}_{\rm P}(\kvec, t+\tau)\right]^*\right\rangle,\\
    \mathcal{D}^{(n)}_{\perp\|}(\kvec,\tau) &= \left\langle \left[E^{(n)}_{\rm T}(\kvec, t)+\Icplx\vprlhat B^{(n)}_{\rm P}(\kvec, t)\right] \left[E^{(n)}_{z}(\kvec, t+\tau)-\Icplx\vprphat B^{(n)}_{\rm P}(\kvec, t+\tau)\right]^*\right\rangle,\\
    \mathcal{D}^{(n)}_{\|\|}(\kvec,\tau) &= \left\langle \left[E^{(n)}_{z}(\kvec, t)-\Icplx\vprphat B^{(n)}_{\rm P}(\kvec, t)\right] \left[E^{(n)}_{z}(\kvec, t+\tau)-\Icplx\vprphat B^{(n)}_{\rm P}(\kvec, t+\tau)\right]^*\right\rangle.
\end{align}
\end{subequations}
The full expansion of these in terms of the various correlations between fields is unenlightening (see, e.g., \citealp{Schlickeiser1993-if}).
The diffusion coefficients \eqref{eq:OuterDiffusionCoeff} are intractably complex for general use; to gain physical insight into how they affect the heating of ions in turbulence with different levels of imbalance, we take the RMHD limit as described in Section~\ref{sec:RMHDLimitCoeffs}, which simplifies them dramatically.

For completeness, we elucidate the steps outlined in Section~\ref{sec:RMHDLimitCoeffs} used to transform the diffusion coefficients \eqref{eq:InnerDiffusionCoeffs} to their RMHD form.
For fluctuations with $\rhoiscales\ll 1$, we can write the diffusion coefficients in terms of the turbulent velocity field via Ohm's law, $\Evec_1 = -(\Bmeanmag/c)\vel_1 \crossprod \unitvector{z}$, which gives $E_\pm(\kvec,t)=\pm (\Icplx\Bmeanmag/c)u_\pm(\kvec,t)$ and $E_z(\kvec, t)=0$.
Assuming an Alfv\'enic polarisation, where fluctuations with perpendicular (unit) wavevector $\unitvector{k}_\perp = \cos(\psi)\unitvector{x}+\sin(\psi)\unitvector{y}$ are polarised in the $\unitvector{e}_{\kvec}\equiv \unitvector{k}_\perp\crossprod\unitvector{z}=\sin(\psi)\unitvector{x}-\cos(\psi)\unitvector{y}$ direction, allows the velocity and magnetic perturbations to be written as $\vel_1(\kvec,t)=A_{\vel}(\kvec,t)\unitvector{e}_{\kvec}$ and $\B_1(\kvec, t)=A_{\B}(\kvec,t)\unitvector{e}_{\kvec}$.
This then allows us to write $u_\pm(\kvec, t)=\mp iA_{\vel}(\kvec,t)/2$ (and thus $E_\pm(\kvec, t)=\Bmeanmag A_{\vel}(\kvec, t)/2c$) and $B_\pm(\kvec,t)=\mp iA_{\B}(\kvec, t)/2$.
Inserting the expressions for $E_\pm$ and $B_\pm$ into \eqref{eq:GyroAvgComponents} allows us to simplify the coefficients in \eqref{eq:InnerDiffusionCoeffs} as
\begin{subequations}
    \label{eq:InnerDiffusionCoeffsRMHD}
\begin{align}
    \mathcal{D}^{(n)}_{\perp\perp}(\kvec,\tau) &= \left\langle \left[\frac{\Bmeanmag}{c}u^{(n)}_{\kvec \rm P}(\kvec, t)+\vprlhat B^{(n)}_{\kvec\rm P}(\kvec, t)\right]\left[\frac{\Bmeanmag}{c}u^{(n)}_{\kvec \rm P}(\kvec, t+\tau)+\vprlhat B^{(n)}_{\kvec\rm P}(\kvec, t+\tau)\right]^*\right\rangle,\\
    \mathcal{D}^{(n)}_{\|\perp}(\kvec,\tau) &= -\left\langle \left[\vprphat B^{(n)}_{\kvec\rm P}(\kvec, t)\right]\left[\frac{\Bmeanmag}{c} u^{(n)}_{\kvec \rm P}(\kvec, t+\tau)+\vprlhat B^{(n)}_{\kvec\rm P}(\kvec, t+\tau)\right]^*\right\rangle,\\
    \mathcal{D}^{(n)}_{\perp\|}(\kvec,\tau) &= -\left\langle \left[\frac{\Bmeanmag}{c} u^{(n)}_{\kvec \rm P}(\kvec, t)+\vprlhat B^{(n)}_{\kvec\rm P}(\kvec, t)\right] \left[\vprphat B^{(n)}_{\kvec\rm P}(\kvec, t+\tau)\right]^*\right\rangle,\\
    \mathcal{D}^{(n)}_{\|\|}(\kvec,\tau) &= \left\langle \left[\vprphat B^{(n)}_{\kvec\rm P}(\kvec, t)\right] \left[\vprphat B^{(n)}_{\kvec\rm P}(\kvec, t+\tau)\right]^*\right\rangle,
\end{align}
\end{subequations}
where $u^{(n)}_{\kvec\rm P}\equiv \veckcomp{u}{+}\BesselJ{n-1}-\veckcomp{u}{-}\BesselJ{n+1}$.
These coefficients are then used to obtain \eqref{eq:QLDiffusionCoefficientsAlfvenicPol} using \eqref{eq:OuterDiffusionCoeff}.

\subsubsection{Reduction to the \citet{Kennel1966-rx} and \citet{Stix1992-bo} form}\label{app:KandEReduction}

If instead of general electromagnetic fluctuations we assume waves satisfying a dispersion relation $\omega(\kvec)=\omega_{\kvec r}+\Icplx\gamma_{\kvec}$ (with frequency $\omega_{\kvec r}$ and damping rate $\gamma_{\kvec}$) such that $\Evec_1(\ppos,t) = \Eveck e^{\Icplx(\kvec\dotprod\ppos - \omega(\kvec)t)}$ and
\begin{equation}
\Evec^*_1(\kvec,t)\Evec_1(\kvec,t-\tau) = \norm{\Eveck}^2 e^{\Icplx\omega(\kvec)\tau},\label{eq:WaveCorrel}
\end{equation}
we are able to recover the classic quasi-linear diffusion of \citet{Kennel1966-rx,Stix1992-bo}.
Using this choice, Faraday's law allows us to write $\B_1$ in terms of $\Evec_1$ via $\omega\Bveck=c\kvec\crossprod\Eveck$; inserting this into \eqref{eq:InnerDiffusionCoeffs} expresses the diffusion coefficients entirely in terms of the electric-field Fourier components $\Eveck$.

With the $\tau$-dependence of the correlators contained entirely within the exponential in \eqref{eq:WaveCorrel}, the integral in \eqref{eq:OuterDiffusionCoeff} is now $\intsingle{\tau}{0}{\infty} e^{-\Icplx(\omega(\kvec)-\kprl\vprl-n\Omegai)\tau}=-\Icplx(\omega(\kvec)-\kprl\vprl-n\Omegai)^{-1}$; after using the Plemelj relation and taking the limit of zero wave damping ($\gamma_{\kvec}\to 0$) \citep{Stix1992-bo}, this reduces to the resonance condition $\upi\delta(\omega_{\kvec r}-\kprl\vprl-n\Omegai)$.

After some further algebraic manipulation, the evolution of $f_0$ can be written in the familiar form\footnote{This is also equivalent to Equation 4.1 of \citet{Kennel1966-rx} with the understanding that they do not write the Vlasov equation in divergence form initially (as done in \ref{eq:VlasovDivergence}).}
\begin{equation}
    \parfrac{f_0}{t} = \lim_{V\to\infty}\frac{\upi q^2}{m^2}\sum_{n=-\infty}^{\infty}\int\frac{\romand\kvec}{V\vprp}\mathcal{G}\left[\vprp\delta(\omega(\kvec)-\kprl\vprl-n\Omegai)\norm{\psi_{n,\kvec}}^2\mathcal{G}f_0\right].
\end{equation}
Here, $\psi_{n,\kvec} = \Eveckcomp{+}\BesselJ{n-1}+\Eveckcomp{-}\BesselJ{n+1}+(\vprl/\vprp)\Eveckcomp{z}\BesselJ{n}$ is the complex amplitude of the wave modes, with $\kappa=\kprp\vprp/\Omegai$ the argument of the Bessel functions, and the operator
\begin{equation}
    \mathcal{G} \equiv \left(1-\frac{\vprl}{\vph(\kprl)}\right)\parfrac{}{\vprp}+\frac{\vprp}{\vph(\kprl)}\parfrac{}{\vprl},\label{eq:QLDiffusionOperator}
\end{equation}
where $\vph(\kprl) \equiv \omega_{\kvec r}/\kprl$ is the parallel phase speed of the wave.
This operator constrains resonant particles to diffuse in velocity along curves of constant energy in the frame of the wave \citep{Kennel1966-rx,Stix1992-bo}.

\section{Contribution to heating from $k_\perp\rhoi\gtrsim 1$ fluctuations}\label{app:SubRhoHeating}

In Section~\ref{sec:HeatingRates}, we explicitly study the contribution of RMHD turbulent fluctuations at scales $\kprp\rhoi\ll 1$ to the perpendicular heating of ions.
In our model, we assume that $\kprp\rhoi\gtrsim 1$ fluctuations do not contribute at all to ion heating and can thus be neglected by setting the energy spectrum to zero for $\kprp\rhoi > 1$.
To test this approximation, in this Appendix we compute $\Qprp$ using the ideas of Section~\ref{sec:HeatingRates} but now including a sub-$\rhoi$ contribution.

We assume balanced turbulence and use the 2D strong turbulence spectrum of \citet{Isenberg2019-oy}, which models an Alfv\'enic cascade at scales $\kprp\rhoi\ll 1$ and a kinetic-Alfv\'en-wave (KAW) cascade at scales $\kprp\rhoi\gg 1$.
Using $\kprptilde\equiv\kprp\rhoi$ and $\kprltilde \equiv \kprl \vA/\Omegai$, this spectrum is
\begin{equation}
\energyspecEtildetwoD(\kprptilde,\kprltilde) = \frac{1}{2}\stocfracth\kprptilde^{-7/3} \left[\frac{1+\kprptilde^{5/3}}{1+\kprptilde^2}\right]\Theta\left(\kprltildeCB(\kprptilde) - |\kprltilde|\right)
\end{equation}
where 
\begin{equation}
\kprltildeCB(\kprptilde) = \stocfracth\kprptilde^{2/3} \left[\frac{1+\kprptilde^{5/3}}{1+\kprptilde^2}\right]
\end{equation}
is the boundary of the CB cone and $\Theta$ is the Heaviside step function, corresponding to a choice to only consider the effects of strong turbulence in this part.
The smoothing factor $(1+\kprptilde^{5/3})/(1+\kprptilde^2)$ goes to 1 in the limit $\kprptilde \ll 1$ and $\kprptilde^{-1/3}$ in the limit $\kprptilde \gg 1$.
When integrated over $\kprltilde$, the spectrum has the standard scalings of $\kprp^{-5/3}$ for Alfv\'enic turbulence and $\kprp^{-7/3}$ for KAW turbulence.
As in the 2D RMHD model outlined in Section~\ref{sec:SchekochihinSpectrum}, this spectrum is normalised such that
\begin{equation}
\frac{1}{\vthi^2}\int^{e^{1/2}/\rhoi}_{e^{-1/2}/\rhoi} \romand\kprp \int_{-\kprl^{\rm CB}}^{\kprl^{\rm CB}}\romand\kprl \energyspecE_{\rm 2D} \approx \stocfracth^2;
\end{equation}
i.e. such that the energy in $\rhoi$-scale modes is given by the standard stochastic heating parameter.

\begin{figure}
    \centering
    \includegraphics[width=0.75\textwidth]{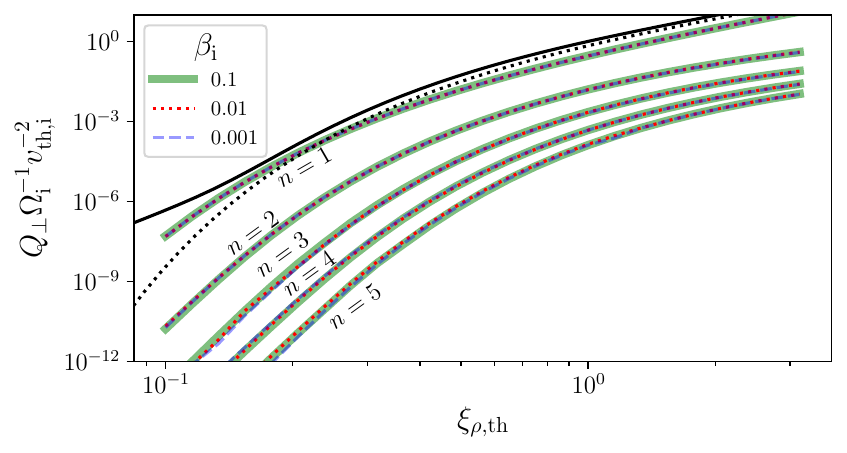}
    \caption{Perpendicular heating rates $\Qprp^{(n)}$ for $1\leq n\leq 5$, calculated using \eqref{eq:Qprpn}. The solid black line shows the contribution to $\Qprp$ from all modes calculated numerically using \eqref{eq:QprpGeneralCase} with $\crosshel=0$ and $\betai=0.1$, while the dotted line shows only the contribution from modes below the CB cone.}
    \label{fig:SubRhoiQprp}
\end{figure}

Because we are now including the effects of $\kprp\rhoi\gtrsim 1$ fluctuations, we cannot apply the RMHD limit of the Bessel functions used in the calculations of Section~\ref{sec:HeatingRates} (namely $\BesselJ{1}^2(\kappa)/ \kappa^2 \approx 1/4$ and $n^2\BesselJ{n}^2(\kappa)/\kappa^2\approx 0$ for $n>1$, where $\kappa=\kprp\vprp/\Omegai$ is the argument of the Bessel functions).
We assume for simplicity that the KAWs are polarised similarly to Alfv\'enic fluctuations and that the $\ExB$ flow is the dominant contribution to their electric field via Ohm's law (these are true for $\betai \ll 1$; \citealp{Schekochihin2009-qo}), and that the ions are in an initial Maxwellian distribution.
With this, we calculate the heating rate as a sum over the individual Bessel function terms as $\Qprp = \sum_{n>0}\Qprp^{(n)}$, where 
\begin{equation}
    \Qprp^{(n)} \equiv \frac{4}{\sqrt{\upi\betai^5}}\intsingle{\vprltilde}{-\infty}{\infty}\intsingle{\vprptilde}{0}{\infty}e^{-(\vprptilde^2+\vprltilde^2)/\betai}\vprptilde^3\mathcal{D}^{(n)}\label{eq:Qprpn}
\end{equation}
is the perpendicular heating rate \eqref{eq:QprpDprp}, and
\begin{align}
    \frac{\mathcal{D}^{(n)}}{\Omegai\vthi^2}&\equiv \frac{\upi}{4}\sum_{\nu=\pm1}\times\nonumber\\
    &\intsingle{\kprptilde}{0}{\infty}\frac{n^2\BesselJ{n}^2(\kappa)}{\kappa^2}\intsingle{\kprltilde}{-\kprltildeCB(\kprptilde)}{\kprltildeCB(\kprptilde)}\frac{\energyspecEtildetwoD(\kprptilde,\kprltilde)}{2\omeganltilde(\kprptilde)}\sech\left(\frac{\upi}{2\omeganltilde(\kprptilde)}\left[\kprltilde(\vprltilde+\nu)+1\right]\right)
\end{align}
is the generalised diffusion coefficient \eqref{eq:DppGeneralIntegrand} with $\crosshel=0$ and the Bessel functions (and thus $\kprp\rhoi\gtrsim 1$ behaviour) restored.

Figure~\ref{fig:SubRhoiQprp} shows $\Qprp^{(n)}$, calculated numerically using \eqref{eq:Qprpn}, for $1\leq n\leq 5$ and $\betai = 0.001$, $0.01$, and $0.1$.
The heating rate shows little to no dependence on $\betai$, which is also seen in the balanced RMHD turbulence case of Section~\ref{sec:balancedlimit} (cf. figure~\ref{fig:QprpBalanced}).
Additionally, we clearly see that the main contributor to $\Qprp$ is the $n=1$ Bessel function term $\BesselJ{1}^2(\kappa)/\kappa^2$, as it is approximately $1/4$ for $\kprp\rhoi\ll 1$ and quickly drops to 0 for $\kprp\rhoi\gtrsim 1$, biasing the integral over the energy spectrum $\energyspecEtildetwoD$ to the larger-amplitude $\kprp\rhoi\ll 1$ fluctuations.
In contrast, the $n>1$ Bessel function terms $n^2\BesselJ{n}^2(\kappa)/\kappa^2$ are approximately 0 for $\kprp\rhoi\ll 1$ and peak at $\kprp\rhoi\gtrsim 1$, biasing the integral over the energy spectrum $\energyspecEtildetwoD$ to the smaller-amplitude $\kprp\rhoi\gg 1$ fluctuations.
To compare with the main-text calculation where only $\kprp\rhoi <1$ modes are used, we also show $\Qprp$ calculated numerically using \eqref{eq:QprpGeneralCase} with $\crosshel=0$ and $\betai=0.1$ (as in figure~\ref{fig:DppGeneral}), with the contribution from all modes as well as that from modes below the CB cone taken into account.
This is qualitatively similar to $Q^{(1)}_\perp$, with the smaller values of $Q^{(1)}_\perp$ at large $\stocfracth$ arising due to the variation of the Bessel function near $\kprp\rhoi\sim 1$, and a weaker suppression at small $\stocfracth$ due to the inclusion of modes with $\kprp\rhoi\gtrsim 1$; however, the contribution from the $\kprp\rhoi\gtrsim 1$ modes at $\stocfracth \ll 1$ to the heating rate is overall smaller than that from $\kprp\rhoi< 1$ fluctuations.
These results show that our approach of neglecting the contribution of $\kprp\rhoi\gtrsim 1$ fluctuations to ion heating and taking the RMHD limit in Section~\ref{sec:HeatingRates} is valid.

We only consider balanced turbulence in this calculation as imbalanced turbulence can introduce a helicity barrier at $\kprp\rhoi\sim 1$ scales in the low-$\betai$ plasmas we consider in this paper, which blocks the cascade of energy to small perpendicular scales and strongly modifies the energy spectrum of turbulent fluctuations \citep{Meyrand2021-ix,Squire2022-dm,Squire2023-jn,Adkins2024-iw,Adkins2025-by,Johnston2025-ss}.
Because this modification depends on the level of imbalance and other parameters of the turbulence, further work would be required to generalise the approach of this Appendix to different levels of imbalance.

\section{Dependence of heating rate on choice of temporal correlation function}\label{app:TimeCorrCompare}

As discussed in Section~\ref{sec:HeatingRates}, the choice of the function $f(\tau\omeganl)$, which characterises the temporal correlations of the fluctuations, can significantly influence the resulting heating rate.
To ensure consistency with observations and theoretical expectations, this function must therefore be chosen with care.
A formal renormalisation group calculation of space–time correlations in hydrodynamic turbulence, as well as phenomenological arguments, show that $f(\tau\omeganl)$ approaches an exponential form, $e^{-|\tau\omeganl|}$, for $\tau\omeganl \gg 1$, and a Gaussian form, $e^{-(\tau\omeganl)^2}$, for $\tau\omeganl \ll 1$ \citep{Gorbunova2021-bh}.
Thus, this provides important constraints on the physically allowable forms of $f(\tau\omeganl)$.
In this Appendix, we explicitly illustrate the impact of this choice by calculating $\Qprp$ in the balanced limit (Section~\ref{sec:balancedlimit}) using three different temporal correlation functions that capture different aspects of these limiting behaviours.

\begin{figure}
    \centering
    \includegraphics[width=\textwidth]{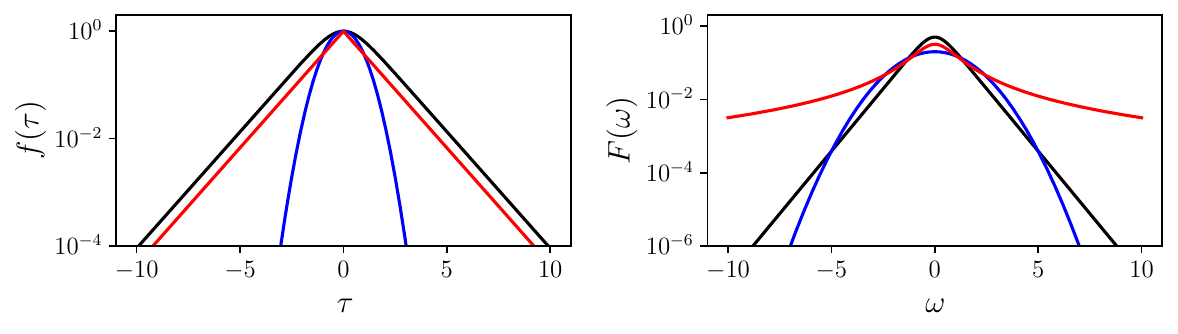}
    \caption{Time correlation functions $f(\tau)$ (left) and their temporal Fourier transforms $F(\omega)$ (right), with $f(\tau)=\sech(\alpha\tau)$ (black; \ref{eq:sechCorrel}), $\exp(-(\alpha\tau)^2)$ (blue; \ref{eq:gaussCorrel}), $\exp(-|\alpha\tau|)$ (red; \ref{eq:expCorrel}); $\alpha$ is set to 1 for all functions.}
    \label{fig:TimeCorrelFTs}
\end{figure}

Using the Fourier transform convention $F(\omega)= \frac{1}{2\upi}\intsingle{\tau}{-\infty}{\infty}e^{-\Icplx\omega\tau}f(\tau)$, we use the following correlation functions with their Fourier transform pairs
\begin{align}
    f(\tau)=\sech(\alpha\tau)&\to F(\omega)=\frac{1}{2\alpha}\sech\left(\frac{\upi\omega}{2\alpha}\right)\label{eq:sechCorrel},\\
    f(\tau)=e^{-(\alpha\tau)^2}&\to F(\omega)=\frac{1}{2\alpha\sqrt{\upi}}e^{-\omega^2/(4\alpha^2)},\label{eq:gaussCorrel}\\ 
    f(\tau)=e^{-|\alpha \tau|}&\to F(\omega) = \frac{\alpha}{\upi(\alpha^2+\omega^2)},\label{eq:expCorrel}
\end{align}
which are compared in figure~\ref{fig:TimeCorrelFTs}.
The tails of the Fourier transforms of \eqref{eq:sechCorrel} (used in this paper) and \eqref{eq:gaussCorrel} both decay quickly for large frequencies.
In contrast, the tails of the Fourier transform of \eqref{eq:expCorrel} decay much slower, which is a result of the discontinuity in slope of the correlation function at $\tau=0$.

\begin{figure}
    \centering
    \includegraphics[width=0.75\textwidth]{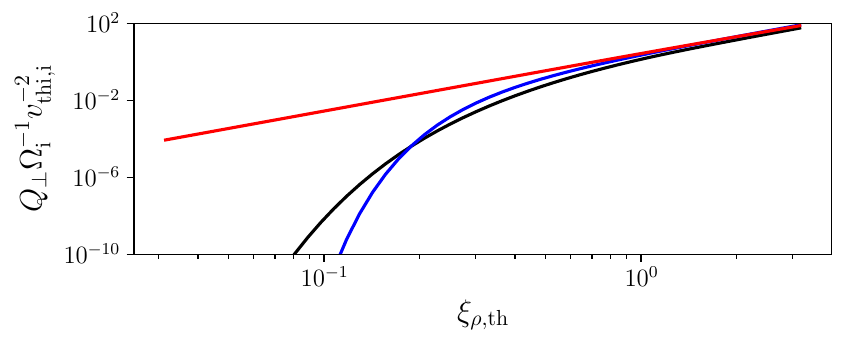}
    \caption{Perpendicular heating rate $\Qprp$ in balanced turbulence with $\betai=0.1$, using the temporal correlation functions $\sech(\omeganl\tau)$ (black; \ref{eq:sechCorrel}), $\exp(-(\omeganl\tau)^2)$ (blue; \ref{eq:gaussCorrel}), $\exp(-|\omeganl\tau|)$ (red; \ref{eq:expCorrel}).
    To better show the suppression, only the contribution from modes beneath the CB cone are considered in the calculation of $\Qprp$.}
    \label{fig:TimeCorrelHeating}
\end{figure}

It is the behaviour of these tails as $|\omega|\to\infty$ that can affect the form of the resultant heating rate.
This is seen in figure~\ref{fig:TimeCorrelHeating}, which shows $\Qprp$ in the balanced limit, numerically calculated using \eqref{eq:QprpGeneralCase} with $\betai=0.1$ and $\crosshel=0$, using the different correlation functions in \eqref{eq:DppGeneralIntegrand}.
Because the Fourier transforms of \eqref{eq:sechCorrel} and \eqref{eq:gaussCorrel} have similar behaviour at large $\omega$ the corresponding heating rates are qualitatively similar, exhibiting a suppression in heating at small $\stocfracth$.\footnote{It should be noted that the heating-rate suppression associated with the Gaussian correlation function---which can be shown to be of the form $\exp{(-1/\stocfracth^2)}$---is considerably stronger than that of \eqref{eq:sechCorrel}, which follows from the fact that the tails of the Fourier transform in \eqref{eq:gaussCorrel} decay much more rapidly than those of \eqref{eq:sechCorrel} (as shown in figure~\ref{fig:TimeCorrelFTs}).
This behaviour is not physically appropriate for modelling heating-rate suppression, which empirical predictions based on magnetic-moment conservation as well as numerical and theoretical results indicate should be of the form $\exp{(-1/\stocfracth)}$ \citep{Chandran2010-ow,Xia2013-ob,Hoppock2018-fv,Johnston2025-ss,Mallet2026-ei}.}

\begin{figure}
    \centering
    \includegraphics[width=\textwidth]{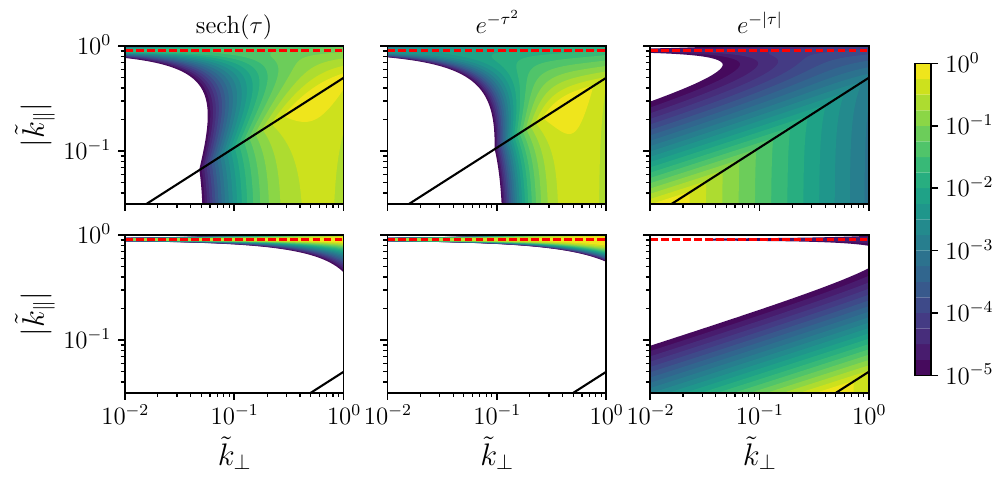}
    \caption{The integrand of \eqref{eq:DppGeneralIntegrand}, normalised to its maximum value, plotted with $\vprltilde=0.1$, $\crosshel=0$, and correlation function $\sech(\omeganl\tau)$ (left; \ref{eq:sechCorrel}), $\exp(-(\omeganl\tau)^2)$ (middle; \ref{eq:gaussCorrel}), $\exp(-|\omeganl\tau|)$ (right; \ref{eq:expCorrel}); the top row sets $\stocfracth=0.5$, and the bottom $\stocfracth=0.05$. The black line corresponds to the CB cone $\kprltilde^{\rm CB} = \stocfracth\kprptilde^{2/3}$. The horizontal red dashed line is $\kprltilde^{(1)}$, the resonant $\kprl$ that ions interact with in the imbalanced limit \eqref{eq:KprlResonance}. Note that we are looking at the $\kprltilde < 0$ modes, corresponding to $\zp$ fluctuations.}
    \label{fig:TimeCorrelKSpace}
\end{figure}

When using the exponential correlation function \eqref{eq:expCorrel}, on the other hand, the heating rate shows no transition in scaling or suppression when only considering modes beneath the CB cone.
The cause of this can be understood by inspecting the integrand of \eqref{eq:DppGeneralIntegrand} for different values of $\stocfracth$, as shown in figure~\ref{fig:TimeCorrelKSpace} (with $\vprltilde=0.1$ and $\crosshel=0$ for all correlation functions).
For $\stocfracth = 0.5$ (top row), the integrand peaks near the CB cone for all correlation functions, allowing the large-amplitude fluctuations to dominate the integral over $\kvec$ and giving rise to a $\stocfracth^3$ scaling in $\Qprp$ (as also seen in Section~\ref{sec:HeatingRates}).
For smaller amplitudes ($\stocfracth = 0.05$; bottom row), the Fourier transforms of the correlation functions \eqref{eq:sechCorrel} and \eqref{eq:gaussCorrel} peak when their argument is 0 (in this case, when $|\kprltilde| = (1+\vprltilde)^{-1}\approx 0.9$) and decay quickly, causing only modes above the CB cone to contribute to the integral and giving rise to the $\stocfracth^6$ scaling (and suppression when only considering modes beneath the CB cone).
Although the Fourier transform of \eqref{eq:expCorrel} also exhibits a peak at $|\kprltilde| = (1+\vprltilde)^{-1}$, its slowly decaying tails also allow it to contribute in regions below the CB cone (where $\kprltilde/\omeganltilde = \kprltilde\stocfracth^{-1}\kprptilde^{-2/3}\lesssim 1$) even at small amplitudes, giving rise to the continuous $\stocfracth^3$ scaling and no heating suppression.
The exponential correlation function is commonly used in modelling the temporal correlation of turbulence; however, its results are clearly unphysical, illustrating the care needed when choosing $f(\tau\omeganl)$ in models of particle heating by turbulent spectra.

\section{Effect of ion-cyclotron dispersion on the resonance location}
\label{app:ICWHeating}

In our quasi-linear calculation we have assumed that the turbulent fluctuations satisfy the MHD Alfvén-wave dispersion relation, $\omega = k_\parallel v_{\rm A}$, throughout the cascade. 
This is an excellent approximation for $\kprp\rhoi\ll1$, but it breaks down at the parallel scales that are expected to produce the strongest cyclotron heating for protons, $k_\parallel d_{\rm p} \sim 1$ (with $d_{\rm p}$ the proton inertial length), where the Alfvén branch becomes dispersive and transitions into the oblique ion-cyclotron wave (ICW).  As a result, using the nondispersive Alfvén dispersion relation misplaces the proton cyclotron resonance in $k_\parallel$ and may bias the heating rate.  In this Appendix we give a simple estimate of how replacing the Alfvén dispersion relation by an oblique ICW dispersion modifies the parallel wavenumber of the resonance.  Our treatment is deliberately schematic; a quantitatively accurate assessment would require a full kinetic description of the turbulence, for example via detailed analysis of hybrid-kinetic simulations.  We also note that this issue is less severe for minor ions: because their cyclotron frequencies $\Omega_{\rm i} = (q_{\rm i}/m_{\rm i}) B_0/c$ are smaller than $\Omega_{\rm p}$, their fundamental cyclotron resonances with $v_\parallel\approx0$ typically occur at $k_\parallel d_{\rm i} \ll 1$, where the Alfvén branch is still only weakly dispersive.

\begin{figure}
\begin{center}
\includegraphics[width=0.6\textwidth]{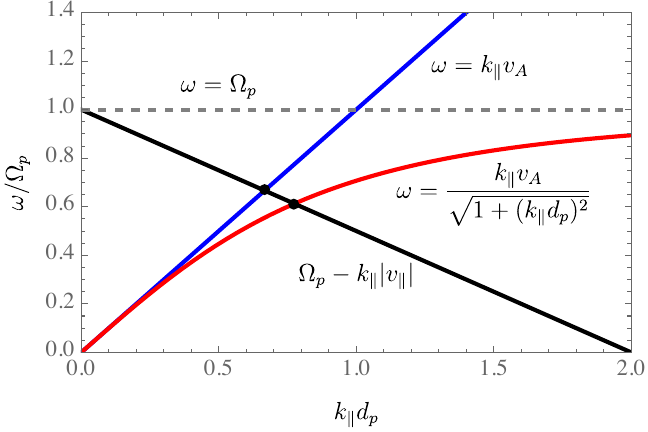}
\caption{Comparison of the resonant parallel wavenumber obtained from the quasi-linear resonance condition \eqref{eq:proton-resonance} when using the Alfv\'en-wave dispersion relation (blue), and the oblique limit of ICWs at $\betai\ll1$ (red).
The intersection between these curves and the line $\omega = \Omegap - \kprl|\vprl|$ (black, illustrated for a proton with $\vprl<0$), where \eqref{eq:proton-resonance} is satisfied, gives the resonant wavenumber.}
\label{fig:ICW-resonance}
\end{center}
\end{figure}

For protons interacting via the fundamental cyclotron resonance, the resonance condition can be written as
\begin{equation}
 k_{\|} v_{\rm ph}(k_{\|}) = k_\parallel v_\parallel +\Omega_{p},
  \label{eq:proton-resonance}
\end{equation}
where $v_\parallel$ is the proton parallel velocity and $v_{\rm ph}(k_{\|})=\omega/k_{\|}$ (we take $n=1$ in \ref{eq:ZpResonance}, which produces by far the strongest resonance).  In figure~\ref{fig:ICW-resonance} we illustrate this condition by plotting the two curves
$k_{\|} v_{\rm ph}(k_{\|})$ and $\Omega_{\rm p} - k_\parallel |v_\parallel|$ as functions of $k_\parallel$ (taking particles with $v_\parallel<0$), so that their intersection gives the resonant parallel wavenumber.  We compare Alfv\'en waves $v_{\rm ph} = v_{\rm A}$ (as assumed throughout the main text), to the oblique limit of ICWs at  $\betai\ll1$, which satisfy 
  \begin{equation}
v_{\rm ph}(k_\parallel) =\pm \frac{ v_{\rm A}}{\sqrt{1 + k_\parallel^2 d_{\rm p}^2}}\label{eq: vicw}
\end{equation}
\citep{Stix1992-bo}. The latter asymptotes to $\omega \to \Omega_{\rm p}$ as $k_\parallel d_{\rm p} \to \infty$.  This illustrates how using the true ICW $v_{\rm ph}$  shifts the resonant wavenumber to larger $k_\parallel d_{\rm p} $. Naively, one would expect less wave power in the higher-$k_{\|}$ region and therefore a reduced quasi-linear heating rate.  However, we now argue that this conclusion   is modified once the critical-balance structure of the turbulence is taken into account, yielding the opposite conclusion.

The key point is that, under critical balance, the parallel structure of the turbulence at a given $k_\perp$ is set by decorrelation of the cascade rather than by linear physics alone.  Writing the linear phase speed as $v_{\rm ph}(k_\parallel)=\omega/k_\parallel$, critical balance implies
\begin{equation}
  k_\parallel v_{\rm ph}(k_\parallel) \sim \tau_{\rm nl}^{-1}(k_\perp),
  \label{eq:cb-vph}
\end{equation}
where $\tau_{\rm nl}(k_\perp)^{-1}\sim k_\perp \delta u_\perp(k_\perp)$ is the nonlinear time.  We assume that the perpendicular spectrum and nonlinearity are largely unchanged by the onset of Hall and ion–cyclotron physics: Hall MHD simulations at low $\betai$ show that a $k_\perp^{-5/3}$  velocity spectrum persists into the Hall range \citep{Meyrand2018-wy}, so that $\delta u_\perp\propto k_\perp^{-1/3}$ and $\tau_{\rm nl}\propto k_\perp^{-2/3}$ likely remain reasonable scalings, even when $k_\parallel d_{\rm p}\sim1$ and the waves are dispersive.

To assess which particles resonate with fluctuations at a given $k_\perp$, we combine \eqref{eq:cb-vph} with the  resonance condition \eqref{eq:proton-resonance}. 
For nondispersive Alfvén waves, $v_{\rm ph}=v_{\rm A}$ and $k_\parallel = 1/(v_{\rm A}\tau_{\rm nl})$.  Substituting into \eqref{eq:proton-resonance} then yields
\begin{equation}
  \frac{|v_\parallel^{\rm (AW)}|}{v_{\rm A}}
    =  \Omega_{\rm p}\,\tau_{\rm nl}(k_\perp)-1.
  \label{eq:vpar-aw}
\end{equation}
Thus thermal protons with $|v_\parallel|\lesssim v_{\rm th}\lesssim v_{\rm A}$ can only resonate where $\Omega_{\rm p}\tau_{\rm nl}(k_\perp)$ approaches unity, i.e. in a relatively narrow band of $k_\perp$ at small scales.  This behaviour is encoded in the main-text calculation of Section~\ref{sec:imbalancedlimit} and the sharp cutoff of the heating rate in the imbalanced case when fluctuations above the critical balance cone are excluded (see figure~\ref{fig:QprpImbalanced}).

For oblique ICW waves,  $v_{\rm ph}<v_{\rm A}$, so that the $k_{\|}$ shifts higher for a given $k_{\perp}$.  Using $  v_{\rm ph}(k_\parallel) $ from \eqref{eq: vicw}, equations \eqref{eq:cb-vph} and \eqref{eq:proton-resonance}  give
\begin{equation}
  k_\parallel d_{\rm p} = \frac{1}{\sqrt{(\Omega_{\rm p}\tau_{\rm nl})^2-1}},
  \qquad
  \frac{|v_\parallel^{\rm (ICW)}|}{v_{\rm A}}
    = \bigl(\Omega_{\rm p}\tau_{\rm nl}-1\bigr)\sqrt{1-(\Omega_{\rm p}\tau_{\rm nl})^{-2}}.
  \label{eq:vpar-icw}
\end{equation}
Thus, the resonant velocity for ICWs is
\begin{equation}
  \bigl|v_\parallel^{\rm (ICW)}\bigr|
    = \bigl|v_\parallel^{\rm (AW)}\bigr|\,
     \sqrt{1-(\Omega_{\rm p}\tau_{\rm nl})^{-2}},
\end{equation}
which is always \emph{smaller} than in the purely Alfvénic case at the same $\tau_{\rm nl}(k_\perp)$.  Equivalently, with dispersive waves, thermal protons with $|v_\parallel|\lesssim v_{\rm th}$ can resonate with fluctuations at somewhat lower $k_\perp$ (larger $\tau_{\rm nl}$, where more power resides), even though the cyclotron resonance itself has moved to larger $k_\parallel$.  In this sense, the increased power at high $k_\parallel$ implied by critical balance in the ICW case more than compensates for the shift of the resonance to higher $k_\parallel$, suggesting that neglecting dispersive effects may in fact {\em underestimate} the proton heating rate.

The above discussion assumes that the ion heating is purely perpendicular to the magnetic field.
However, the dispersive nature of the ICWs modifies the quasi-linear resonance contours along which ions diffuse in velocity space \citep{Chandran2010-ny,Squire2022-dm,Zhang2025-yd}, which may lead to more parallel heating (or cooling) and reduces perpendicular heating.
It is unclear whether this effect would dominate over the mechanism discussed with our schematic treatment; a kinetic description of the turbulence would be required to fully understand the effect of these processes on ion heating.

\bibliographystyle{jpp}
\bibliography{main}

\end{document}